\documentclass[fleqn,usenatbib]{mnras}

\usepackage{newtxtext,newtxmath}

\usepackage[T1]{fontenc}

\DeclareRobustCommand{\VAN}[3]{#2}
\let\VANthebibliography\thebibliography
\def\thebibliography{\DeclareRobustCommand{\VAN}[3]{##3}\VANthebibliography}

\usepackage{graphicx}	
\usepackage{amsmath}	
\usepackage{subfig}
\usepackage{array}
\hypersetup{colorlinks=true,citecolor=blue,linkcolor=black, urlcolor=black}

\title[GRB data processing for magnetic propellers]{An analysis of the effect of data processing methods on magnetic propeller models in short GRBs}

\author[T. R. L. Meredith et al.]{
Tomos R. L. Meredith,$^{1}$\thanks{E-mail: trlm1@le.ac.uk (TRLM)}
Graham A. Wynn,$^{2}$
Philip A. Evans$^{1}$
\\
$^{1}$School of Physics and Astronomy, University of Leicester, University Road, Leicester, LE1 7RH, UK
\\
$^{2}$Department of Mathematics, Physics and Electrical Engineering, Northumbria University, Newcastle upon Tyne, NE1 8ST, UK
}

\date{Accepted XXX. Received YYY; in original form ZZZ}

\pubyear{2022}

\begin{document}
\label{firstpage}
\pagerange{\pageref{firstpage}--\pageref{lastpage}}
\maketitle

\begin{abstract}
We present analysis of observational data from the \textit{Swift} Burst Analyser for a sample of 15 short gamma-ray bursts with extended emission (SGRBEEs) which have been processed such that error propagation from \textit{Swift}'s count-rate-to-flux conversion factor is applied to the flux measurements. We apply this propagation to data presented by the Burst Analyser at 0.3-10 keV and also at 15-50 keV, and identify clear differences in the morphologies of the light-curves in the different bands. In performing this analysis with data presented at both 0.3-10 keV, at 15-50 keV, and also at a combination of both bands, we highlight the impact of extrapolating data from their native bandpasses on the light-curve. We then test these data by fitting to them a magnetar-powered model for SGRBEEs, and show that while the model is consistent with the data in both bands, the model's derived physical parameters are generally very loosely constrained when this error propagation is included and are inconsistent across the two bands. In this way, we highlight the importance of the \textit{Swift} data processing methodology to the details of physical model fits to SGRBEEs.
\end{abstract}




\section{Introduction}

Gamma-ray bursts (GRBs) are the brightest electromagnetic phenomena in the universe, with durations between fractions of a second and thousands of seconds \citep{Horvath2016Durations}. GRBs are observed to have a bimodal distribution in both temporal extent and their spectra, such that they are considered to be two separate populations, referred to as short GRBs (SGRBs) and long GRBs (LGRBs), or even as short-hard GRBs and long-soft GRBs. The classification of GRBs is usually made based on their $T_{90}$ - the interval in which 90\% of the flux is detected, with SGRBs typically having $T_{90}<2$s and LGRBs having $T_{90}>2$s.

In reality, there is some overlap between these populations, in part caused by a subclass of SGRBs that have a long-lasting late-time plateau or late-time rebrightening in their lightcurves, termed extended emission (EE) \citep{Norris_2005}. \citet{BostanciBATSESGRBEE} found evidence of EE in around 7\% of SGRBs detected by BATSE, while \citet{KanekoSGRBEE2015} estimated that EE was present in around 15\% of SGRBs detected by \textit{Swift}/BAT, and around 5\% of SGRBs detected by \textit{Fermi}/GBM. While the $T_{90}$ values of GRBs exhibiting extended emission are typically consistent with those of LGRBs, their spectrally hard prompt emission is consistent with SGRBs with an additional, softer component \citep{Norris2006SGRBEE}. The total fluence of EE can be comparable (or greater) than that of the prompt emission, thus requiring a substantial energy budget \citep{Perley2009EE}. EE complicates the question of modelling SGRBs, as models proposed must provide a framework to explain the origin of EE, but only in a small fraction of GRBs.

With SGRBs widely belived to to be produced during the course of the merger of binary neturon-star systems, as evidenced by GW170817/GRB170817A \citep{Abbott_2017}, some models for the late-time emission from SGRBs assume that some of these merger events produce a remnant magnetar (e.g. \citet{Gompertz1,SuvorovPrecessingMag2020}). In such models, the magnetar's dipole spin-down is invoked to inject additional energy post-merger, as a magnetar rotating with a period of $\sim 1$ ms generally has sufficient energy budget to match EE fluence, where the isotropic-equivalent total fluences are estimated at roughly $10^{51}$ erg.

Magnetar spin-down can generally explain long-lasting plateaus, however a great deal of investment has been made into developing models that can also explain late-time rebrightenings observed in GRB light-curves such as flares. Works such as \citet{Gompertz2} and \citet{Gibson1} for example proposed magnetar models where the magnetic field interacted with a disc of surrounding ejecta to allow these models to incorporate rebrightenings into EE.

However, as observational data for GRBs are often very noisy, it is pertinent to first consider the question of whether or not these flares truly are real features in the lightcurve as opposed to artefacts of the way in which data are processed. GRB light-curves often have poorly constrained spectral data and the errors must be accounted for, lest we obtain highly volatile light-curves. Previous fits to observational data often found themselves tightly constrained in some regions of parameter space by the need to match rapidly-varying data which are sensitive to the algorithm used for processing the data. They additionally featured data from different observational instruments with different bandpasses which were extrapolated to the same band. As such, previous results may be under-constrained and over-extrapolated since they lacked consideration for error propagation and the effects of the extrapolation process.

In this work, we present an analysis of the processing done to the observational data to which the models of \citet{Gompertz2,Gibson1} were fit. We do this by using data which has been extrapolated to different bandpasses and altering the processing algorithm before fitting the models to the observational data. We highlight the dependence of the fit quality and best-fitting parameters on the method by which the data is processed. From this, we show that tightly fine-tuned parameters with extreme values within permitted parameter space as identified by previous works may no longer be necessary to accurately fit the data, perhaps rendering such models of EE more feasible, but less well constrained. We also demonstrate that the statistical significance of fluctuations in the light-curve of GRBs may be overstated due to the way in which the data is processed, thus leading to potentially spurious features in lightcurves which may have contributed to an overestimation of the prevalence of substructure in late-time GRB data.

\subsection{The \textit{Swift} Burst Analyser}

The data we use were taken by the Neil Gehrels Swift Observatory (\textit{Swift}, \citet{Gehrels_2004}), launched in 2004 as a dedicated multiwavelength platform for rapid observations of GRBs. \textit{Swift} features three instruments for capturing a GRB between hard X-rays and the optical band - the Burst Alert Telescope (BAT, \citet{Barthelmy2005BAT}), which has a bandpass of 15-150 keV; the X-Ray Telescope (XRT, \citet{Burrows2005XRT}), with a bandpass of 0.3-10 keV; and the Ultra-Violet/Optical Telescope (UVOT, \citet{Roming2005UVOT}). All flux and photon index data were taken from the Burst Analyser from the UK \textit{Swift} Science Data Centre (UKSSDC, \citet{EvansSwift2010}). The UKSSDC presents separate unabsorbed flux curves for the BAT data and XRT data, as well as observed flux conversion factors and photon index measurements $\Gamma$ corresponding to each flux data point.

Natively, light curves produced by X-ray satellites are in count-rate units, i.e. they report the rate at which photons are detected over the (broad) band-pass of the instrument. In contrast, physical models predict energy flux. To convert the measured count-rate into flux requires knowledge of both the source spectrum and the instrumental response; from this one can construct a time-evolving energy conversion factor (ECF), a number by which one multiplies the count-rate to get the flux. Measuring the ECF accurately requires much more data than simply measuring the count-rate, thus one has to either use large time bins, losing information about temporal variability; or allow the ECF to be measured with lower time resolution than the count-rate light curve, bringing complications into the propagation of uncertainties. 

The Swift Burst Analyser\footnote{\url{https://www.swift.ac.uk/burst_analyser}.} \citep{EvansSwift2010} presents flux light curves of GRBs detected by \textit{Swift}, created using the latter philosophy. A time series of the ECF (and its uncertainty) is created with lower time resolution than the count-rate light curve, and the ECF for a given light-curve bin is found by interpolation on this time series. The ECF time series is created using hardness ratios data, converted to ECF using a look-up table generated in XSPEC \citep{Arnaud1996XSPEC}. This allows a greater time-resolution than would be possible by generating and fitting spectra, at the expense of larger uncertainties.

As noted above, the drawback to decoupling the binning of the ECF and the count-rate light curve lies in the error propagation. This is best demonstrated by an (exaggerated) example: consider two light curve bins, with count-rates differing by a factor of two (and small errors), indicating a genuine change in source intensity. Both bins fall within the same ECF time bin, and the ECF has a factor of two uncertainty. If the ECF uncertainty is propagated into the flux light curve, there would be no evidence of this significant temporal variability in the flux. At the other extreme, take two light curve bins with identical count rates (and small errors), but in this case falling in separate ECF bins. Imagine the two ECFs differ by a factor of two, but with large errors so that they are easily consistent with each other. In this case, the ECF error must be propagated otherwise the two light curve bins will appear to have flux differences with strong statistical significance. As discussed by \citet{EvansSwift2010}, the ECF uncertainties are not propagated onto the flux bins shown in the burst analyser plots (so as to retain visibility of genuine features), but are provided so that users can propagate them or not, as appropriate for their use cases. In this paper we demonstrate that the decision in previous works not to apply this propagation artificially increases the significance of certain physical measurements.

This work is structured as follows. Section 2 introduces the observation data used, outlines our processing routines, as well as how they differ from those of previous works. Section 3 summarises the physical model of \citet{Gibson1} that we are using to test our data processing. Section 4 outlines the fitting routine for the model to the data. Section 5 features the results and discussion of the fits obtained thus far.

\section{Observational data}

We adopted the same sample of SGRBs as \citet{Gibson1}, comprising of SGRBs identified by \citet{Gompertz1,Gompertz2,KanekoSGRBEE2015} as displaying extended emission, as well as GRB150424A and GRB160410A, for a total of 15 SGRBs. The UKSSDC presents separate unabsorbed flux curves for the BAT data and XRT data, as well as observed flux conversion factors and photon index measurements $\Gamma$ corresponding to each flux data point as discussed above. In this work, we utilised both the BAT data and XRT data together. The ECF values provided by the UKSSDC can allow for BAT and XRT data to be extrapolated to the same bandpass, allowing for easy comparison between them.

In addition to providing unabsorbed flux curves for the BAT data and the XRT data, the UKSSDC also provides combined BAT+XRT flux curves, wherein data from both instruments are presented together at either 0.3-10 keV, 15-50 keV, or as flux densities at 10 keV. As such, any data presented in a bandpass differing from that of its instrument's native bandpass (e.g. BAT data presented as 0.3-10 keV flux) must be extrapolated into the new bandpass using the ECF.

Previous works by \citet{Gompertz2,Gibson2} used a combined BAT+XRT dataset extrapolated to 0.3-10 keV, therefore their BAT data were extrapolated from their native bandpass while the XRT data were not. After the data underwent cosmological $k$-correction and absorption correction, the bolometric luminosities found in their work had to be calculated. By treating 1-10000 keV flux as a proxy for the bolometric flux, and by assuming the spectral energy distribution can be described by a single power-law with photon index $\Gamma$ (taken to be the XRT data's average $\Gamma$ value), we can use $\Gamma$ to calculate the fraction of the bolometric flux contained within the 0.3-10 keV bandpass. This fraction, combined with the 0.3-10 keV flux, yields the bolometric luminosity.

The assumption of a single photon index is a significant one - the same assumption of a consistent spectral shape is made when the UKSSDC extrapolates each data point from its native bandpass. The existence of spectral breaks between 0.3-10 keV and 15-50 keV would contaminate the results of this extrapolation process, with similar issues emerging if spectral breaks emerge in the 1-10000 keV range we consider to be the bolometric luminosity proxy. However, for the extrapolation between the 0.3-10 keV and 15-50 keV bandpasses at least, we are able to mitigate this problem for data that are not subjected to that extrapolation (i.e. XRT data presented at 0.3-10 keV or BAT data presented at 15-50 keV).

There exist some (generally small) discrepancies in the combined BAT+XRT data compared to viewing each instrument's datasets separately - BAT datapoints with $t<0$ are discarded in the combined dataset and therefore the raw BAT data is binned slightly differently when generating the flux data \citep{EvansSwift2010}. To retain as much data as possible, we opted to use both the BAT and XRT files, as opposed to the combined BAT+XRT file generated by the UKSSDC.

\begin{figure*}
    \centering
    \addtolength{\tabcolsep}{-0.4em}
    \begin{tabular}{cc}
    \includegraphics[width=0.9\columnwidth]{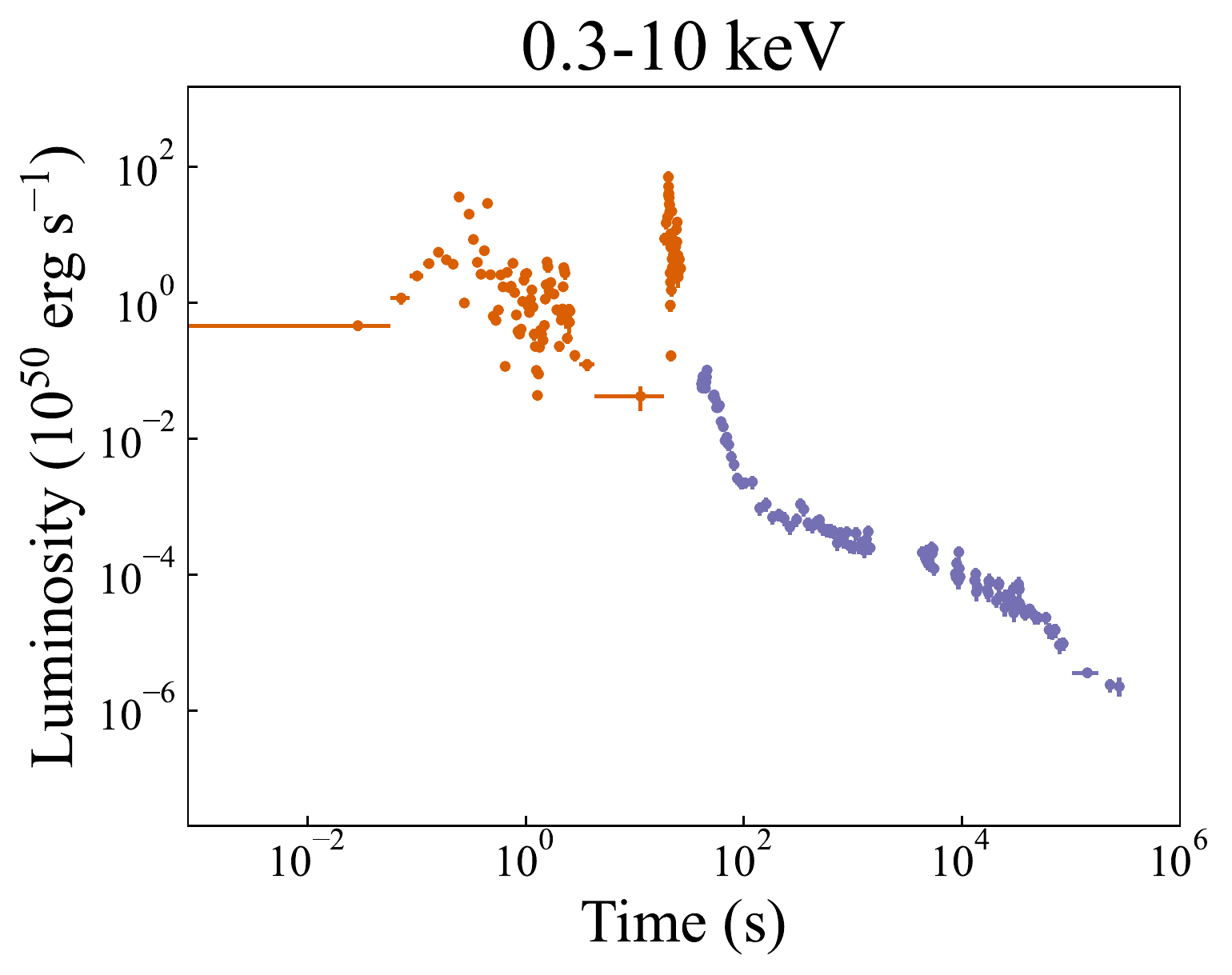} &
    \includegraphics[width=0.9\columnwidth]{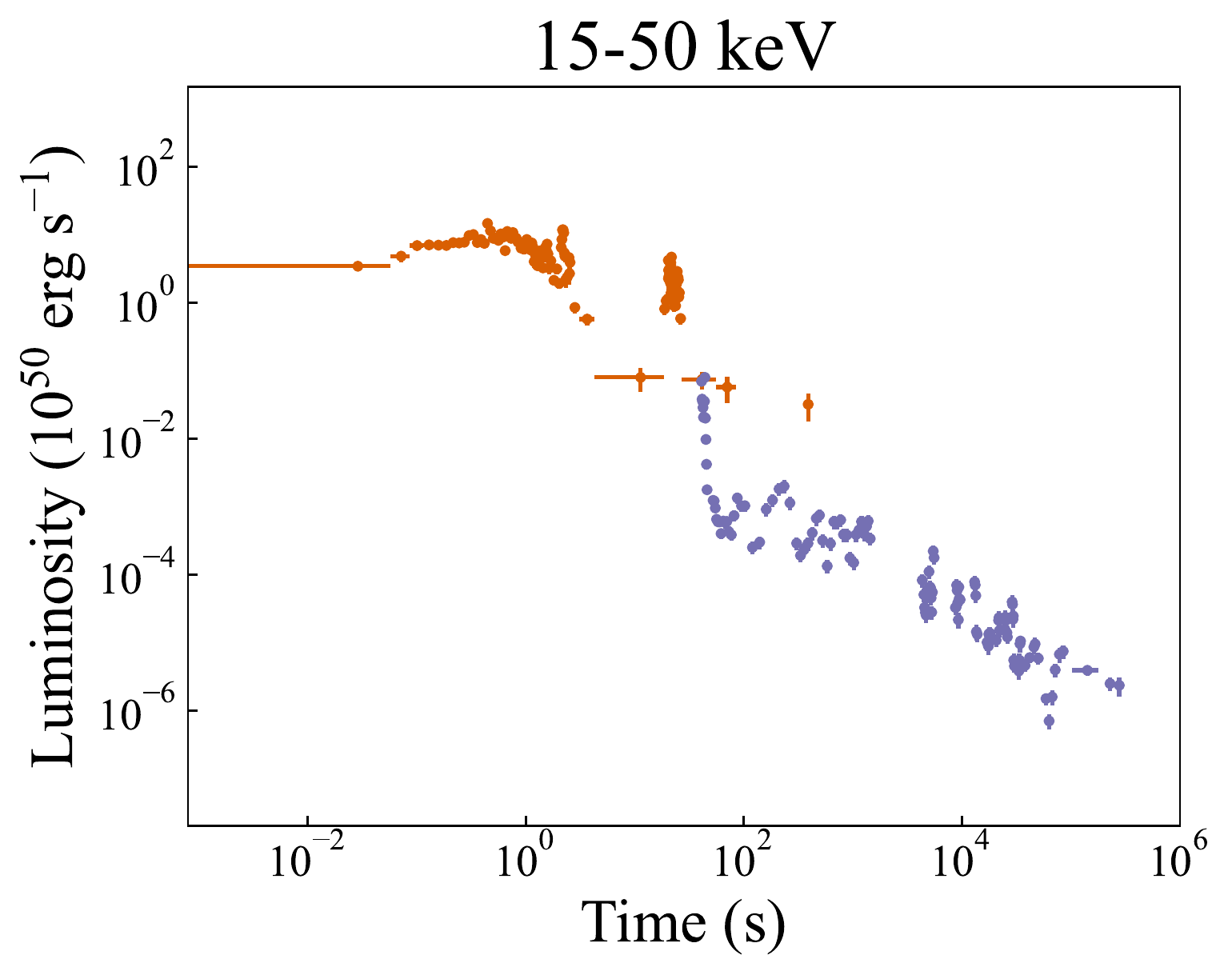} \\
    \includegraphics[width=0.9\columnwidth]{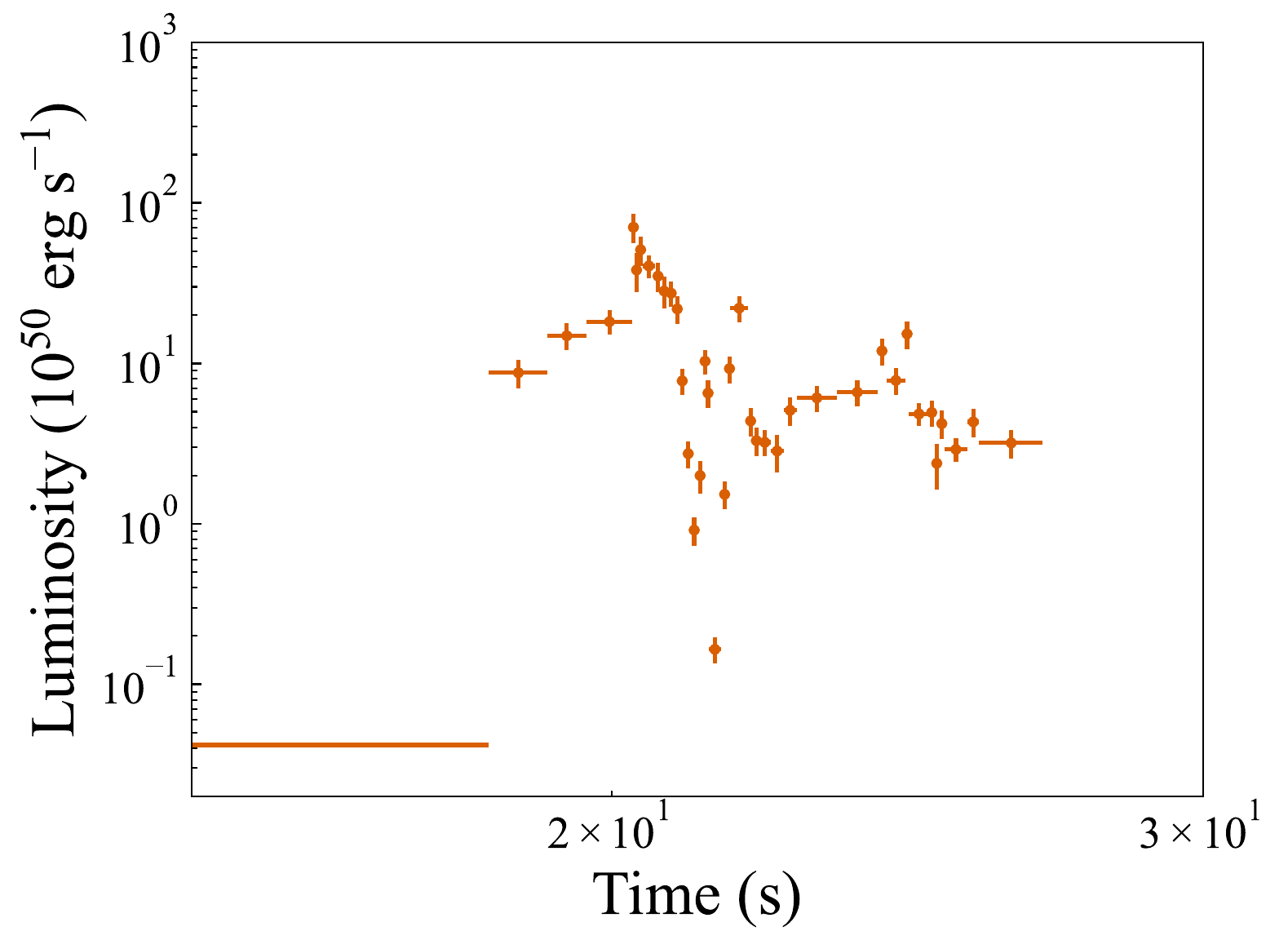} &
    \includegraphics[width=0.9\columnwidth]{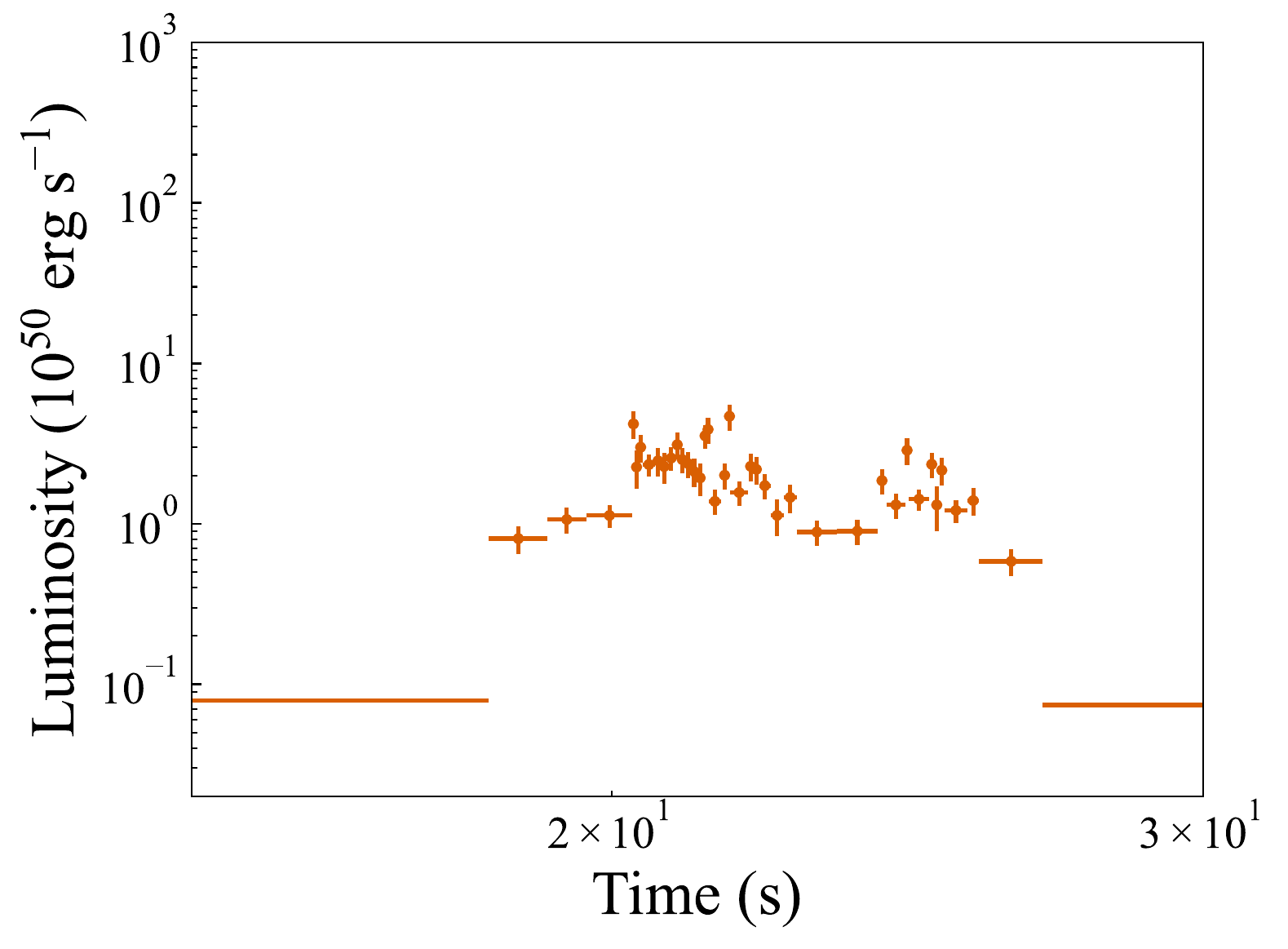}
    \end{tabular}
    \caption{Bolometric luminosity curves for GRB100522A. The figures on the left used the UKSSDC's 0.3-10 keV dataset, whereas the right figures used the 15-50 keV dataset. The lower figures are the same as the above except zoomed in on the $\sim 20$s spike. Orange data were taken with the BAT; purple data were taken with the XRT.}
    \label{fig:100522A_nopropagation}
\end{figure*}

However, the UKSSDC light-curves presented in the different bandpasses can actually appear substantially different due to the extrapolation of the count-rates to flux, since the ECF is less well-constrained when extrapolating further from the instrument's native bandpass. This can substantially increase the volatility of the data in the light-curve for BAT data at 0.3-10 keV or for XRT data at 15-50 keV and can introduce spurious features to the light-curve which are not present in other bandpasses. Without propagating the errors on the ECF, the same data presented in different bands may appear inconsistent. Of our sample of GRBs, this is most evident for GRB100522A, as is shown by Fig. \ref{fig:100522A_nopropagation}. The BAT data at $\sim 20$s show a highly statistically-significant luminosity variation of several orders of magnitude at 0.3-10 keV which is much reduced at 15-50 keV. In the same way, XRT data at 0.3-10 keV show considerably less variation than at 15-50 keV, where the variation might be interpreted as real substructure otherwise.

Here, while the lack of propagation of the ECF errors prevents washing out what may be a real feature of the light-curve and also prevents potentially introducing systematic errors with the interpolated ECF values, we have risked introducing spurious features to the light-curve instead. Given that the light-curve is so substantially different in the different bands, it can be difficult to disentangle which features may be real as opposed to spurious. In this work we utilised a method where the error on the ECF is directly propagated, since this would more closely mimic the algorithm used by the UKSSDC for generating the light-curves in different bandpasses. The count rates $R$ recorded by \textit{Swift} are converted to flux within a bandpass $f$ by that bandpass' ECF $C$,

\begin{equation}
    f=CR,
\end{equation}

with the UKSSDC's flux errors simply calculated as

\begin{equation}
    \Delta f=C\Delta R.
\end{equation}

Given $f$, $\Delta f$, $C$, and $\Delta C$ (all of which are available in the UKSSDC), we can calculate the flux errors including ECF error propagation $\Delta F$ as

\begin{equation}
    (\Delta F)^2 = (\Delta f)^2 + f^2\Big(\frac{\Delta C}{C}\Big)^2.
\end{equation}

After applying this $\Delta C$ error propagation, the average $\Gamma$ value across the GRB's data was then used to provide an estimate for the bolometric luminosity.

Fits to light-curves where the BAT data is extrapolated to the 0.3-10 keV band, as in previous work, tend to fit the BAT data poorly as models lack the high degree of precision and rapid variability required to fit data fluctuating to such an extreme extent. Indeed, most fits in the work of \citet{Gibson1} accurately fitted the late-time observational data (the XRT data in their native bandpass) but poorly matched the volatile early-time BAT data, mostly tending to understate the observed luminosities.

The nature of this extrapolation process means that any data that is extrapolated in this way will be considerably noisier than the unextrapolated data. If ECF errors are not propagated we see a great deal of fluctuation, but if propagation is performed, we find that the data's error bars become much wider than for the unextrapolated data.

In this work, we have performed fits using the same theoretical model as \citet{Gibson1}, but now to \textit{Swift} data with the ECF errors propagated. We aimed to demonstrate that the nature of these fits is sensitive to this error propagation, and also to the extrapolation process in general. We processed the data by mimicking \citet{Gibson1} and considering the data in the 0.3-10 keV band, but we also tested the difference in fits to data in the 15-50 keV band. To assess the impact of the extrapolation process on our results, we also used a regime wherein both instruments' data were kept in their native bandpasses before being converted to bolometric luminosities.

\begin{figure*}
    \centering
    \addtolength{\tabcolsep}{-0.4em}
    \begin{tabular}{cc}
    \includegraphics[width=0.9\columnwidth]{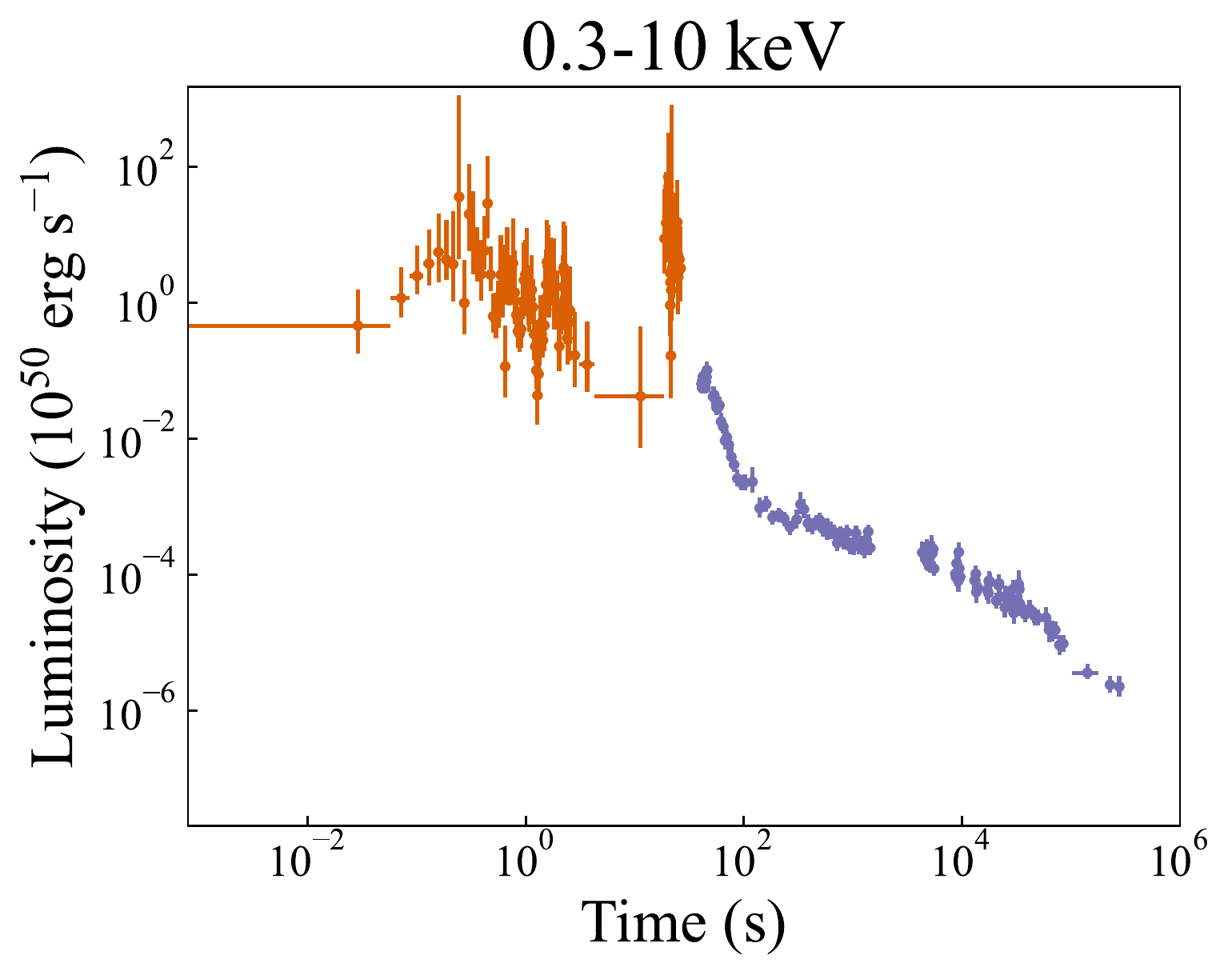} &
    \includegraphics[width=0.9\columnwidth]{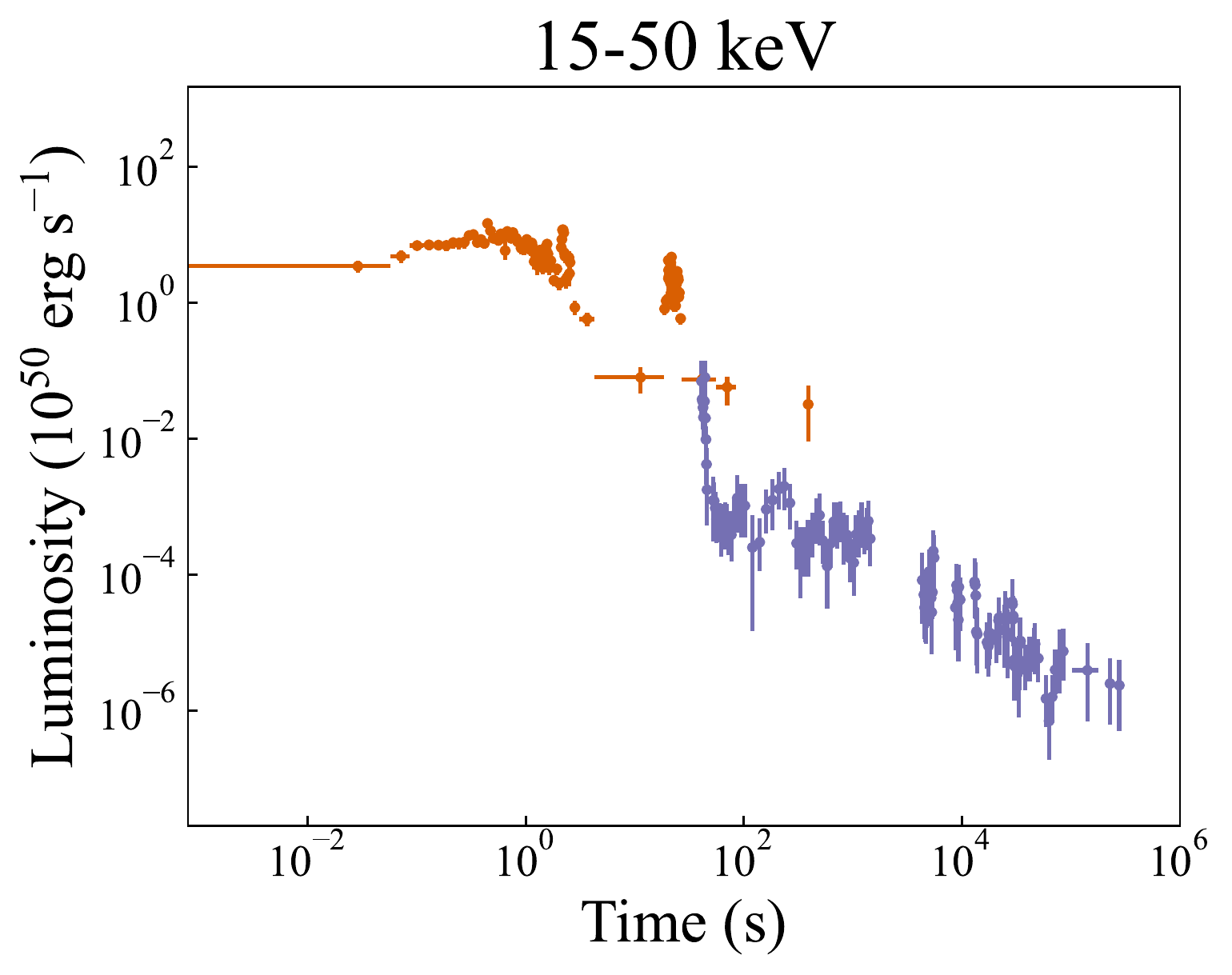} \\
    \includegraphics[width=0.9\columnwidth]{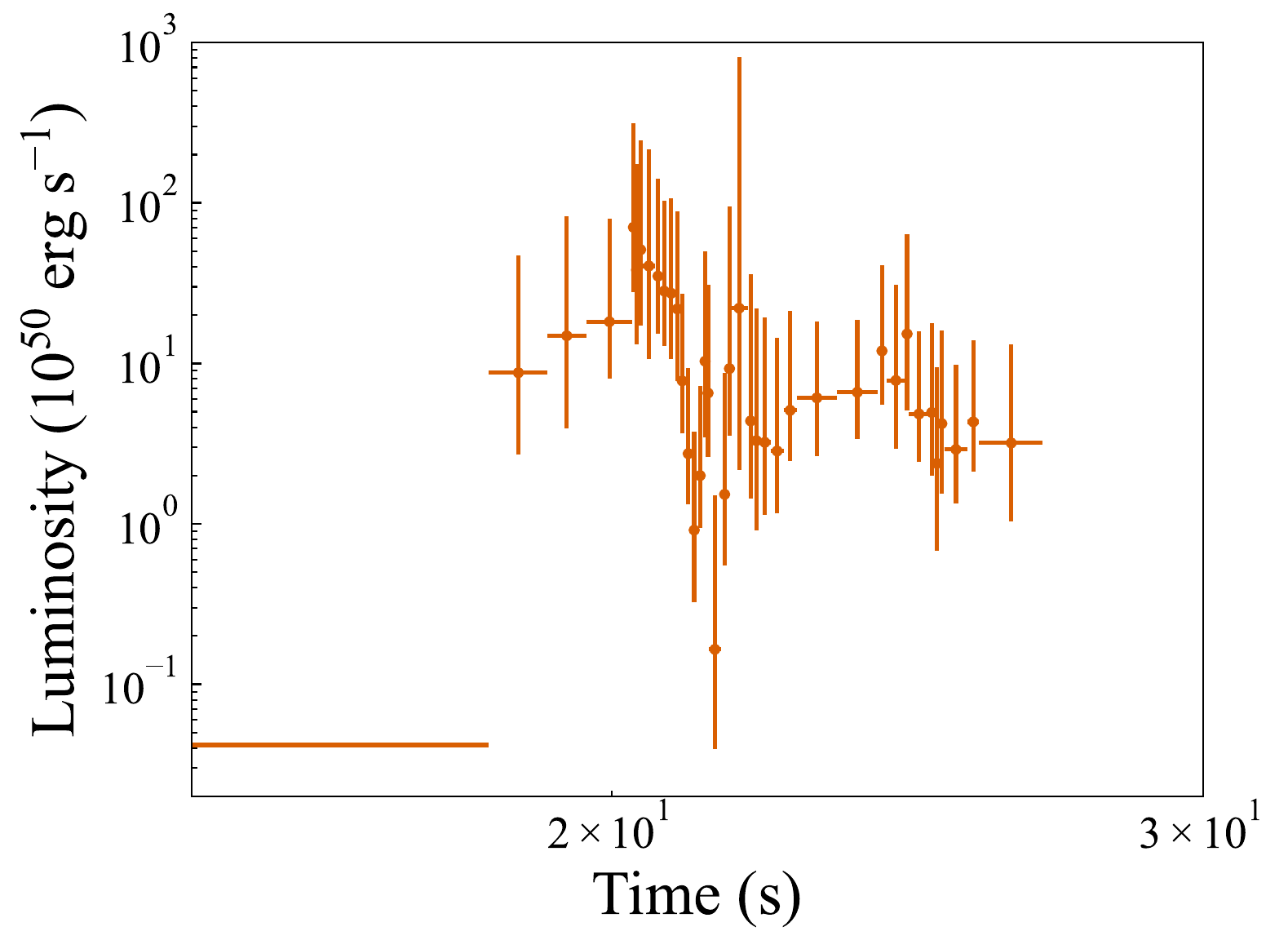} &
    \includegraphics[width=0.9\columnwidth]{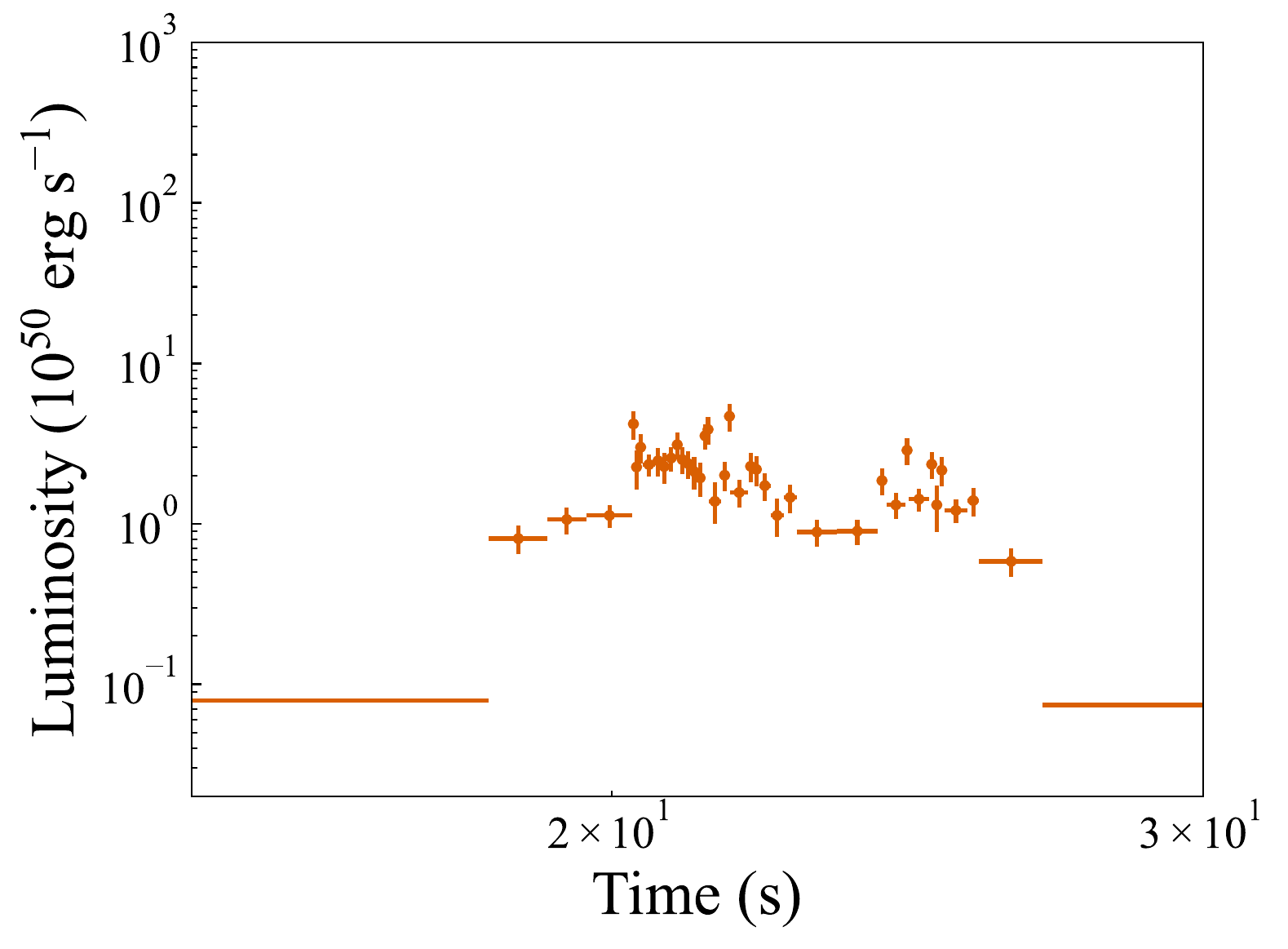}
    \end{tabular}
    \caption{Bolometric luminosity curves for GRB100522A. This figure is the same as Fig \ref{fig:100522A_nopropagation} but now with the ECF errors propagated.}
    \label{fig:100522A_propagation}
\end{figure*}

The effect of including this ECF error propagation is demonstrated by Fig. \ref{fig:100522A_propagation}. The inclusion of these errors significantly reduced the inconsistency of the 0.3-10 keV data and the 15-50 keV data, with the large spike at $\sim 20$s now reduced to nearly a margin-of-error variation. Nevertheless, the light-curves still appear to show significant differences, chiefly that BAT data at 0.3-10 keV have much more variance and also considerably larger errors than the XRT data, whereas the opposite is true at 15-50 keV. It is prudent to investigate how differently models fit to these data and whether or not different features of the light-curves drive the fits when they are presented in different bandpasses. Obtained fits may be substantially altered not only when we incorporate ECF error propagation but also when we change the bandpass of the data.

It may be noted that, even with ECF error propagation incorporated, the light-curves are not always consistent across the two bands. This is because the photon index is not constant - the use of an average $\Gamma$ to extrapolate all data to bolometric luminosities is valid if each data point's $\Gamma$ is close to the average, and this assumption generally holds. However, these discrepancies between $\bar{\Gamma}$ and $\Gamma$ can cause the luminosities to be different when extrapolated from the 0.3-10 keV band and the 15-50 keV band.

In our work, in addition to highlighting the impact of extrapolating data from their native bandpasses, we have also sought to mitigate this impact. Since we can extrapolate flux data presented at either 0.3-10 keV or 15-50 keV to a bolometric luminosity, we can also simply extrapolate each instrument's data from their native bandpass - extrapolating the BAT data from 15-50 keV and the XRT data from 0.3-10 keV, as opposed to requiring both instruments' data be presented in the same bandpass. This should allow us to avoid the increased variance introduced when taking BAT data at 0.3-10 keV and XRT data at 15-50 keV. The results of fitting to these data are included alongside the 0.3-10 keV and 15-50 keV extrapolations for comparison. In this treatment, the BAT data naturally match those in the 15-50 keV light-curves, and the XRT data match the 0.3-10 keV light-curves.

\section{Physical Model}

As we are using the physical model of \citet{Gibson1} to serve as the basis for testing the significance of the data processing algorithm, we briefly summarise their work here.

The binary neutron star merger that produces the SGRB leaves behind a remnant magnetar with a rotating disc of expelled material. Whether the material surrounding the magnetar is accreted or propellered away by the magnetic field depends on the Alfv\'{e}n radius $r_{\mathrm{m}}$, within which the dynamics of the disc material is strongly affected by the magnetic field, and the corotation radius $r_{\mathrm{c}}$, at which disc material's orbital period matches the magnetar surface's rotation period,

\begin{equation}
    r_{\mathrm{m}}=\mu^{4/7}(GM_*)^{-1/7}\dot M^{-2/7},
\end{equation}

\begin{equation}
    r_{\mathrm{c}}=\bigg(\frac{GM_*}{\omega^2}\bigg)^{1/3},
\end{equation}

where $\mu=BR$ is the magnetar's magnetic dipole moment for magnetic field $B$ and magnetar radius $R$, $M_*$ is the magnetar mass, $\dot M$ is the rate of change of the disc mass, and $\omega$ is the magnetar's angular rotation frequency.

When $r_{\mathrm{m}}>r_{\mathrm{c}}$, the magnetar's magnetic field rotates more quickly than the disc at the radius at which the field becomes dynamically important, and thus accelerates disc material within this radius and ejects it from the system. This interaction reduces the magnetar's angular momentum and it spins down. When $r_{\mathrm{m}}<r_{\mathrm{c}}$, however, the disc is orbiting more quickly, and the interaction slows down material within $r_{\mathrm{m}}$ and causes it to accrete, with the magnetar gaining angular momentum.

In addition to the propeller effect, the magnetar also loses angular momentum via magnetic dipole spin-down, whose torque is given by

\begin{equation}
    \tau_{\mathrm{dip}}=-\frac{\mu^2\omega^3}{6c^3}.
\end{equation}

From the torques due to the propeller effect and the dipole, we can find the luminosities from each component,

\begin{equation} \label{LpropEquation}
    L_{\mathrm{prop}}=-\tau_{\mathrm{prop}}\:\omega,
\end{equation}

\begin{equation}
    L_{\mathrm{dip}}=-\tau_{\mathrm{dip}}\:\omega,
\end{equation}

where $\tau_{\mathrm{prop}}$ is the torque exerted on the magnetar by the interaction with the disc. We then find the total luminosity by summing these components together and multiplying by a beaming fraction $1/f_B$ which is related to the jet's half-opening angle, since GRBs are taken to be beamed phenomena,

\begin{equation}
    L_{\mathrm{tot}}=\frac{1}{f_{\mathrm{B}}}(\eta_{\mathrm{dip}}L_{\mathrm{dip}}+\eta_{\mathrm{prop}}L_{\mathrm{prop}}),
\end{equation}

where $\eta_{\mathrm{dip}}$ and $\eta_{\mathrm{prop}}$ are the efficiencies of the conversion of energy to luminosity for the dipole and the propeller, respectively.

In the work of \citet{Gibson1}, an additional term is included in Equation \ref{LpropEquation} to account for the gravitational energy lost by expelled disc material,

\begin{equation}
    L_{\mathrm{prop}}=-\tau_{\mathrm{prop}}\:\omega -\bigg(\eta_2\frac{GM_*M_{\mathrm{D}}}{r_{\mathrm{m}}t_{\mathrm{v}}}\bigg),
\end{equation}

however it was shown by \citet{Gibson_phd} that this can be straightforwardly incorporated into $\eta_{\mathrm{prop}}$ and as such we have not included it.

We also adopt the same expression as \citet{Gibson1} for the rate at which material falls back into the disc $M_{\mathrm{fb}}$,

\begin{equation}
    \dot M_{\mathrm{fb}}=\frac{M_{\mathrm{fb}}}{t_{\mathrm{fb}}}\bigg(\frac{t+t_{\mathrm{fb}}}{t_{\mathrm{fb}}}\bigg)^{-\frac{5}{3}},
\end{equation}

where $M_{\mathrm{fb}}$ is parameterised as a multiplicative factor $\delta$ of the initial disc mass, $M_{\mathrm{fb}}=\delta M_{\mathrm{D,i}}$. Similarly, $t_{\mathrm{fb}}$ is the fallback timescale for material to return to the disc, parameterised as a factor $\epsilon$ of the disc's viscous timescale $t_{\mathrm{fb}}=\epsilon t_{\mathrm{v}}$.

\section{MCMC fitting routine}

We fit the processed \textit{Swift} data using the magnetic propeller model of \citet{Gibson1}, using 9 free parameters that we altered to find the optimal fit to the data. Our free parameters were the magnetar surface field strength $B$, initial magnetar rotation period $P_i$, initial disc mass $M_{\mathrm{D,i}}$, the inner radius of the disc $R_{\mathrm{D}}$, the fallback timescale parameter $\epsilon$, the fallback mass parameter $\delta$, the dipole efficiency factor $\eta_\mathrm{{dip}}$, the propeller efficiency factor $\eta_{\mathrm{prop}}$, and the beaming parameter $1/f_{\mathrm{B}}$. The limits on these parameters were the same as in \citet{Gibson1}, and are quoted in Table \ref{tab:PropLimits}. We also fixed the magnetar mass at 1.4 $M_{\odot}$, the propeller switch-on parameter $n=1.0$, the sound speed in the disc $c_s=10^7$ cm $\mathrm{s}^{-1}$, the disc viscosity prescription $\alpha=0.1$, and the light-cylinder capping fraction $k=0.9$.

\begin{table}
    \centering
    \begin{tabular}{l|cc}
    \hline
        Parameter & Lower limit & Upper limit \\
        \hline
        $B$ $(10^{15}$ G) & $10^{-3}$ & 10\\
        $P_{\mathrm{i}}$ (ms) & 0.69 & $10$ \\
        $M_{\mathrm{D,i}}$ ($M_{\odot}$) & $10^{-3}$ & $10^{-1}$ \\
        $R_{\mathrm{D}}$ (km) & 50 & 2000\\
        $\epsilon$ & 0.1 & 1000\\
        $\delta$ & $10^{-5}$ & 50\\
        $\eta_{\mathrm{dip}}$ & 0.01 & 1\\
        $\eta_{\mathrm{prop}}$ & 0.01 & 1\\
        $1/f_{\mathrm{B}}$ & 1 & 600
    \end{tabular}
    \caption{Limits for each parameter in our MCMC magnetic propeller simulations. Limits taken from \citet{Gibson1}.}
    \label{tab:PropLimits}
\end{table}

We used a Monte Carlo Markov chain (MCMC, \citet{MacKayMCMC2003}) simulation to explore parameter space to find the best-fitting models, as in \citet{Gibson1}. The MCMC is well-suited to search many-dimensional parameter spaces in a computationally efficient way. Our MCMC attempts to find the global maximum of the model's posterior probability distribution, which is the sum of two components,
\begin{equation}
    \mathrm{ln}(p)=\mathrm{ln}(p_{\mathrm{likelihood}})+\mathrm{ln}(p_\mathrm{prior}),
\end{equation}
where $\mathrm{ln}(p_{\mathrm{likelihood}})$ is the log-likelihood function, which we give the same form as \citet{Gibson1},
\begin{equation}
    \mathrm{ln}(p_{\mathrm{likelihood}})=-\frac{1}{2}\sum_{i=1}^N \bigg(\frac{y_i-\hat{y}_i}{\sigma_i}\bigg)^2,
\end{equation}
where $y_i$ is a data point, $\hat{y}_i$ is a model point at the same x-value as $y_i$, $\sigma_i$ is the data point's $y$-uncertainty, and $N$ is the number of data points. $\mathrm{ln}(p_\mathrm{prior})$ is the prior probability, which is flat and is designed to allow the MCMC to reject solutions where not all parameters are within the limits we have set for the acceptable bounds of parameter space,
\begin{equation}
    \mathrm{ln}(p_\mathrm{prior})=\left\{
    \begin{array}{ll}
       0  &  x_{\mathrm{k,l}}<x_{\mathrm{k}}<x_{\mathrm{k,u}} \quad \forall \; k\\
        -\infty & \mathrm{otherwise,}
    \end{array}
    \right.
\end{equation}
where $x_{\mathrm{k,l}}$ and $x_{\mathrm{k,u}}$ are the lower and upper limits on parameter $k$, respectively.

When optimising our model with the MCMC, we use the corrected Akaike Information Criterion (AICc, \citet{BurnhamAnderson}) to identify the best-fitting model, rather than a simple $\chi^2$ goodness-of-fit test. Our use of the AICc is for consistency and ease of comparison with the results of \citet{Gibson1}. The AICc is given by
\begin{equation}
    \mathrm{AICc}=-2\mathrm{ln}(p)+2k+\frac{2k(k+1)}{N-k-1},
\end{equation}
where k is the number of free parameters. Note that the AICc still scales with the $\chi^2$ statistic, as the first term cancels to $\chi^2$, but the other terms allow for models to be penalised for overfitting, as we would naturally expect more free parameters to yield better results.

To extract best-fitting parameters and their uncertainties from the MCMC, we first performed a burn-in phase to allow the MCMC walkers to find the AICc's global minimum. Then a shorter chain phase to the MCMC enabled walkers to walk around the global minimum and sample it sufficiently to give good estimates on the uncertainties on each parameter. We ran a 50,000 step burn-in and a 10,000 step chain for 100 walkers, selecting the top 100 distinct probabilities in the burn-in to serve as the starting positions for the walkers in the chain. The Python package \textit{emcee} \citep{emcee2013} was used to implement the MCMC algorithm.

\begin{figure*}
    \centering
    \addtolength{\tabcolsep}{-0.4em}
    \begin{tabular}{ccc}
        {\Large 0.3-10 keV} & {\Large 15-50 keV} & {\Large Native}\\ \hline \\
        \includegraphics[width=0.61\columnwidth]{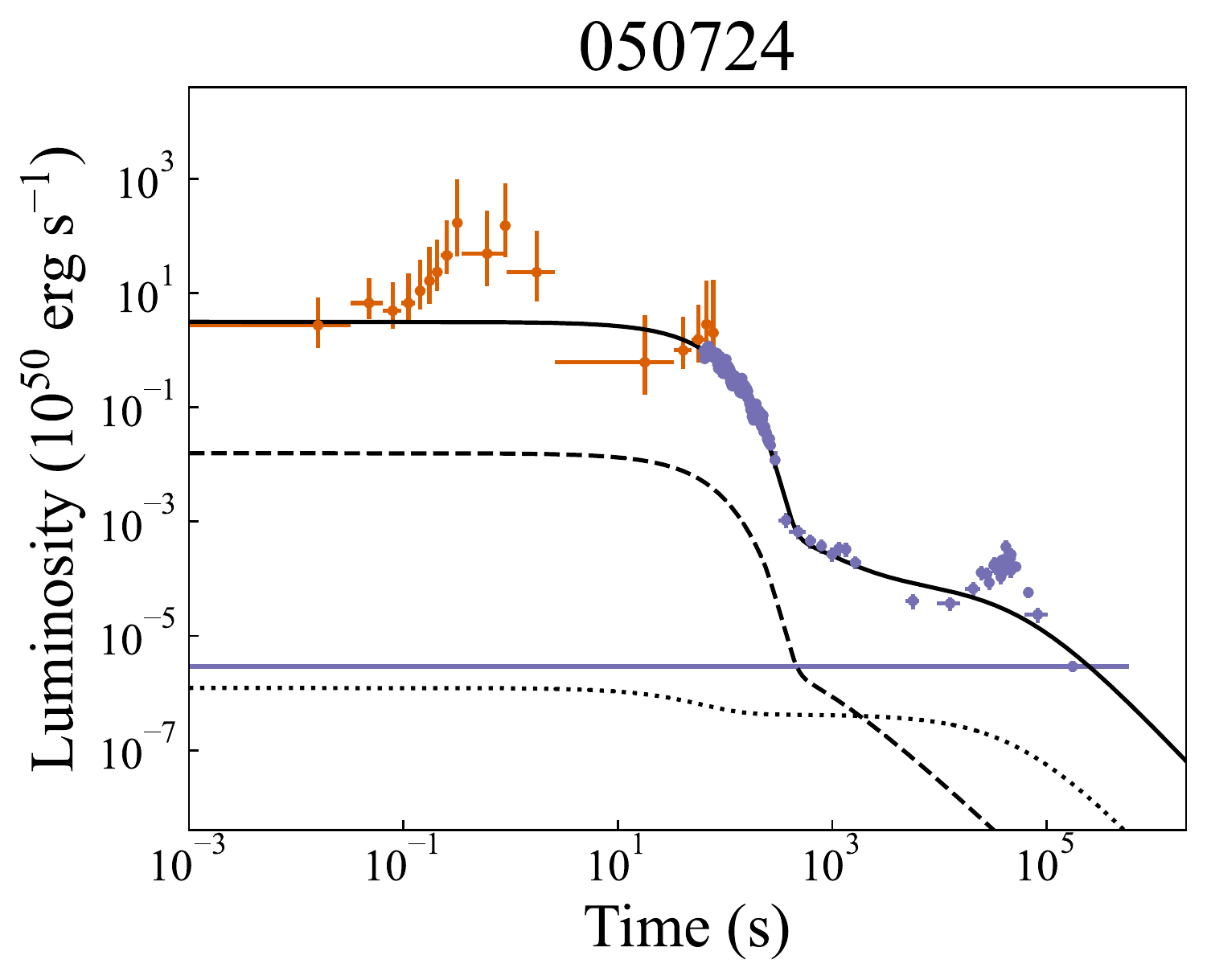} & 
        \includegraphics[width=0.61\columnwidth]{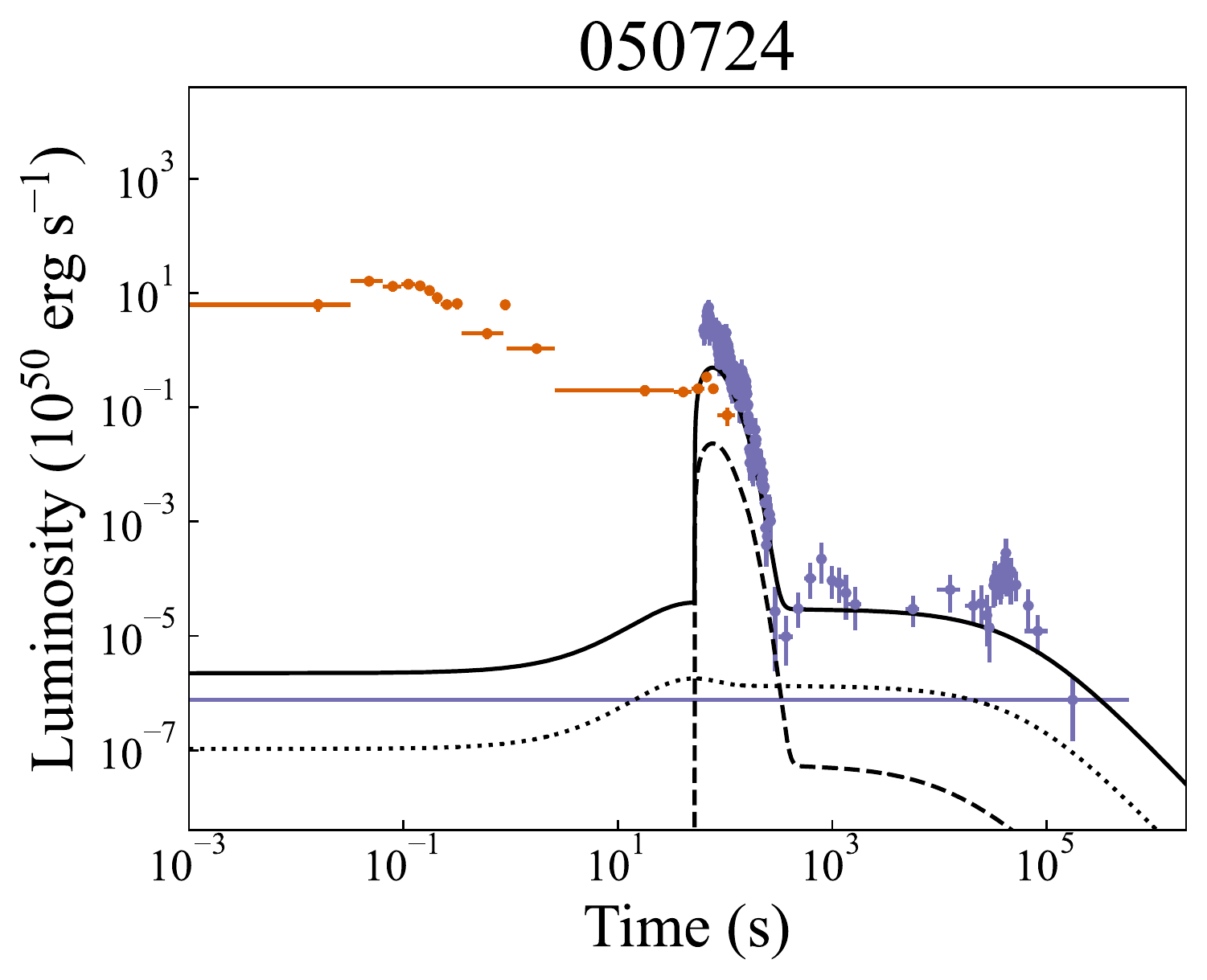} &
        \includegraphics[width=0.61\columnwidth]{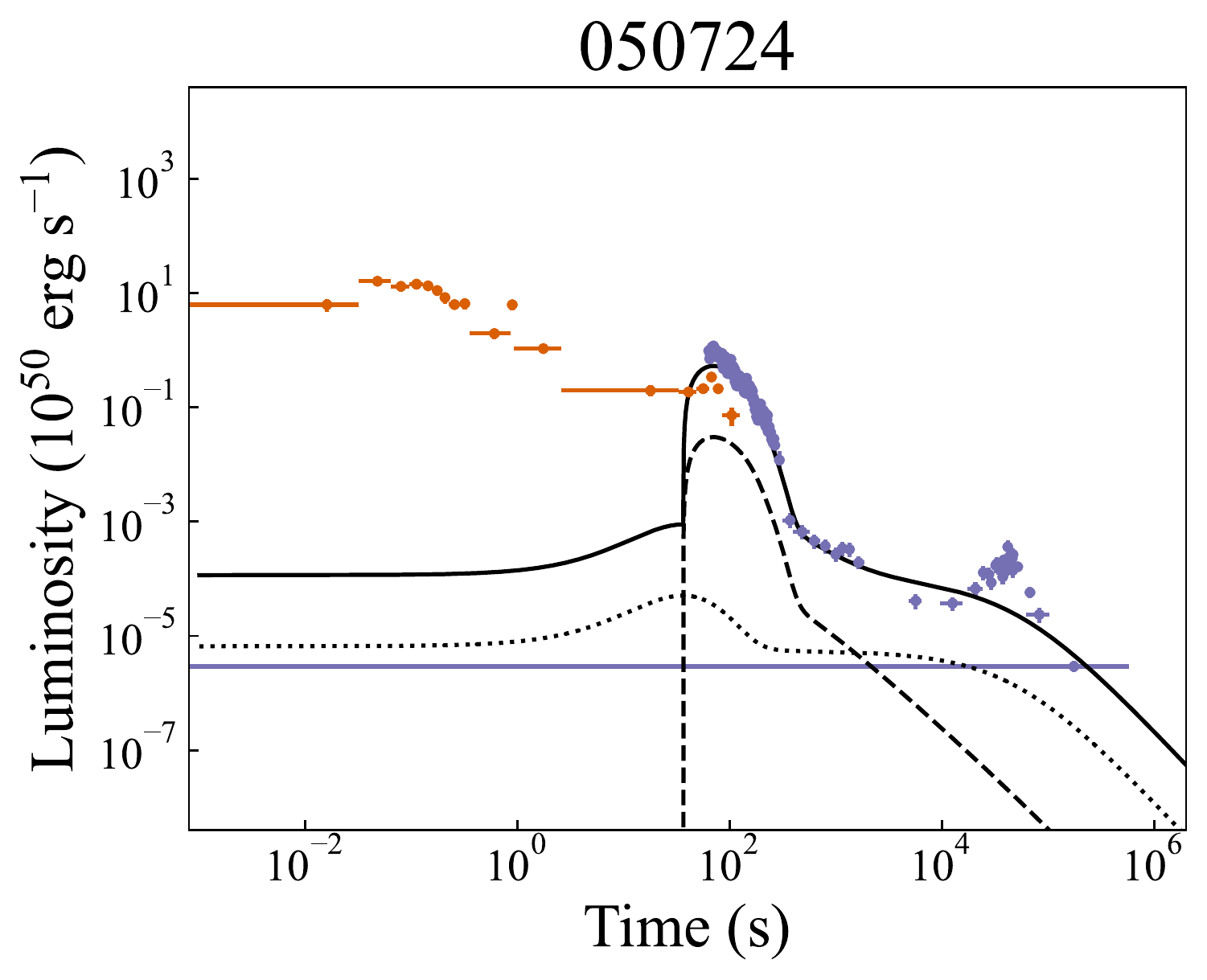}\\ 
        \includegraphics[width=0.61\columnwidth]{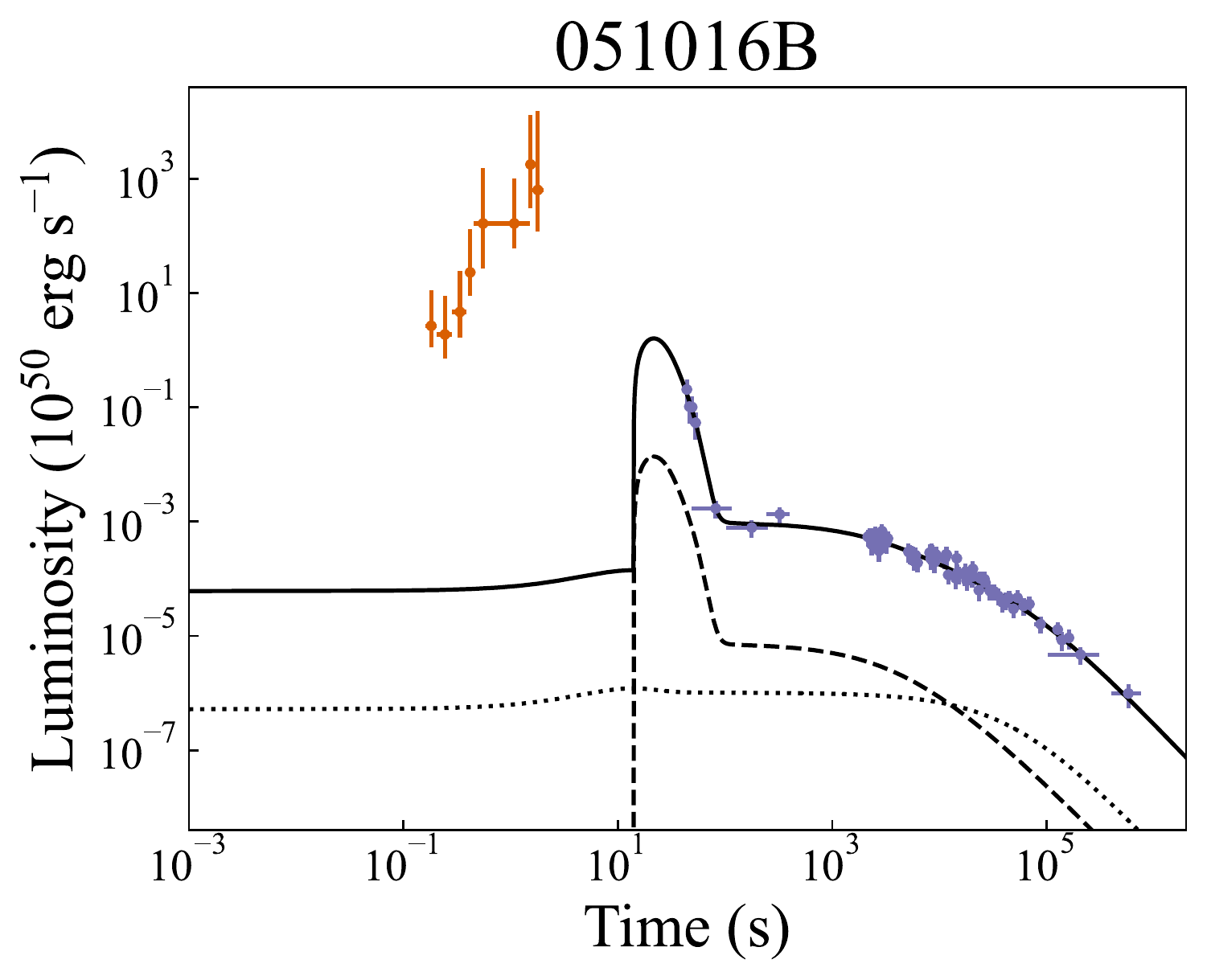} &
        \includegraphics[width=0.61\columnwidth]{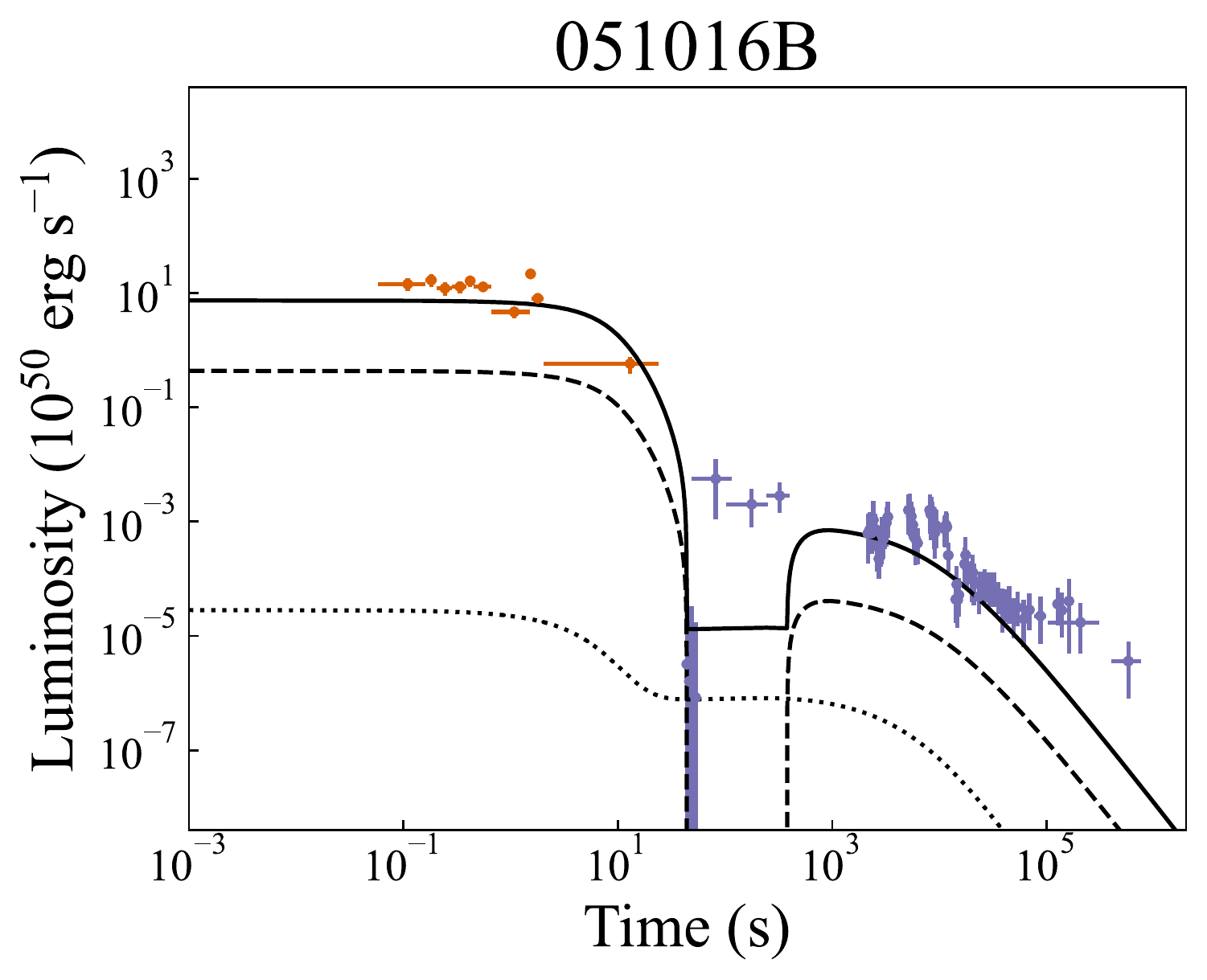} & 
        \includegraphics[width=0.61\columnwidth]{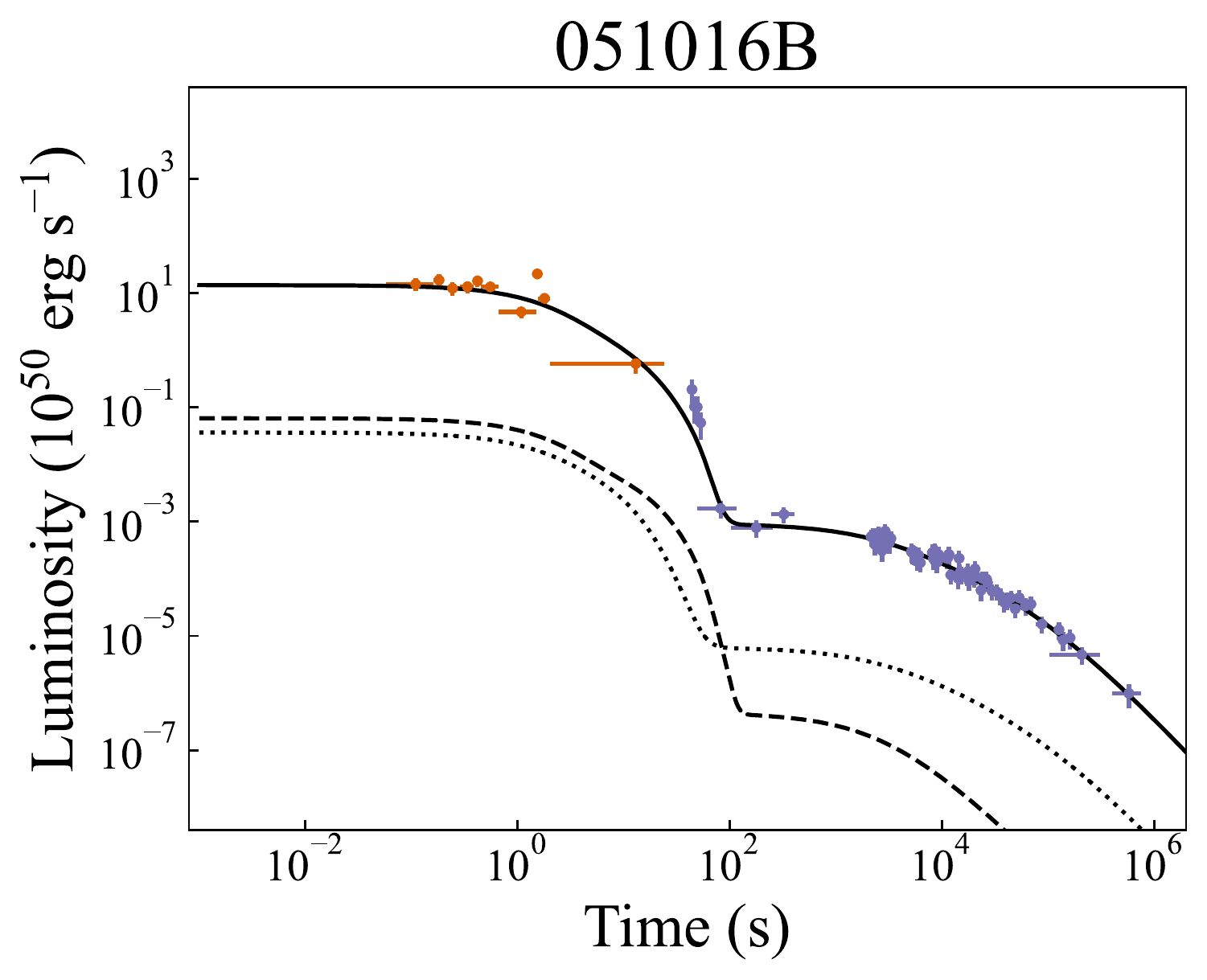}\\ 
        \includegraphics[width=0.61\columnwidth]{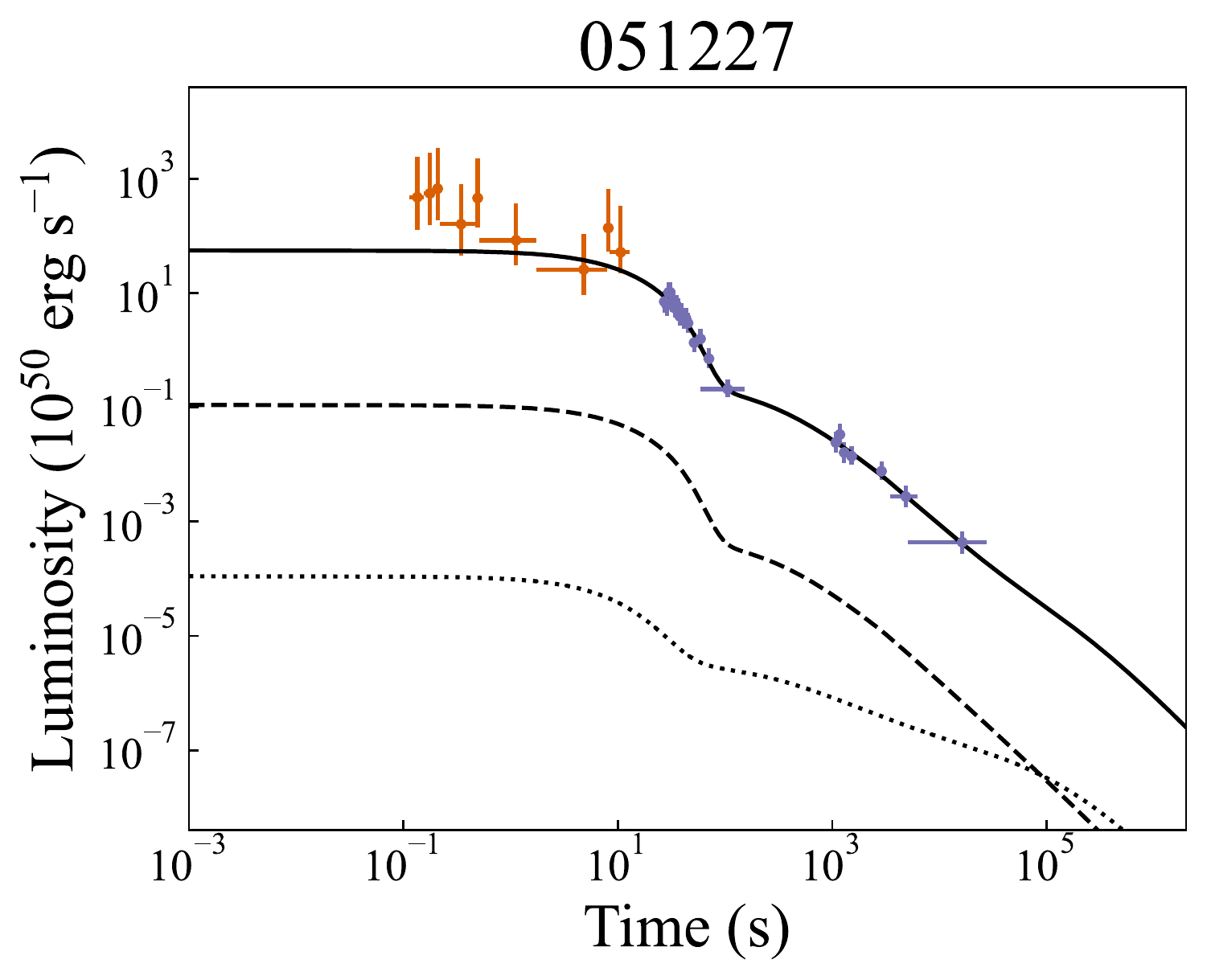} & 
        \includegraphics[width=0.61\columnwidth]{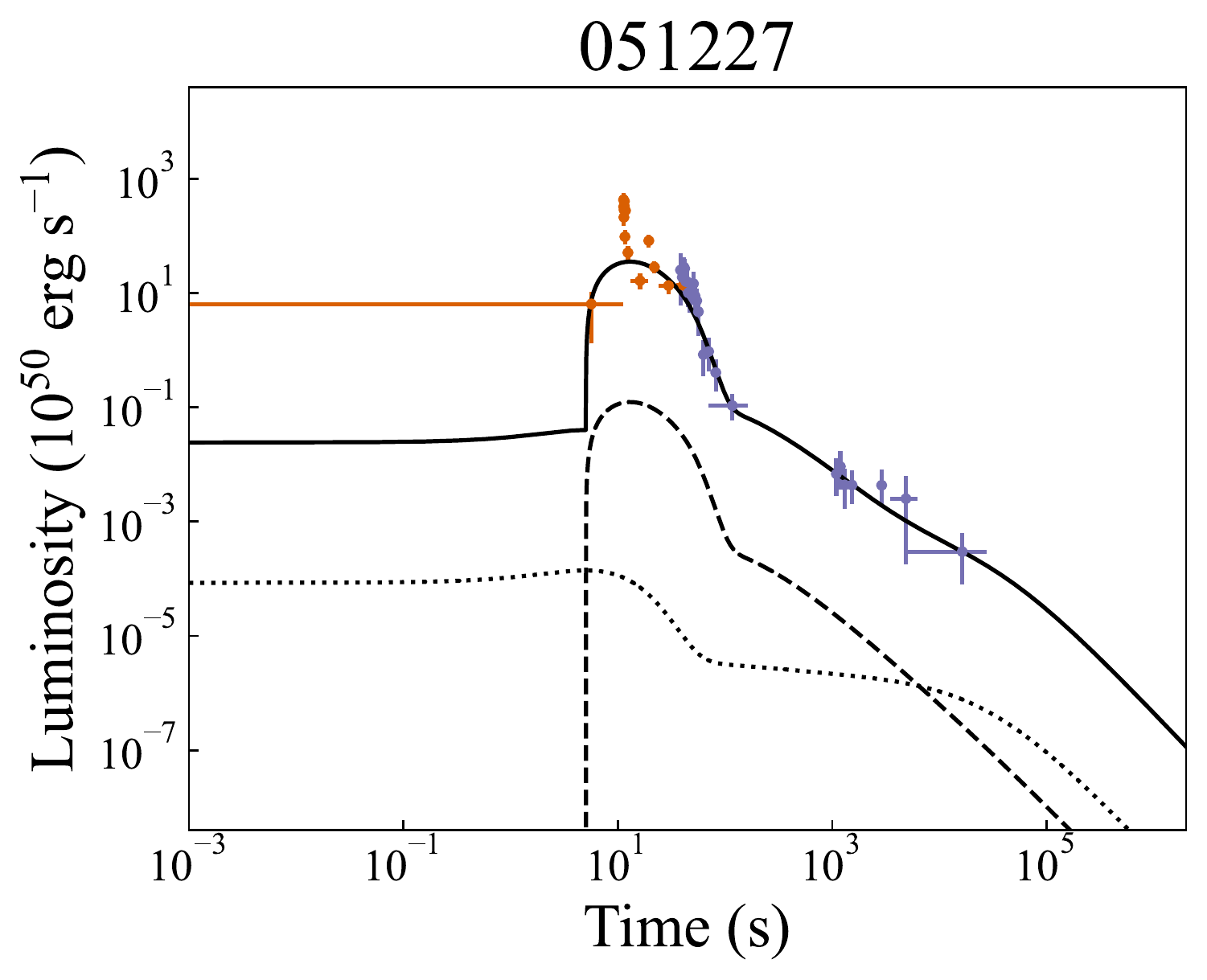} &
        \includegraphics[width=0.61\columnwidth]{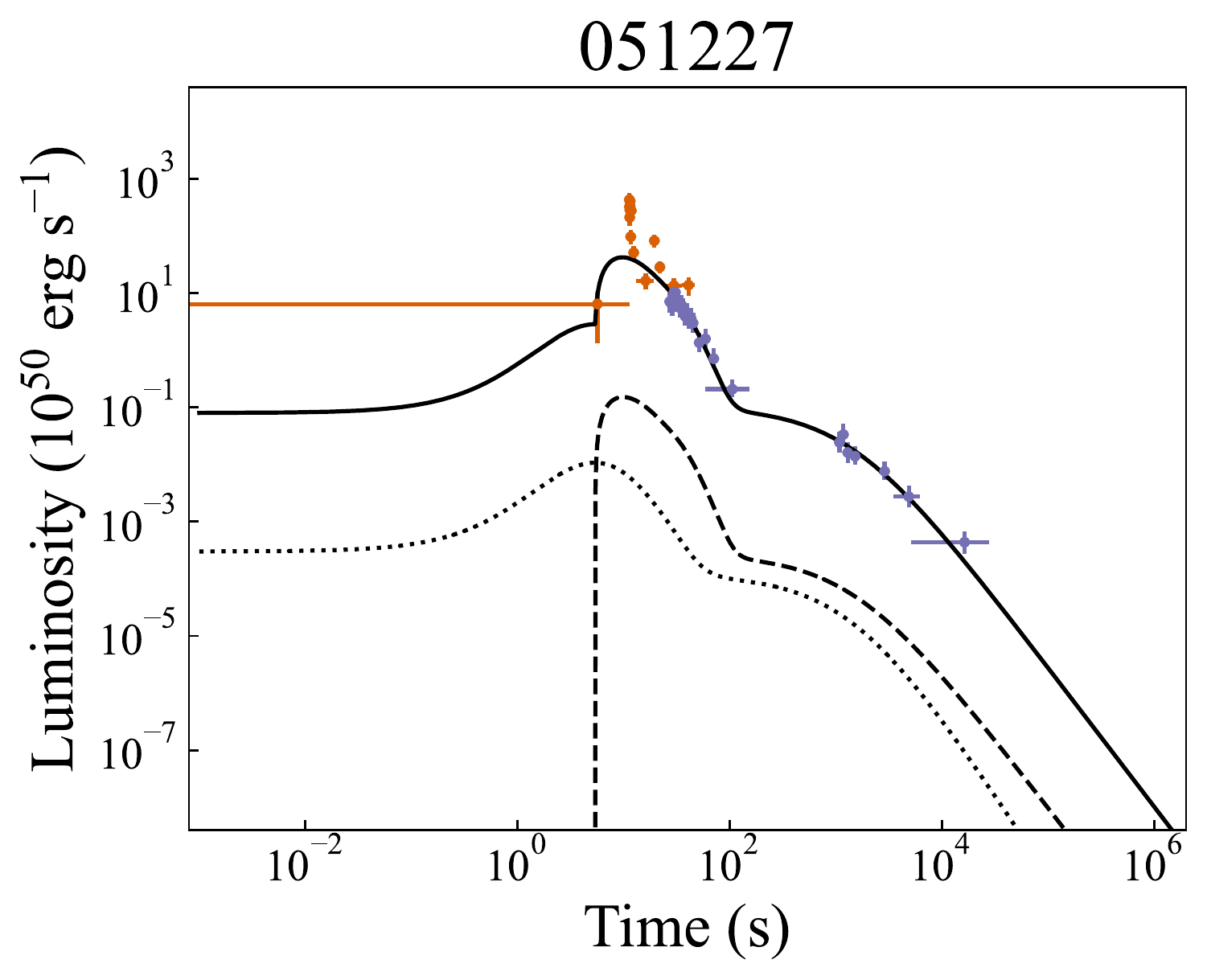}\\
        \includegraphics[width=0.61\columnwidth]{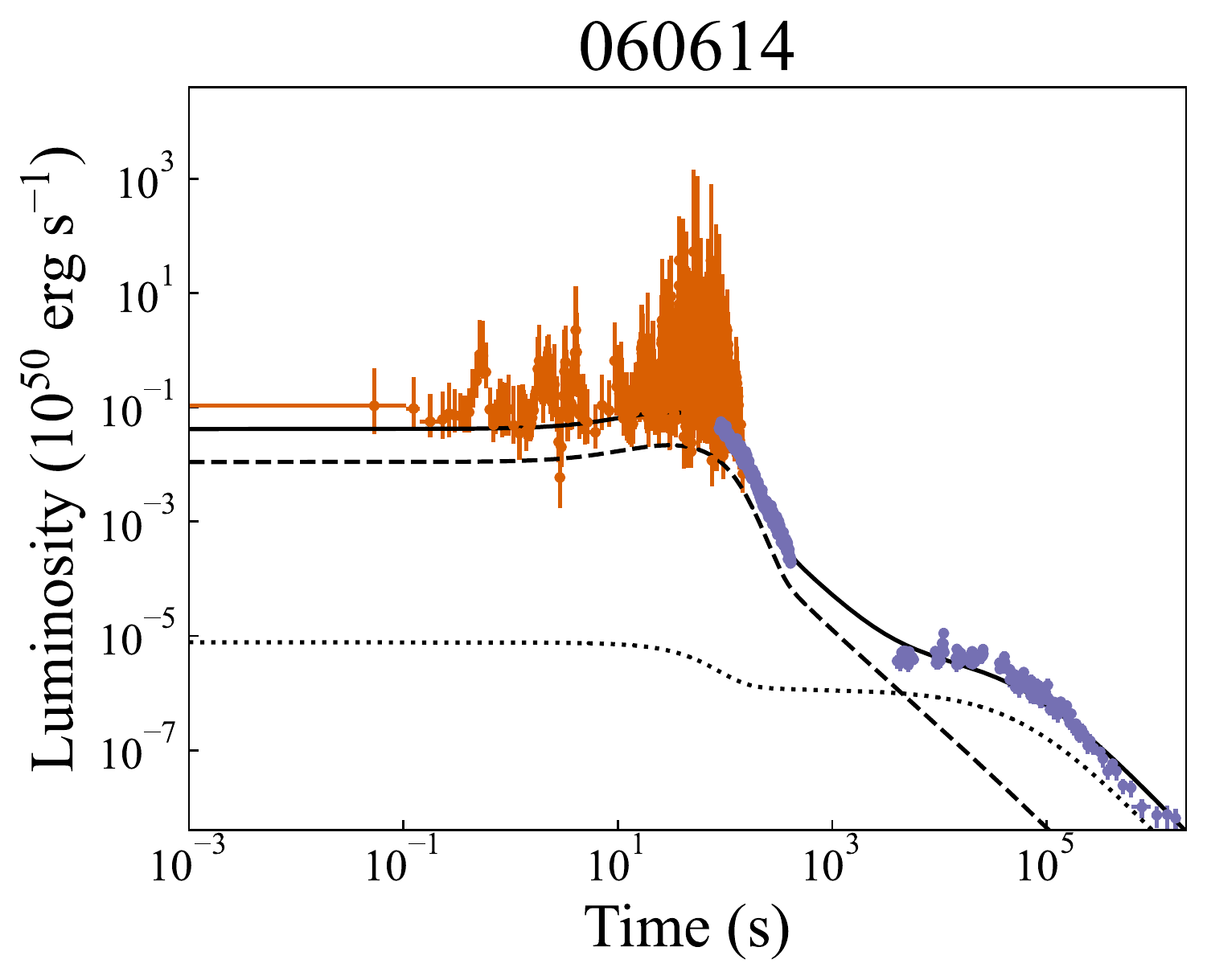} & 
        \includegraphics[width=0.61\columnwidth]{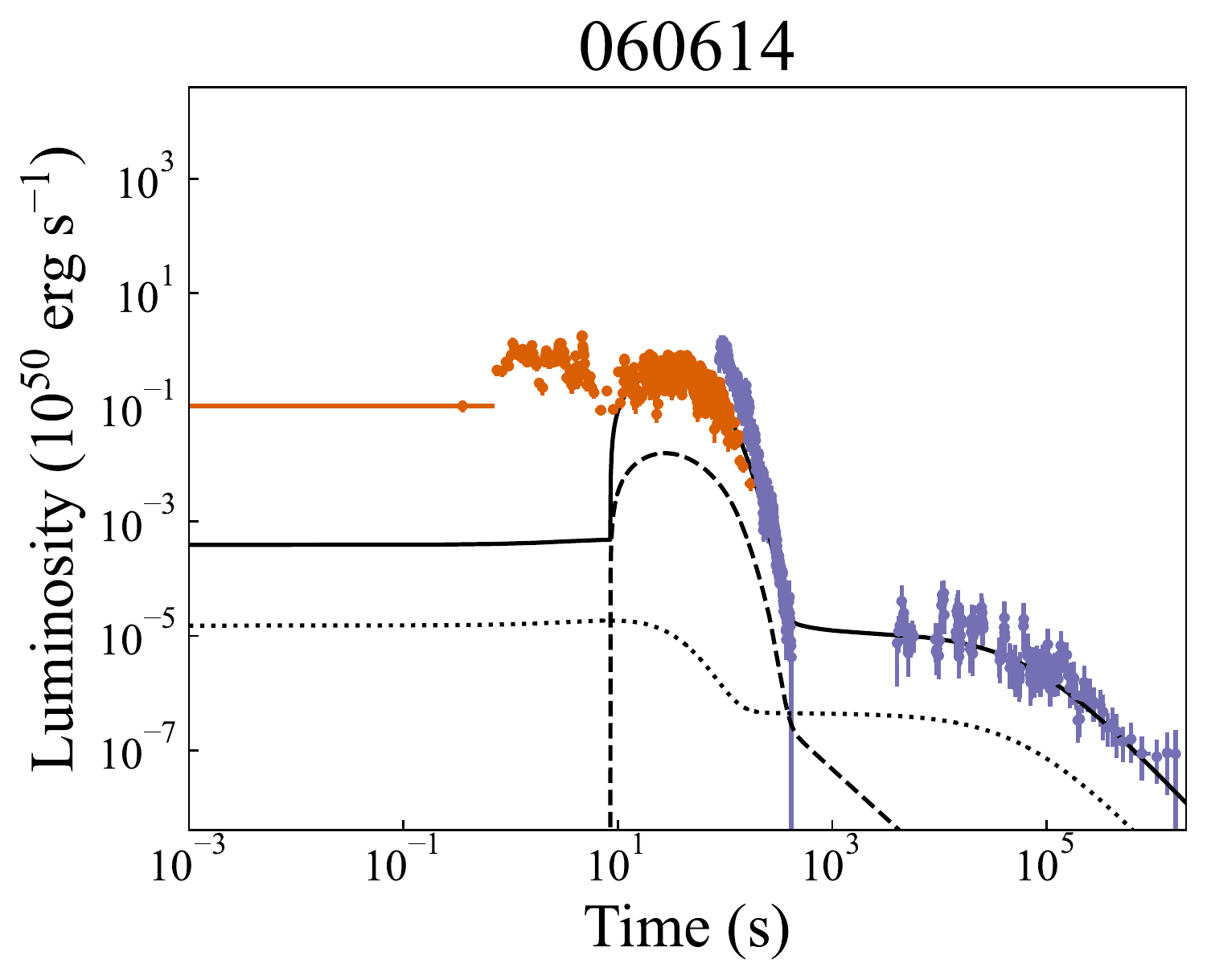}& 
        \includegraphics[width=0.61\columnwidth]{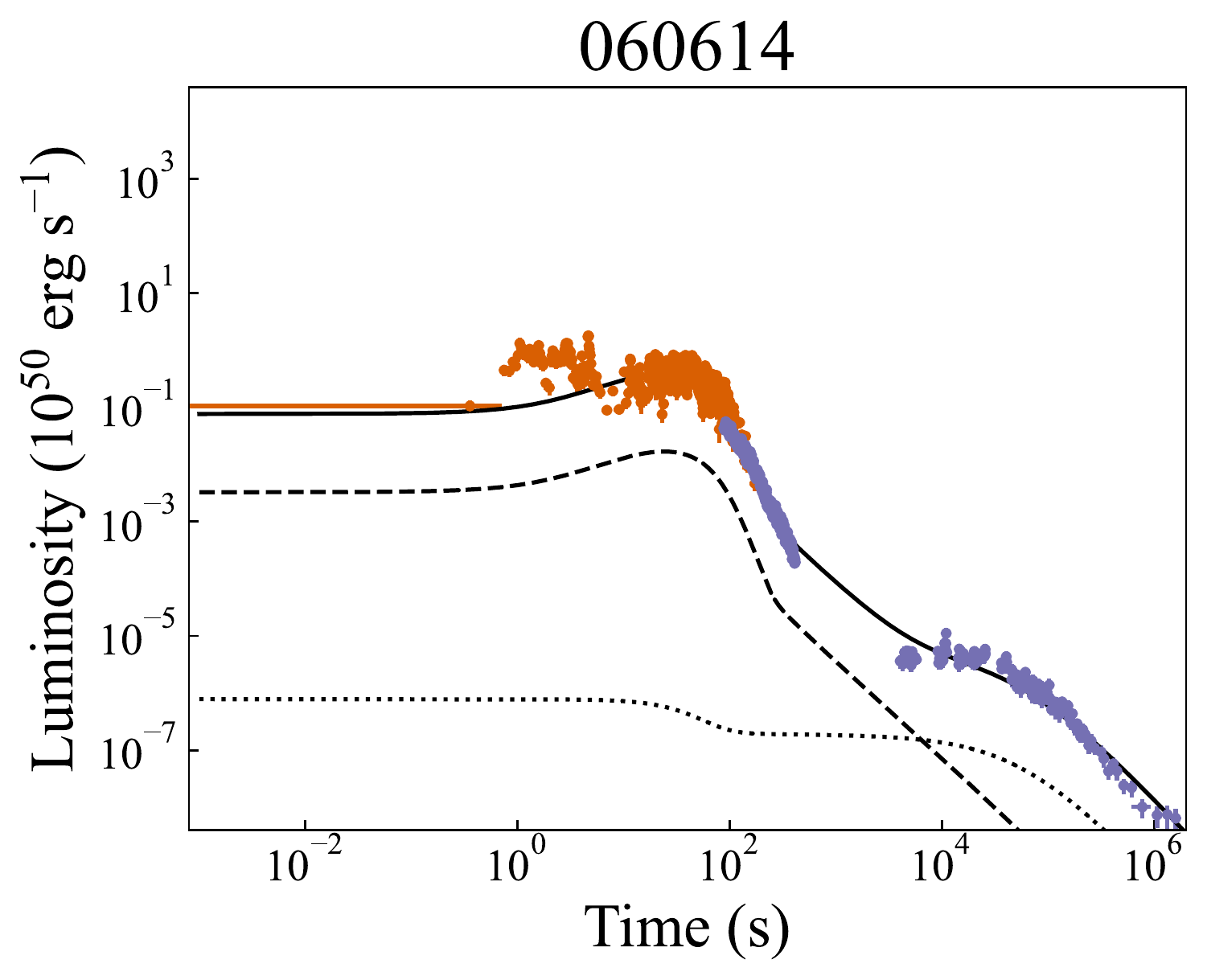}\\
        \includegraphics[width=0.61\columnwidth]{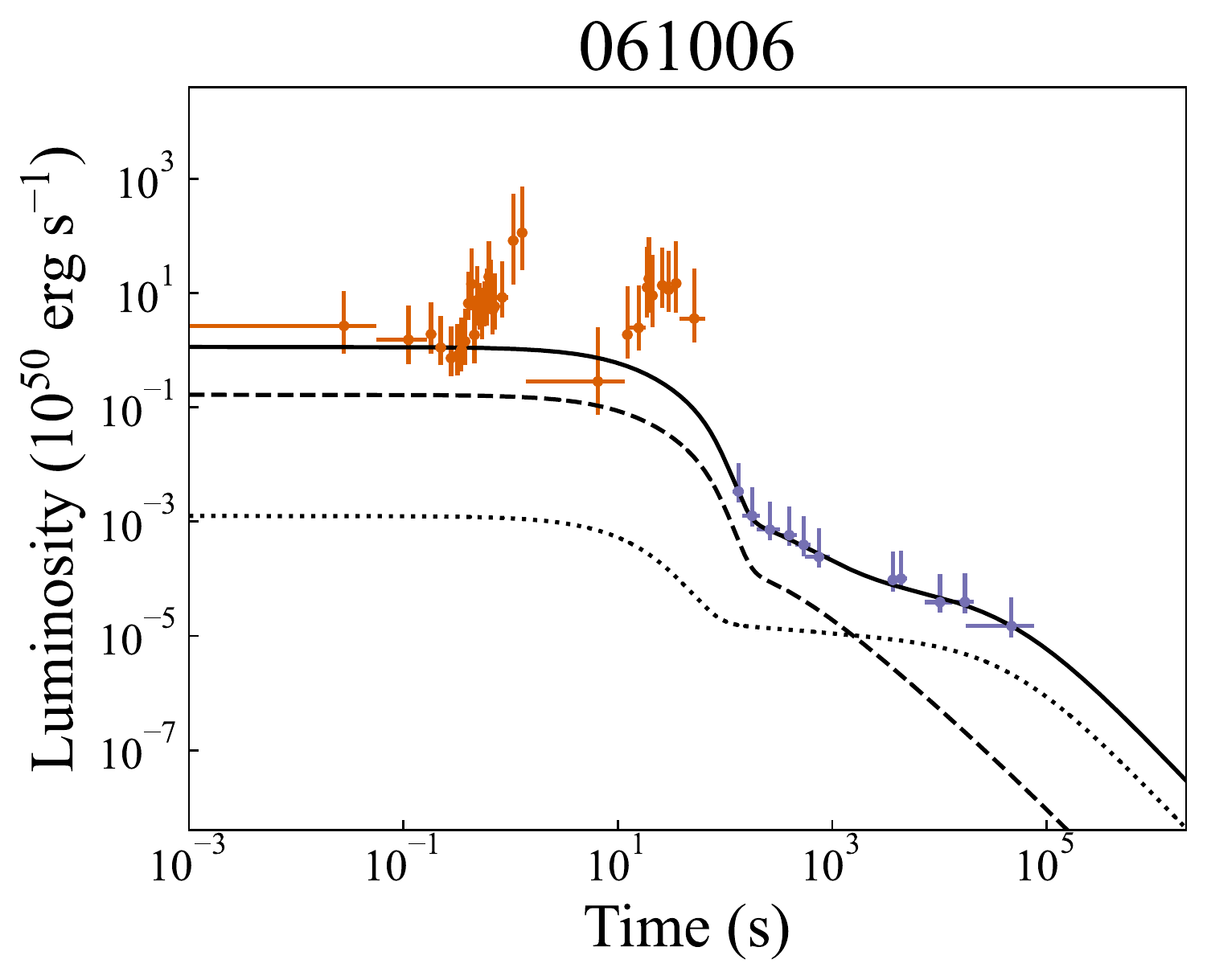} &
        \includegraphics[width=0.61\columnwidth]{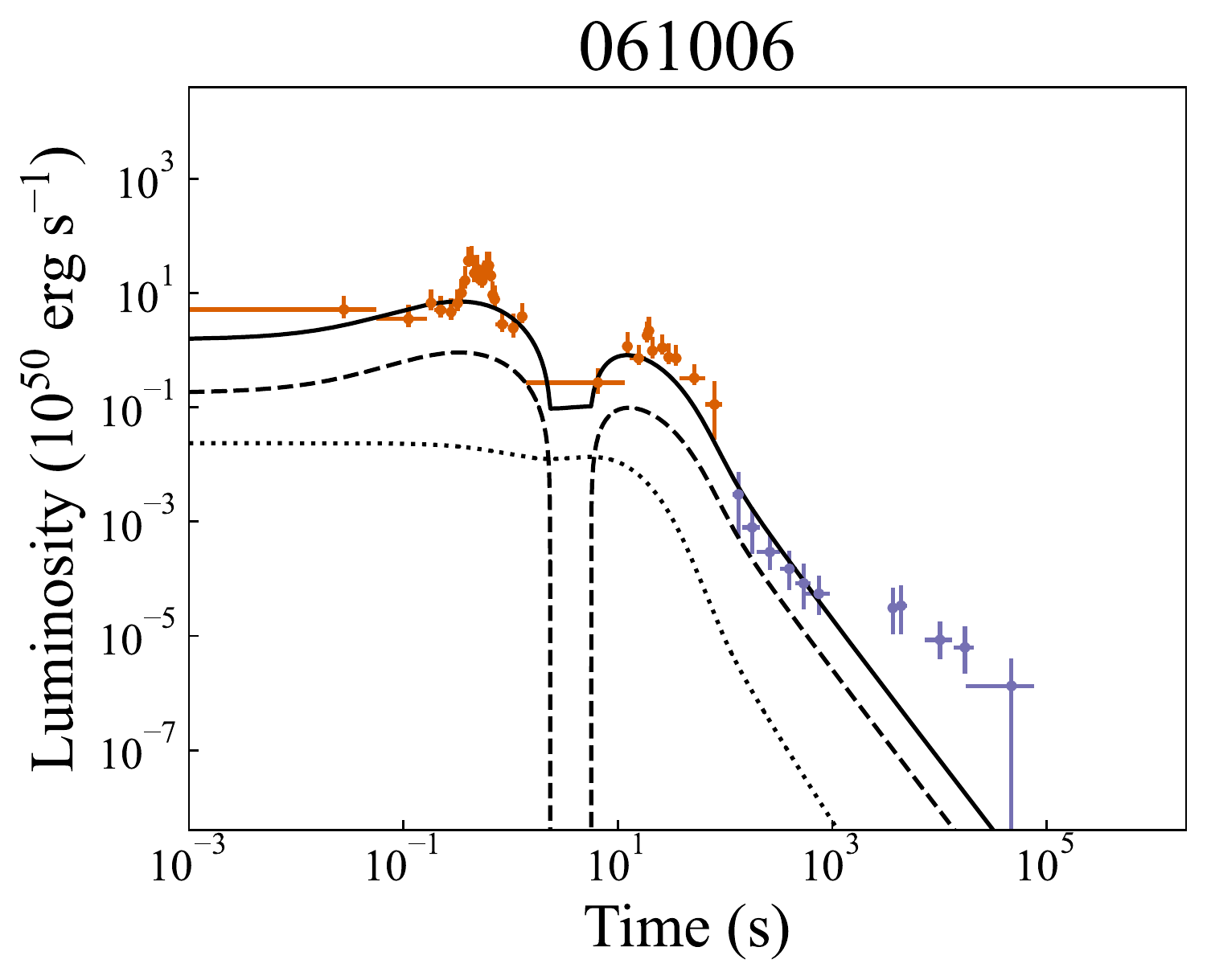} & 
        \includegraphics[width=0.61\columnwidth]{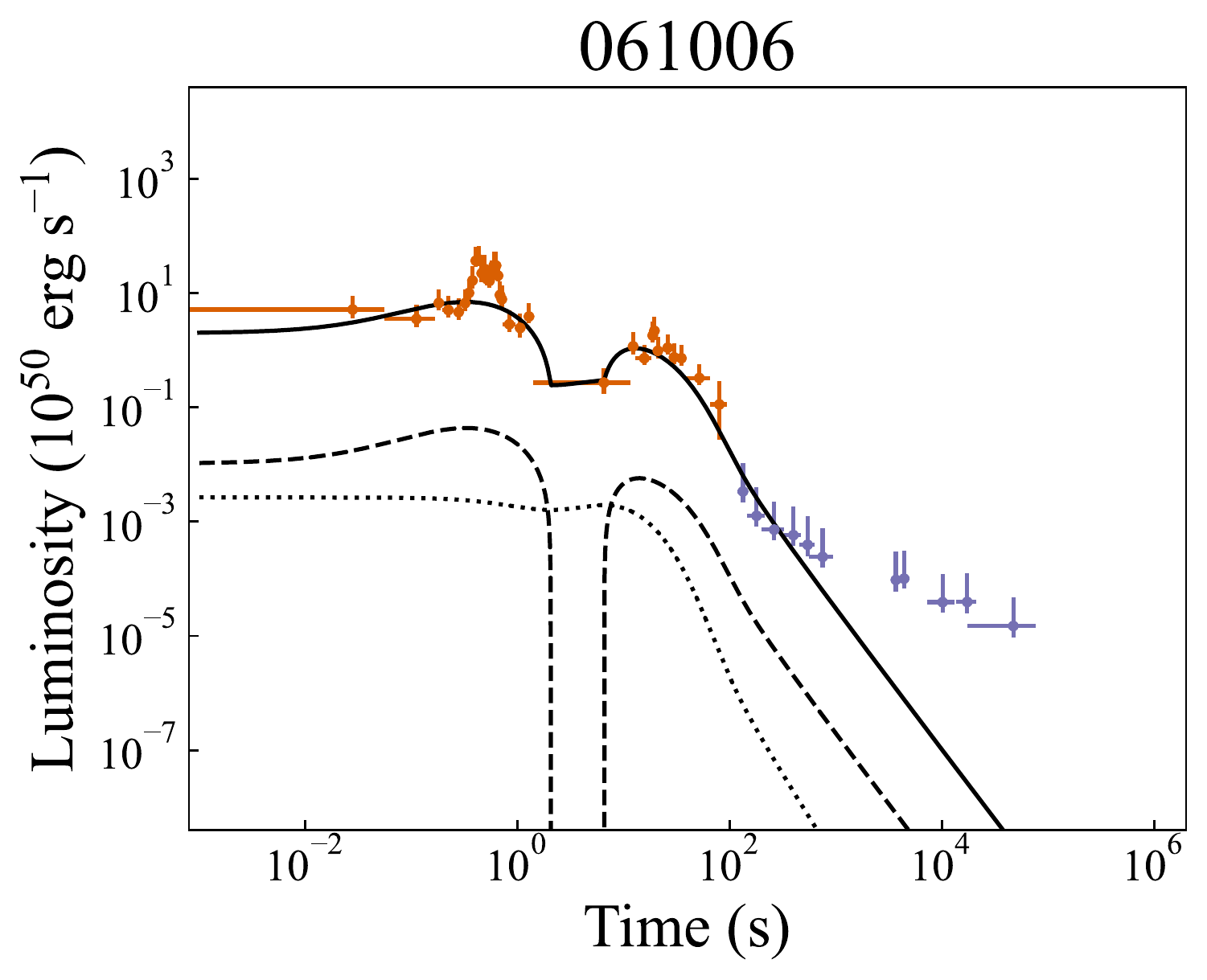}\\
    \end{tabular}
    \caption{Best fits to data with propagated conversion factor errors. The left column features fits to data extrapolated from the 0.3-10 keV band, the central column has data extrapolated from the 15-50 keV band, and the right column has BAT data extrapolated from the 15-50 keV band and the XRT data extrapolated from the 0.3-10 keV band. Orange and purple data points are the BAT and XRT data, respectively. Solid black line - total luminosity of model; dashed line - propeller luminosity; dotted line - dipole luminosity.}
    \label{fig:all_cf_lightcurves}
\end{figure*}
\begin{figure*}\ContinuedFloat
    \centering
    \addtolength{\tabcolsep}{-0.4em}
    \begin{tabular}{ccc}
        {\Large 0.3-10 keV} & {\Large 15-50 keV} & {\Large Native}\\ \hline \\
        \includegraphics[width=0.61\columnwidth]{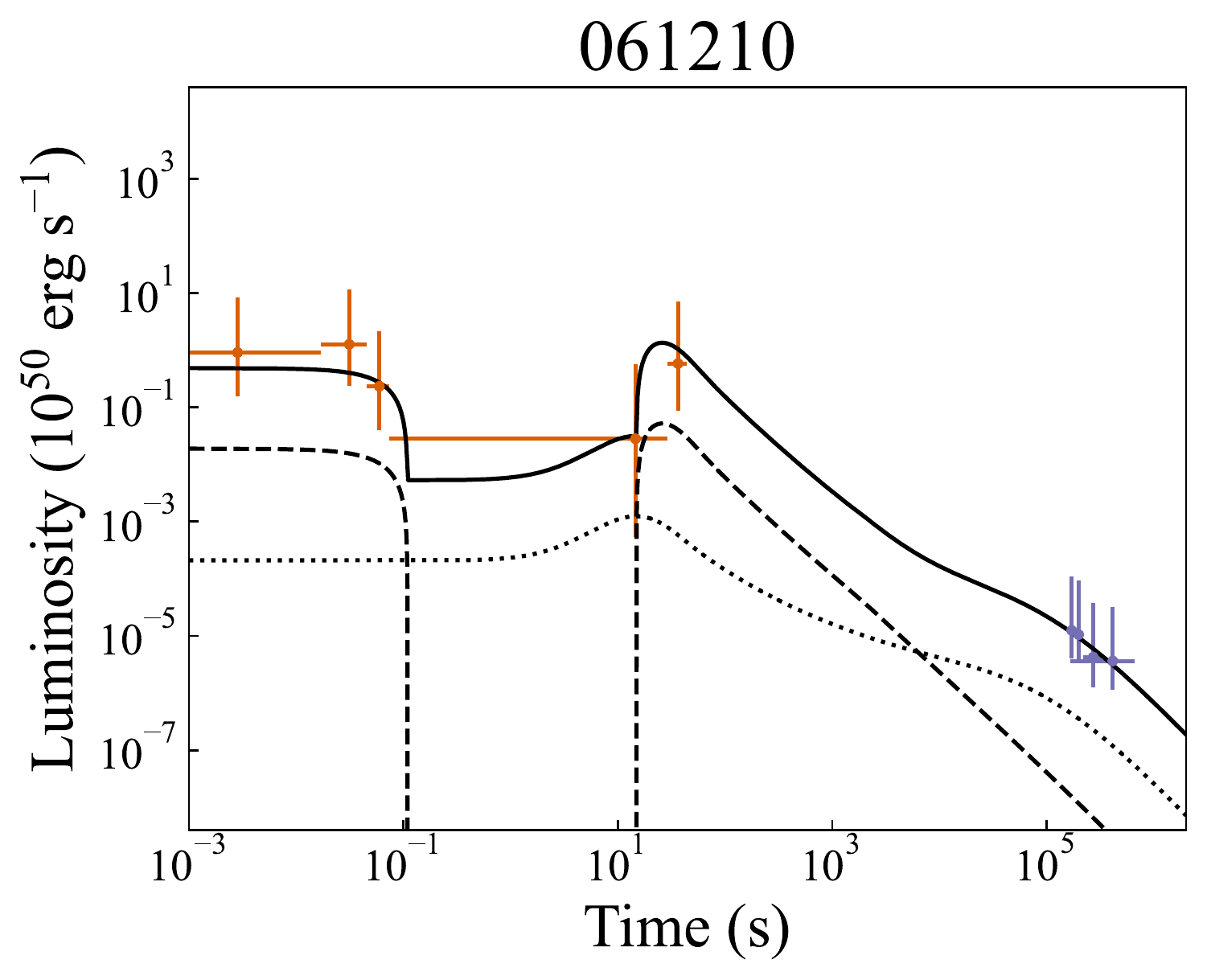} &
        \includegraphics[width=0.61\columnwidth]{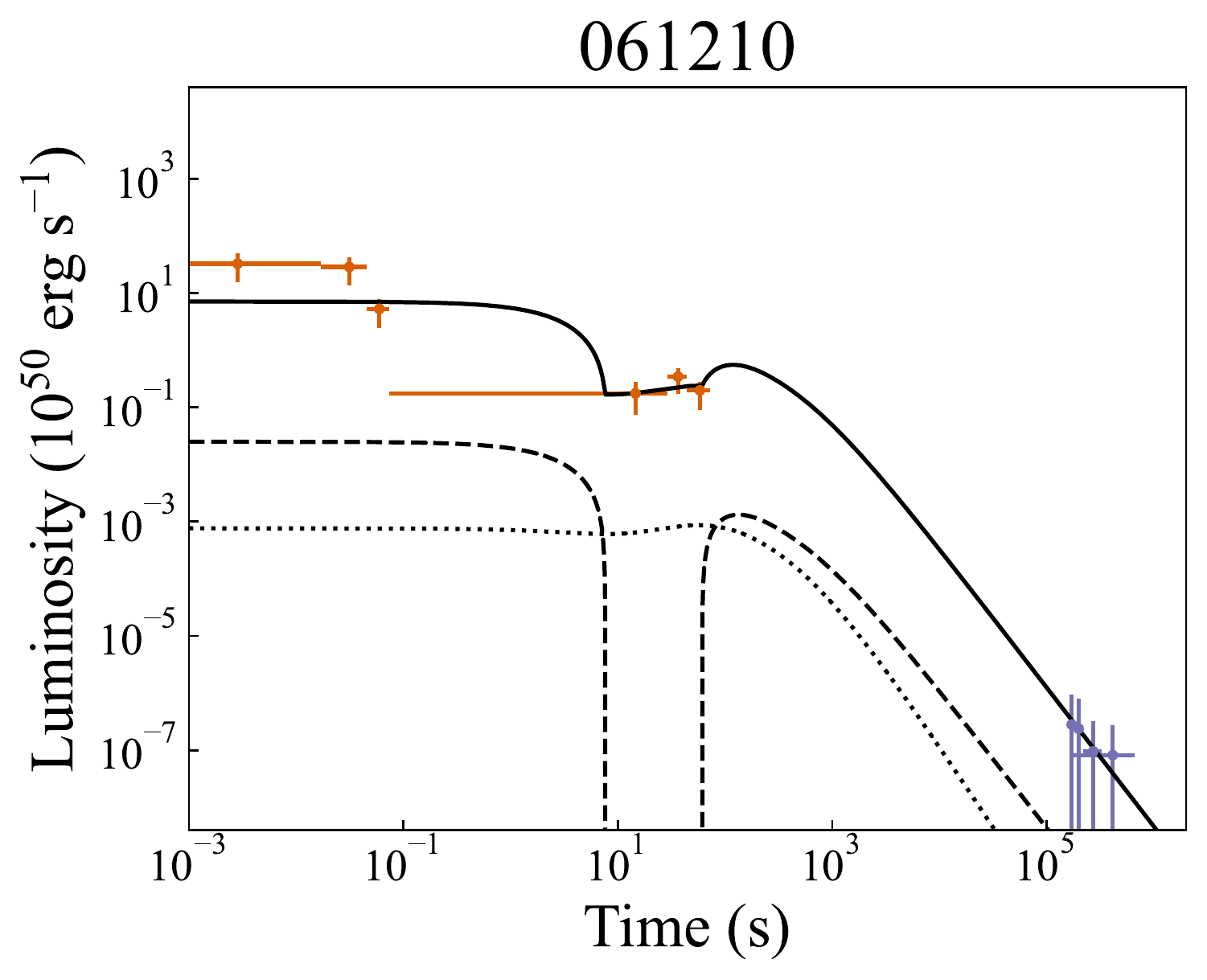} & 
        \includegraphics[width=0.61\columnwidth]{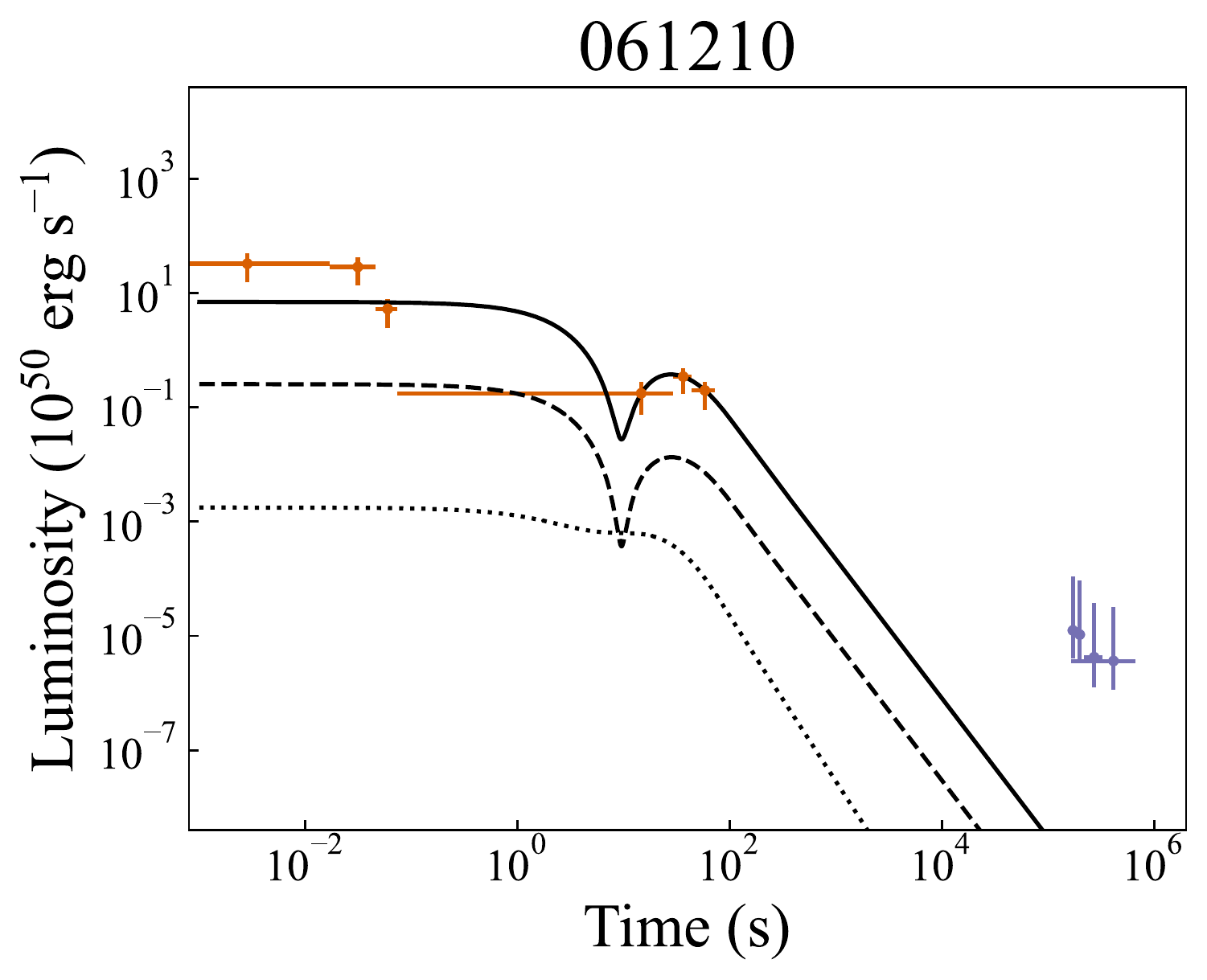}\\
        \includegraphics[width=0.61\columnwidth]{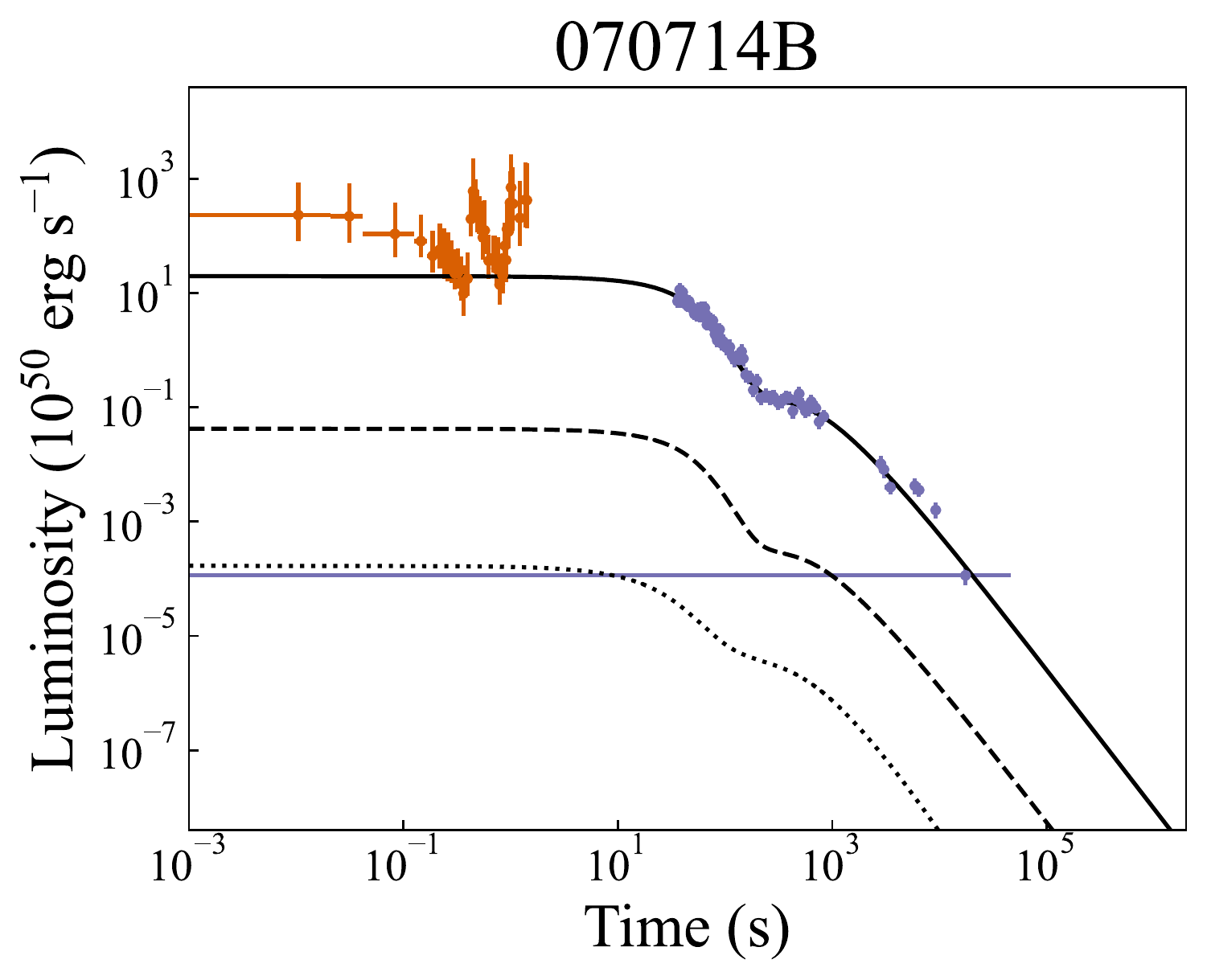} &
        \includegraphics[width=0.61\columnwidth]{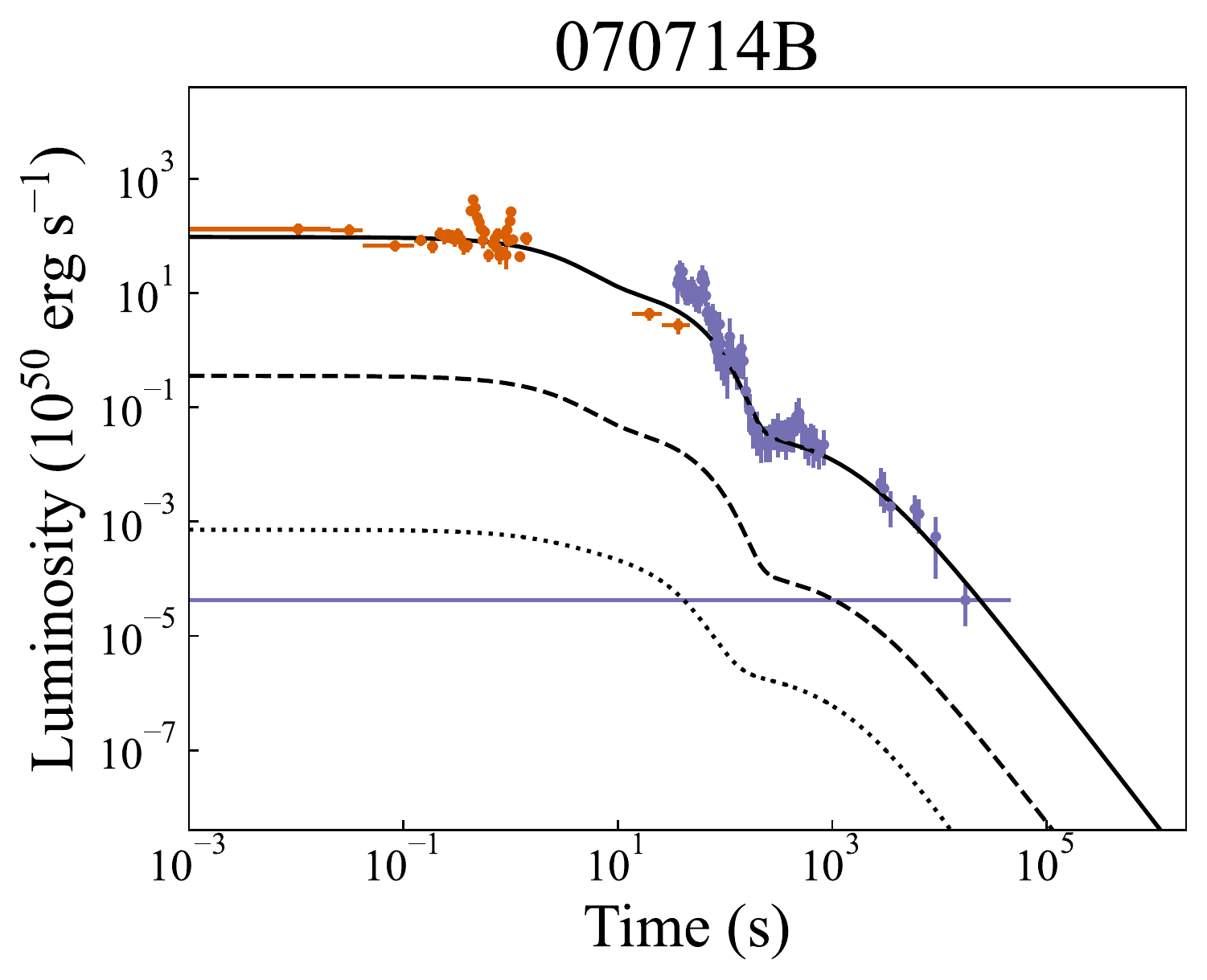} & 
        \includegraphics[width=0.61\columnwidth]{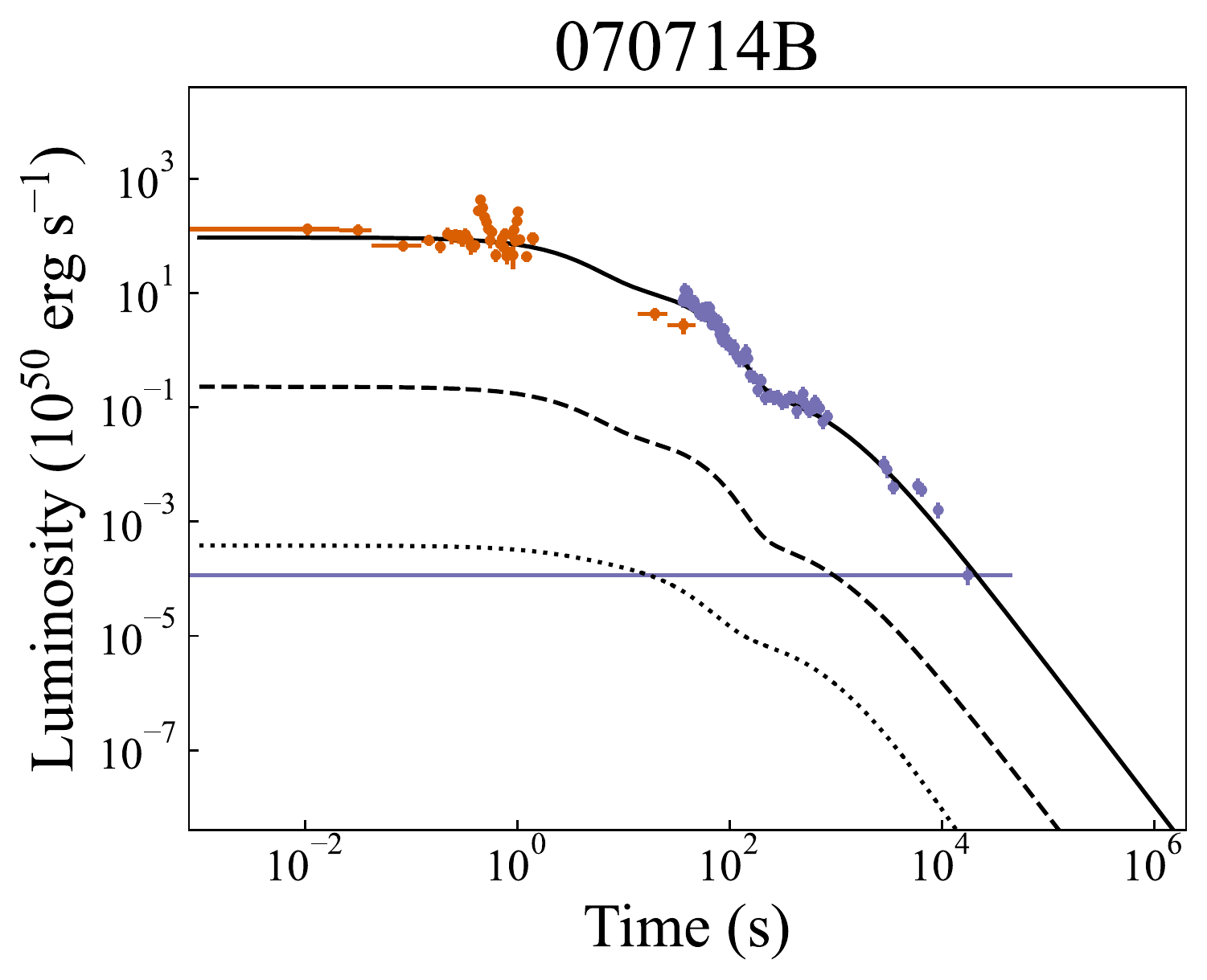}\\
        \includegraphics[width=0.61\columnwidth]{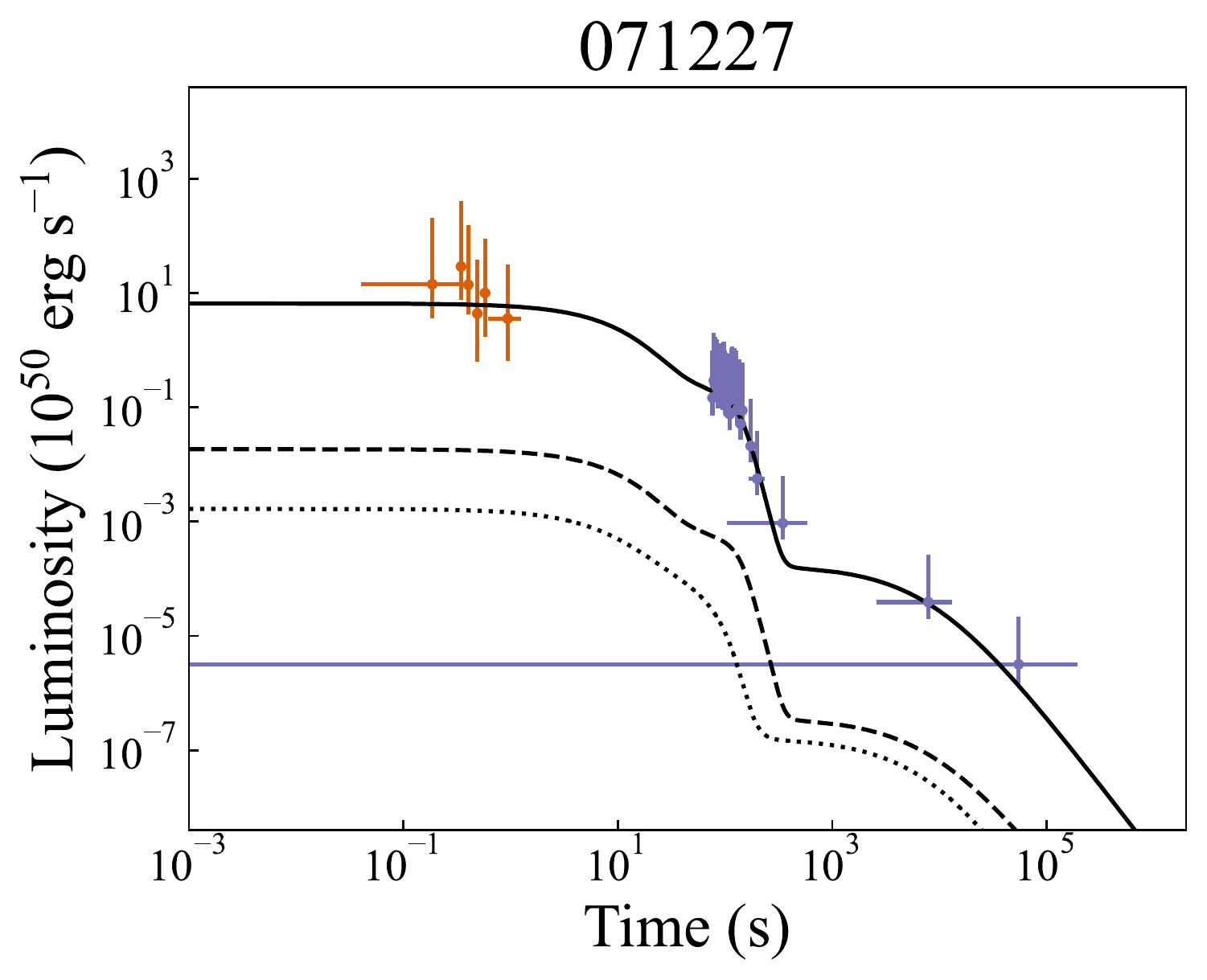} &
        \includegraphics[width=0.61\columnwidth]{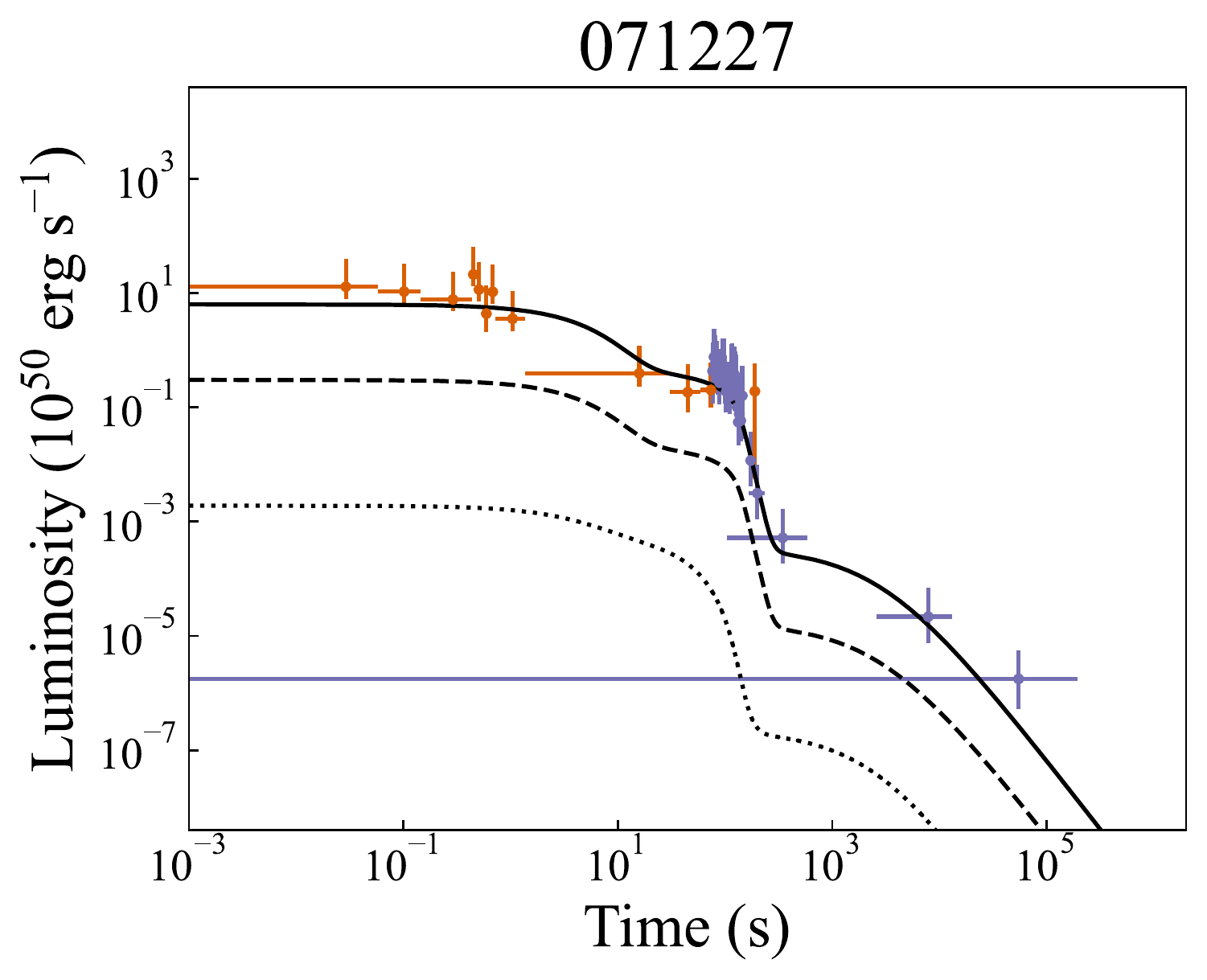} & 
        \includegraphics[width=0.61\columnwidth]{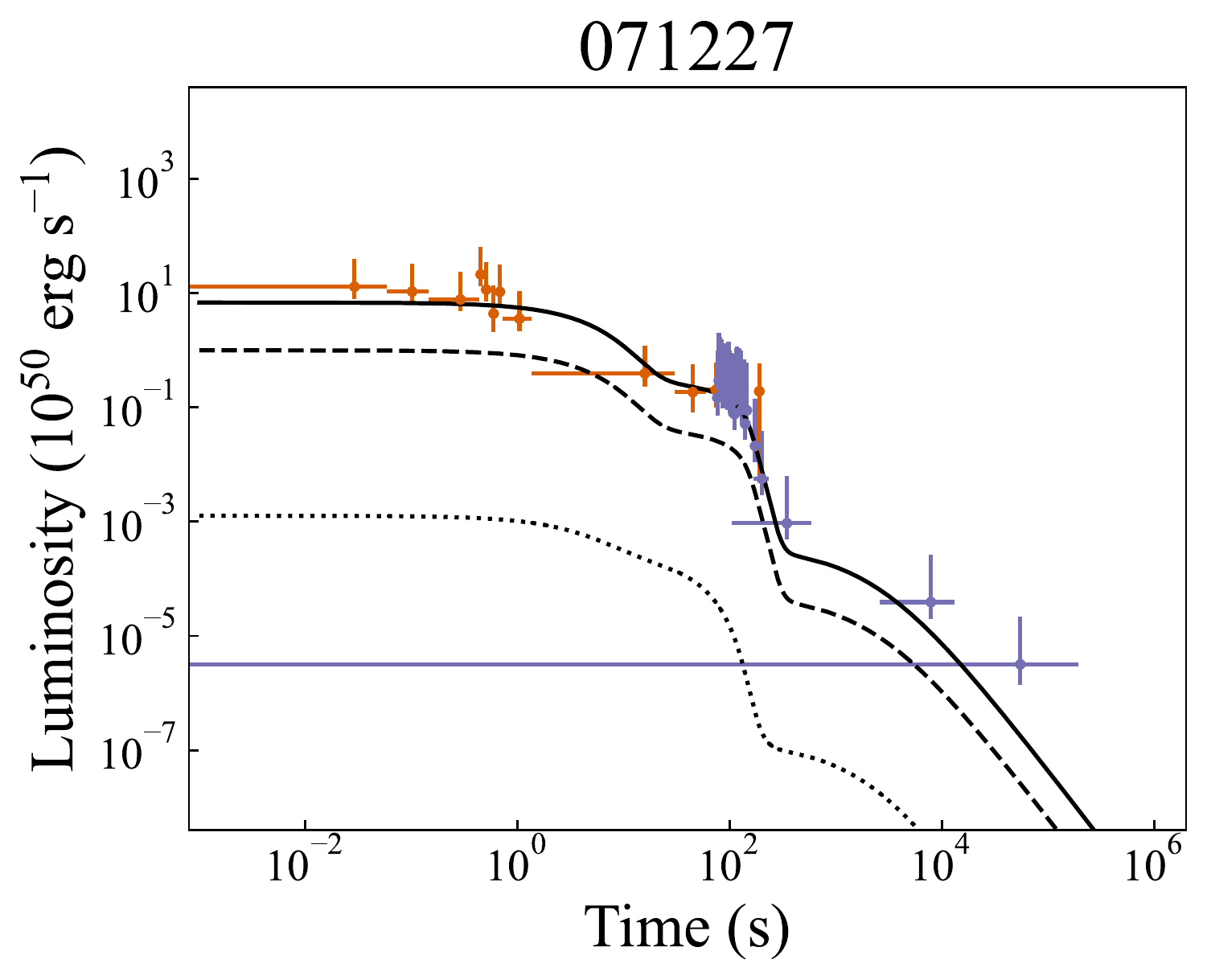}\\
        \includegraphics[width=0.61\columnwidth]{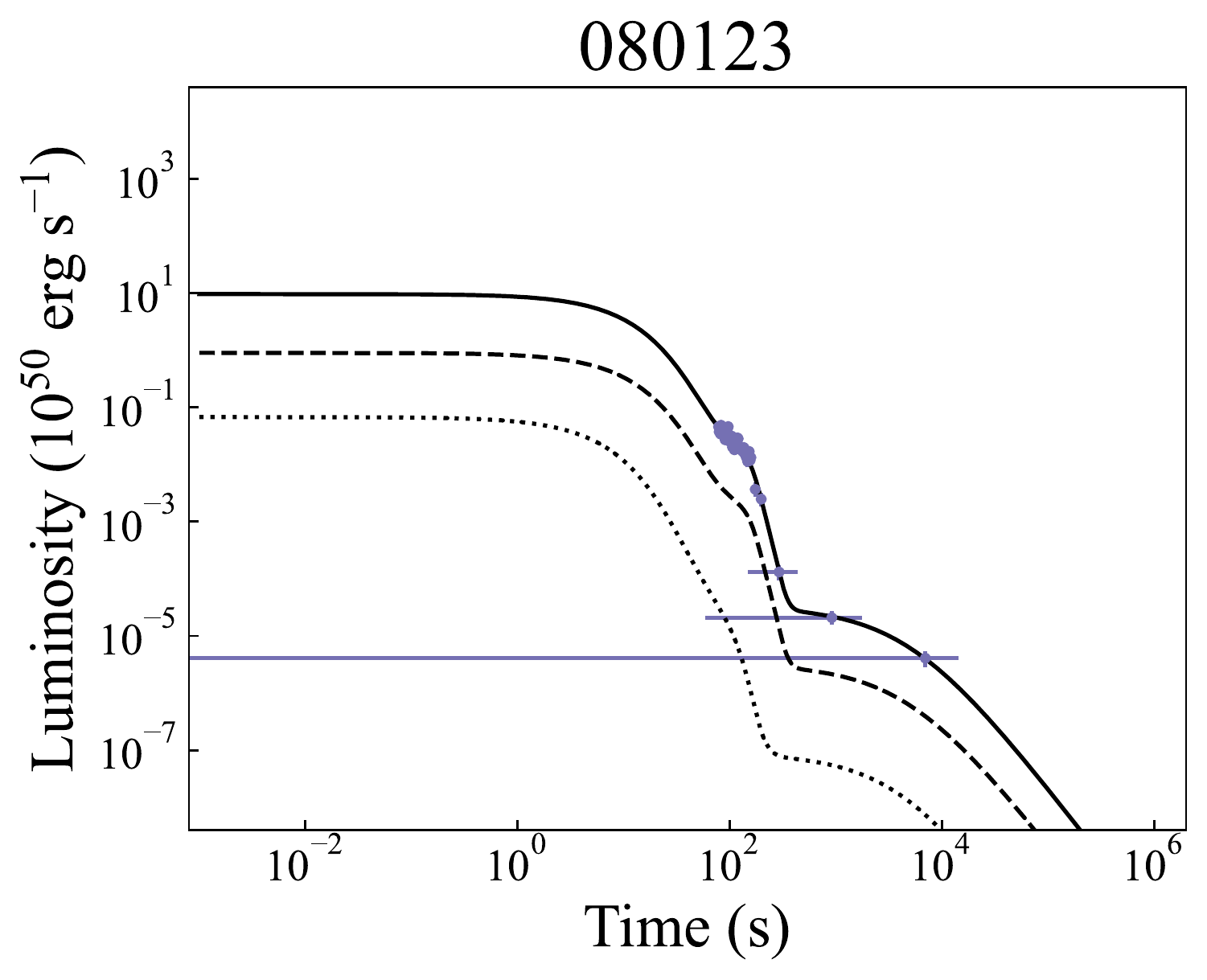} &
        \includegraphics[width=0.61\columnwidth]{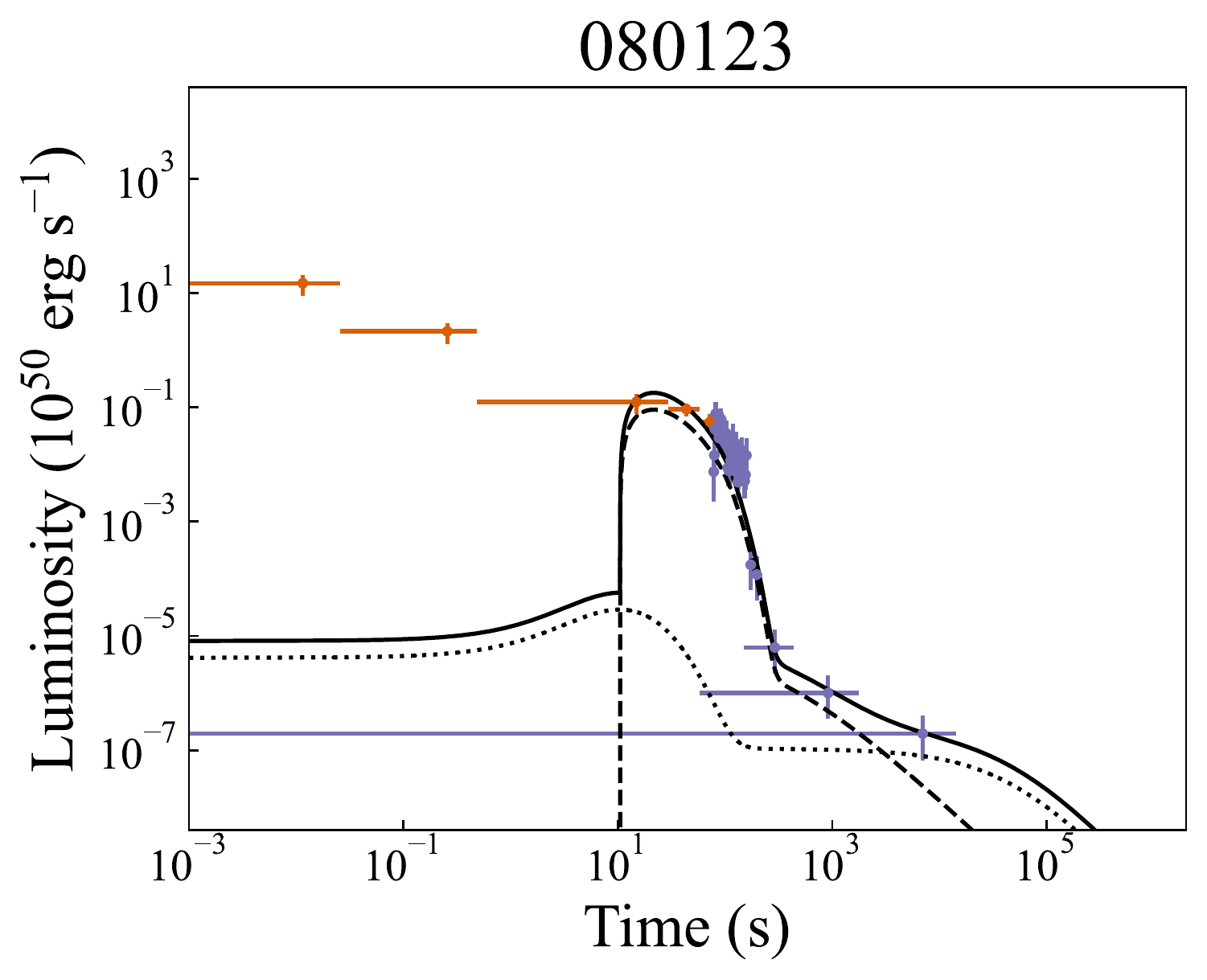} & 
        \includegraphics[width=0.61\columnwidth]{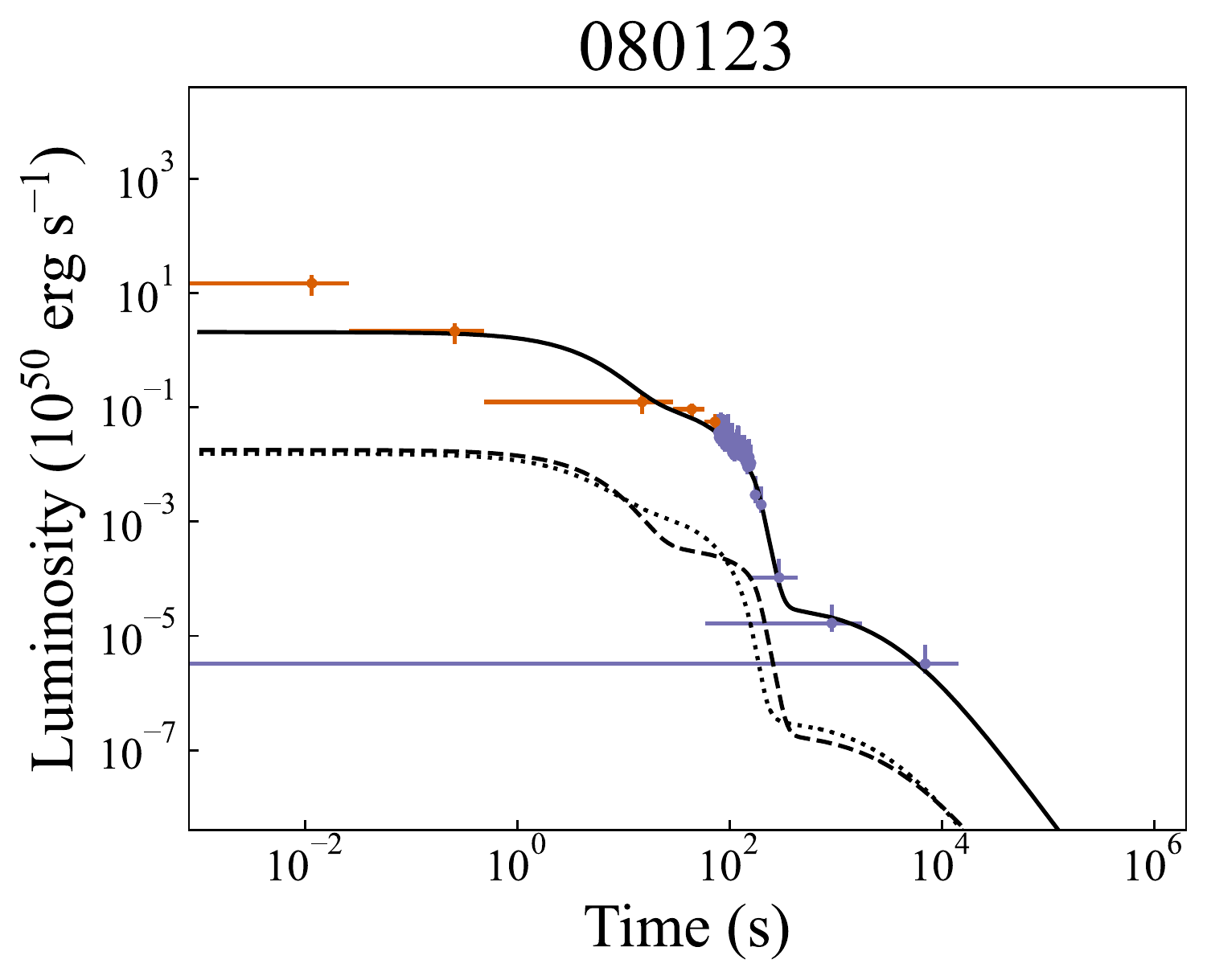}\\
        \includegraphics[width=0.61\columnwidth]{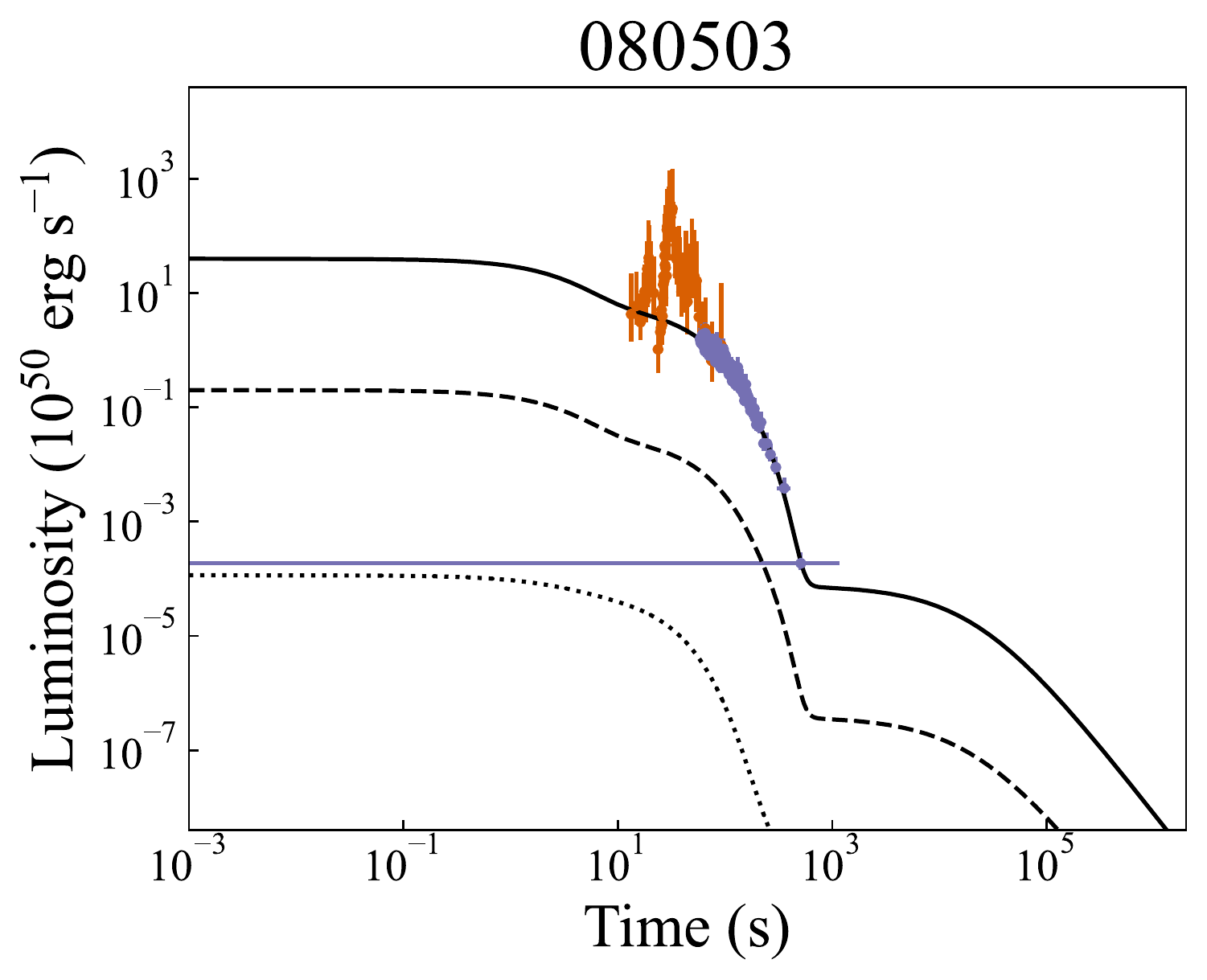} &
        \includegraphics[width=0.61\columnwidth]{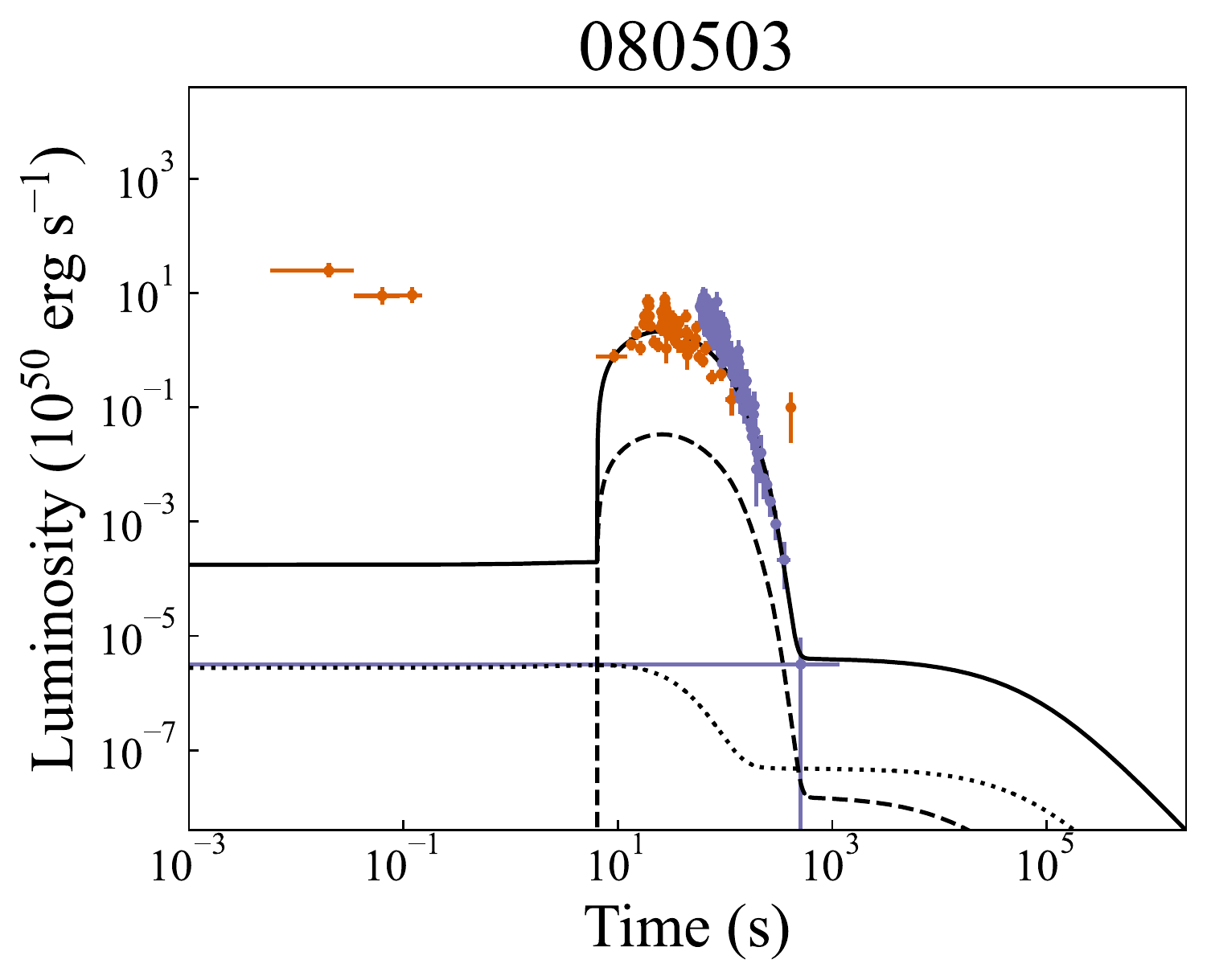} & 
        \includegraphics[width=0.61\columnwidth]{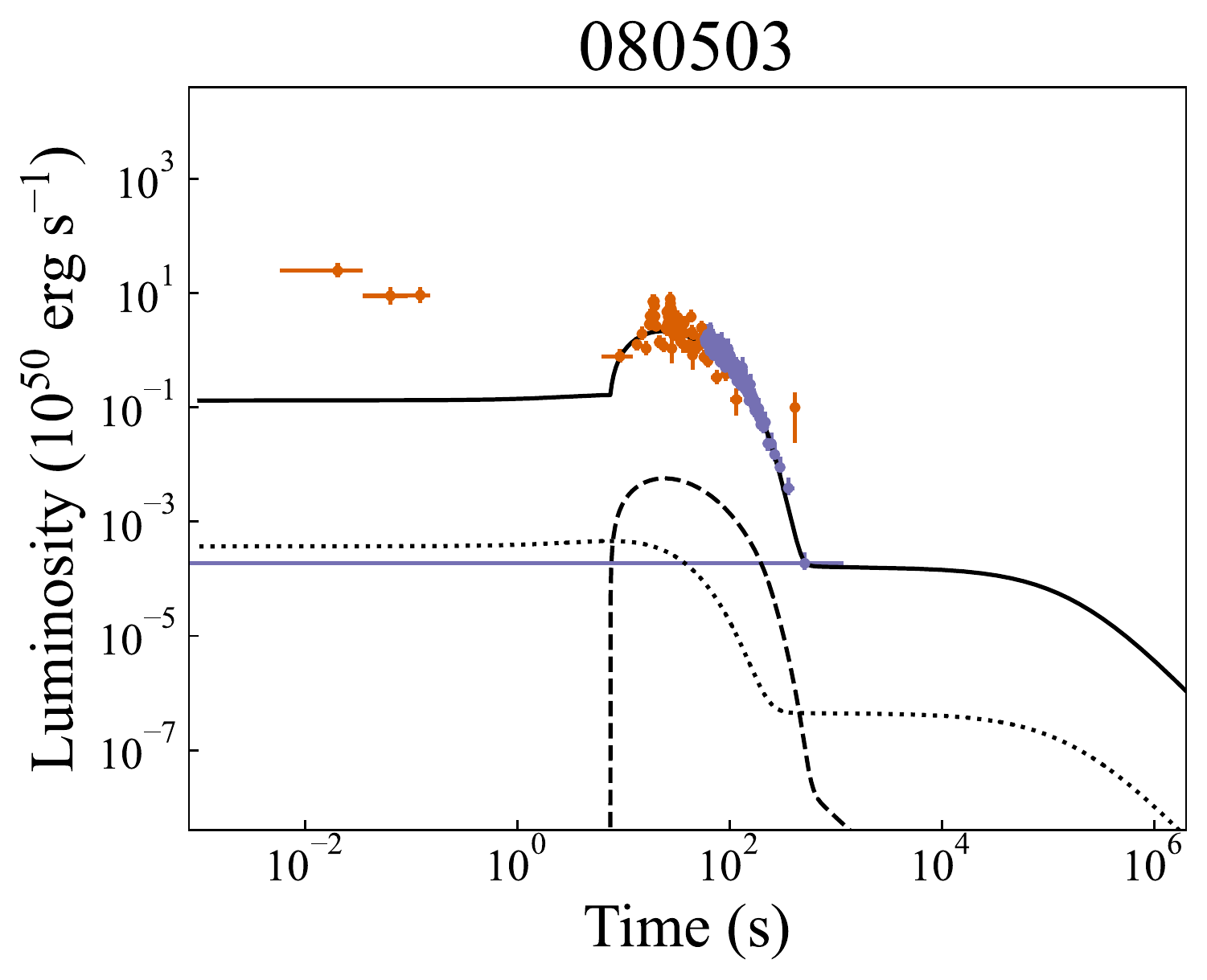}\\
    \end{tabular}
    \caption{Best fits to data with propagated conversion factor errors (cont.).}
    \label{fig:all_cf_lightcurves}
\end{figure*}
\begin{figure*}\ContinuedFloat
    \centering
    \addtolength{\tabcolsep}{-0.4em}
    \begin{tabular}{ccc}
        {\Large 0.3-10 keV} & {\Large 15-50 keV} & {\Large Native}\\ \hline \\
        \includegraphics[width=0.61\columnwidth]{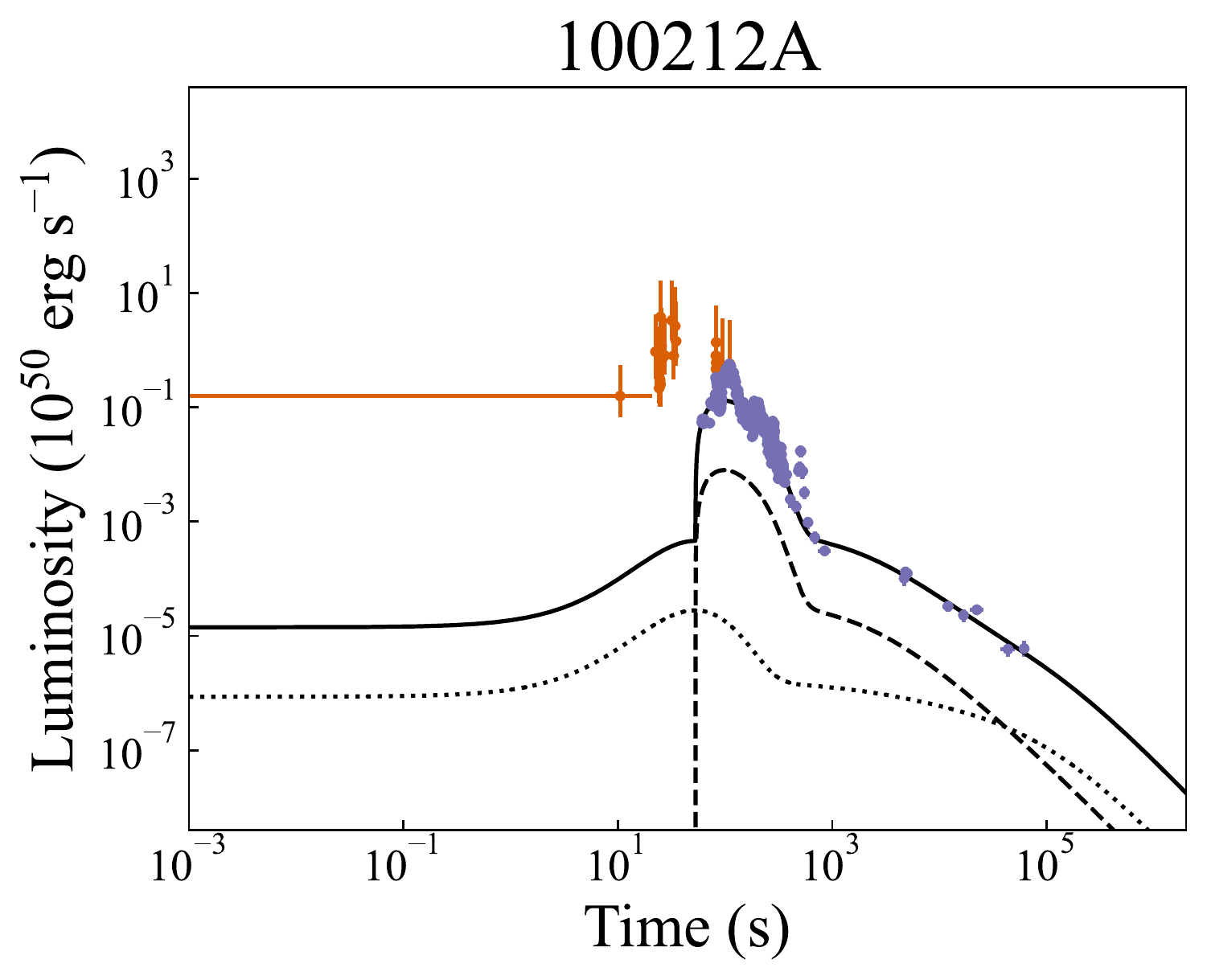} &
        \includegraphics[width=0.61\columnwidth]{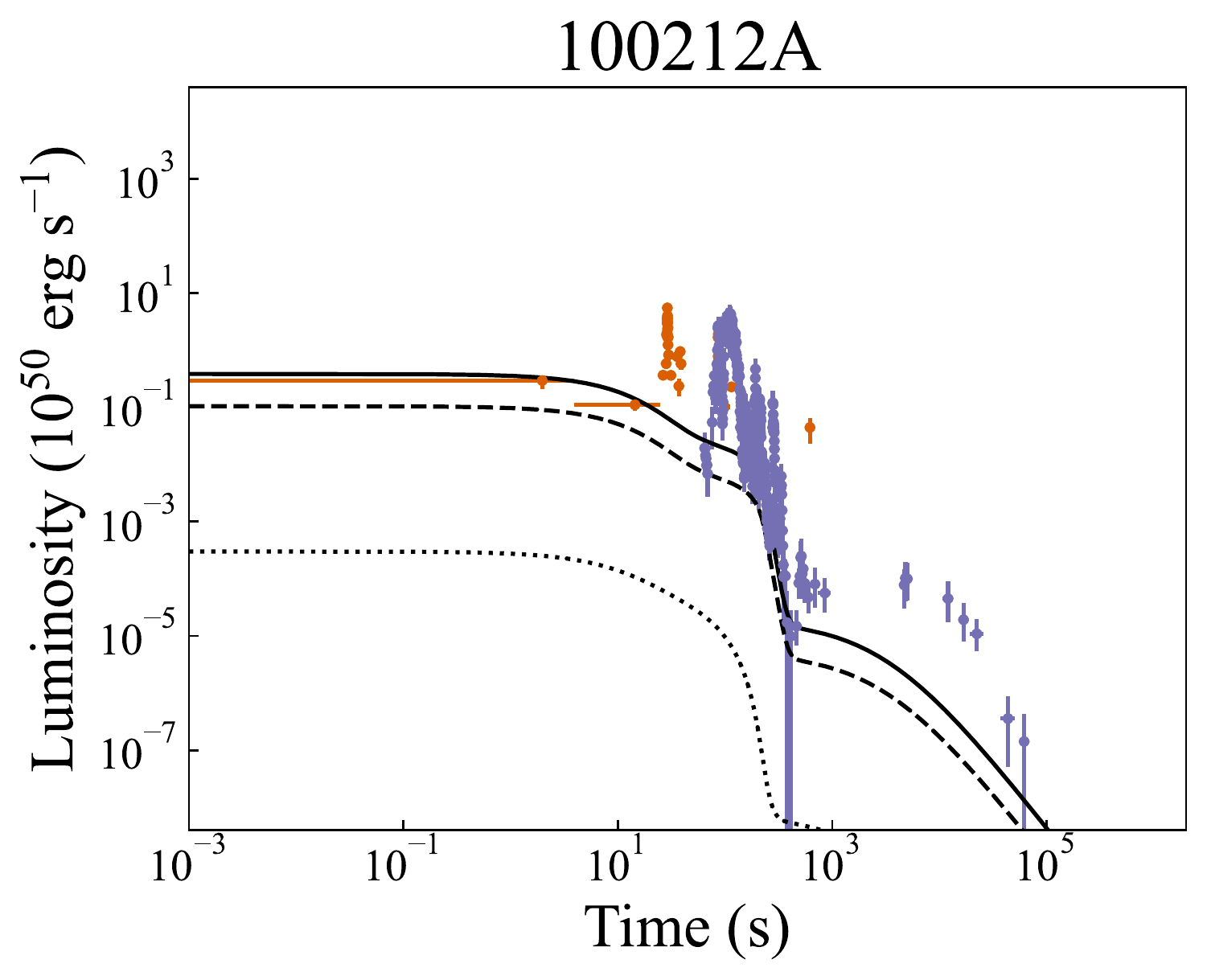} & 
        \includegraphics[width=0.61\columnwidth]{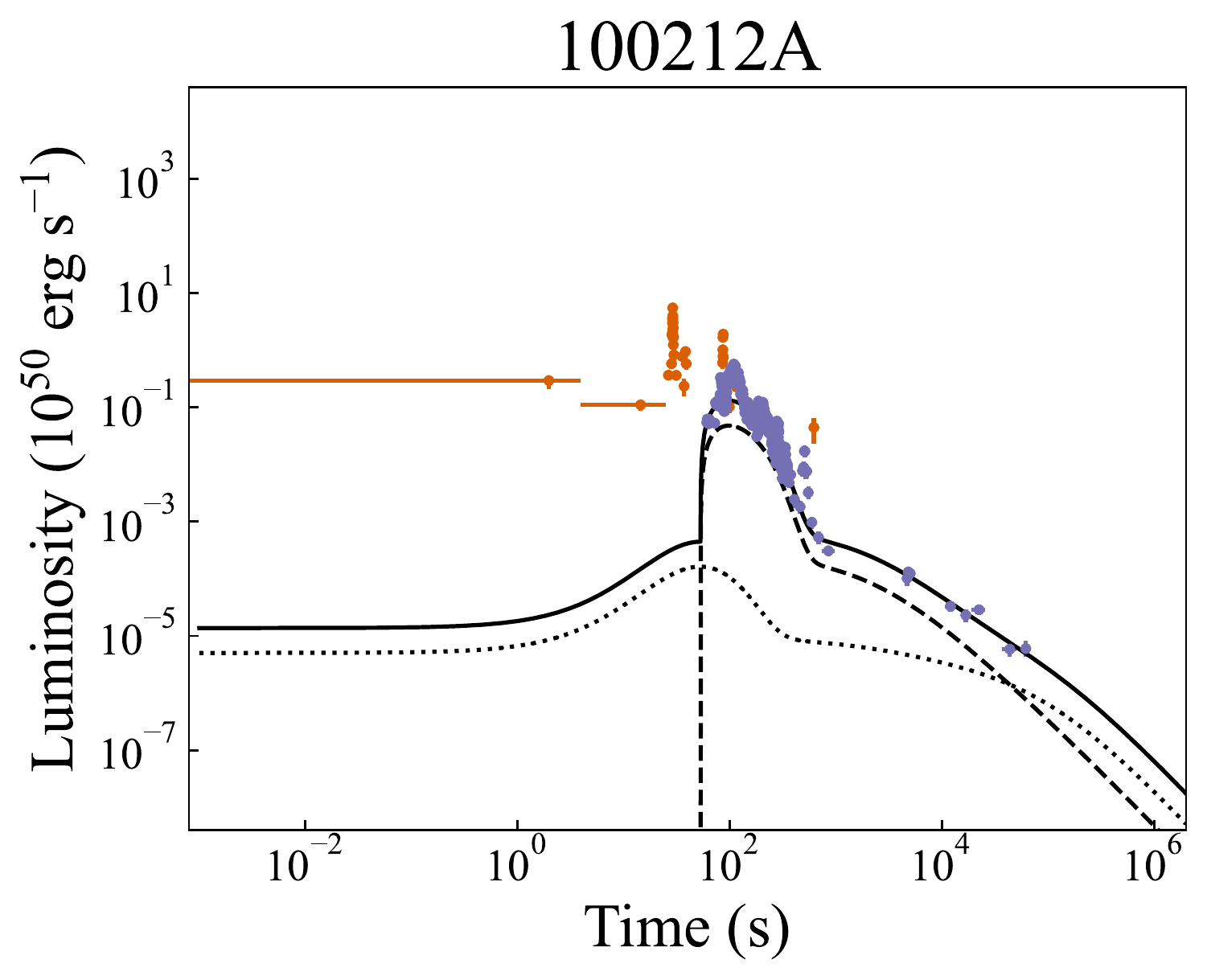}\\
        \includegraphics[width=0.61\columnwidth]{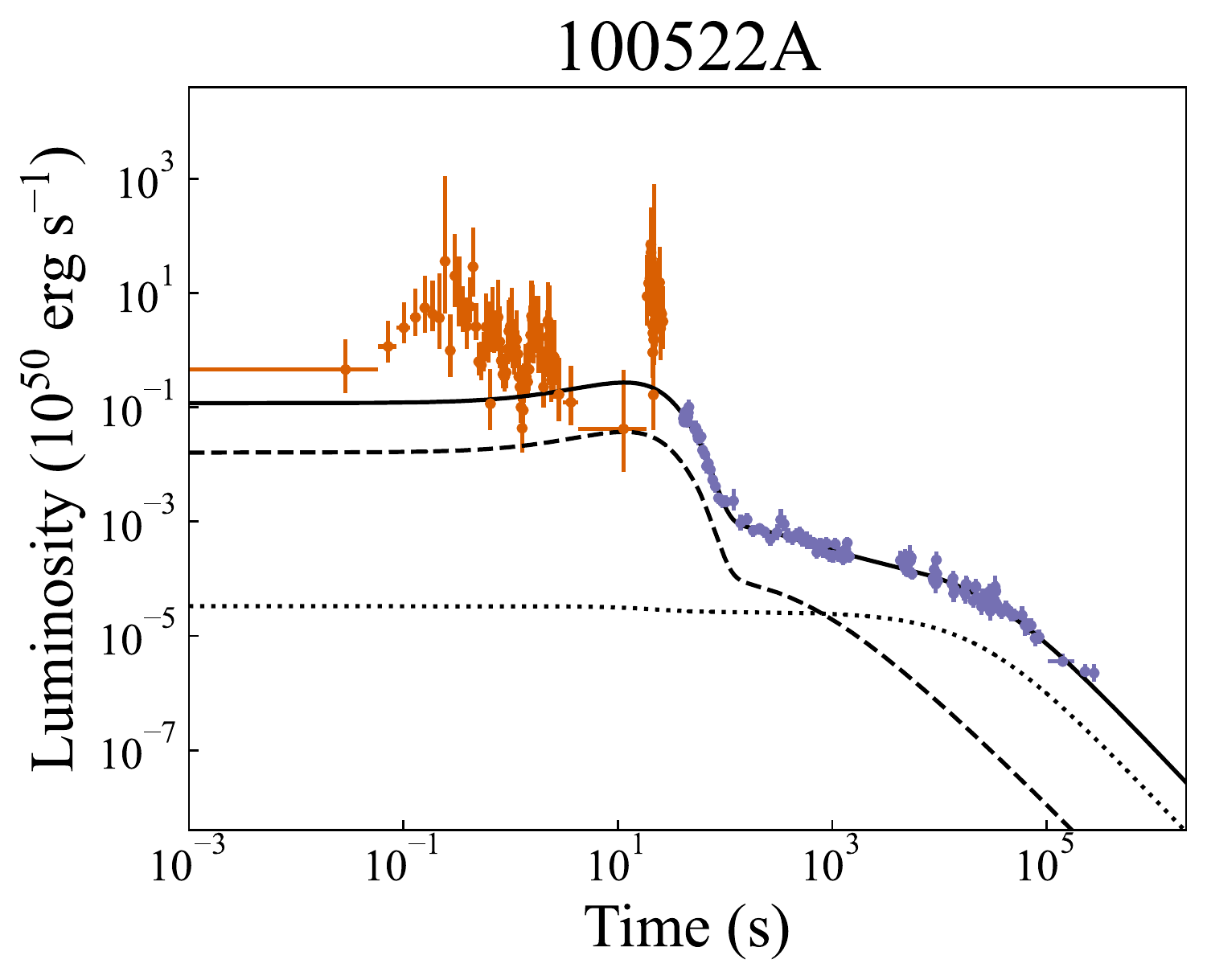} &
        \includegraphics[width=0.61\columnwidth]{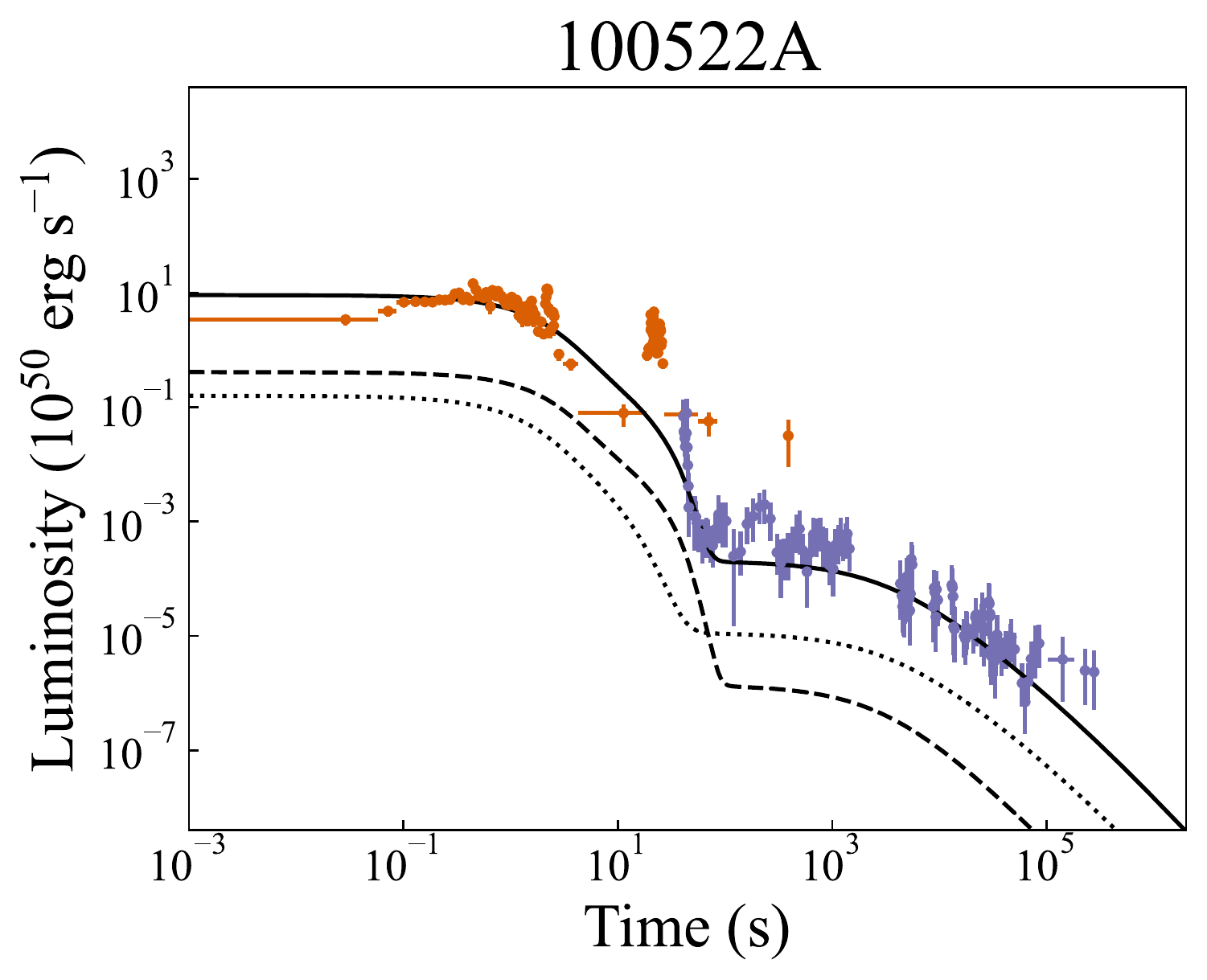} & 
        \includegraphics[width=0.61\columnwidth]{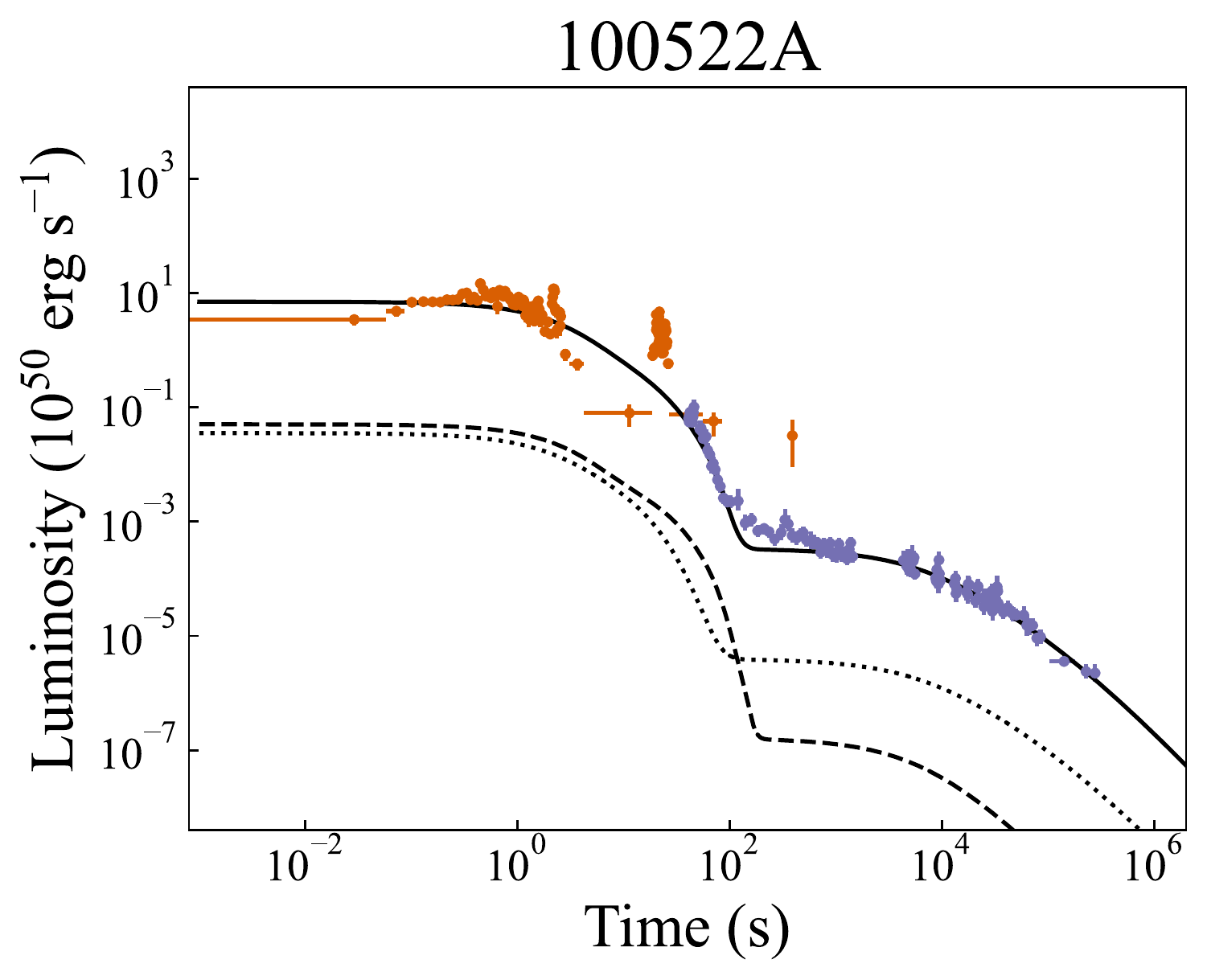}\\
        \includegraphics[width=0.61\columnwidth]{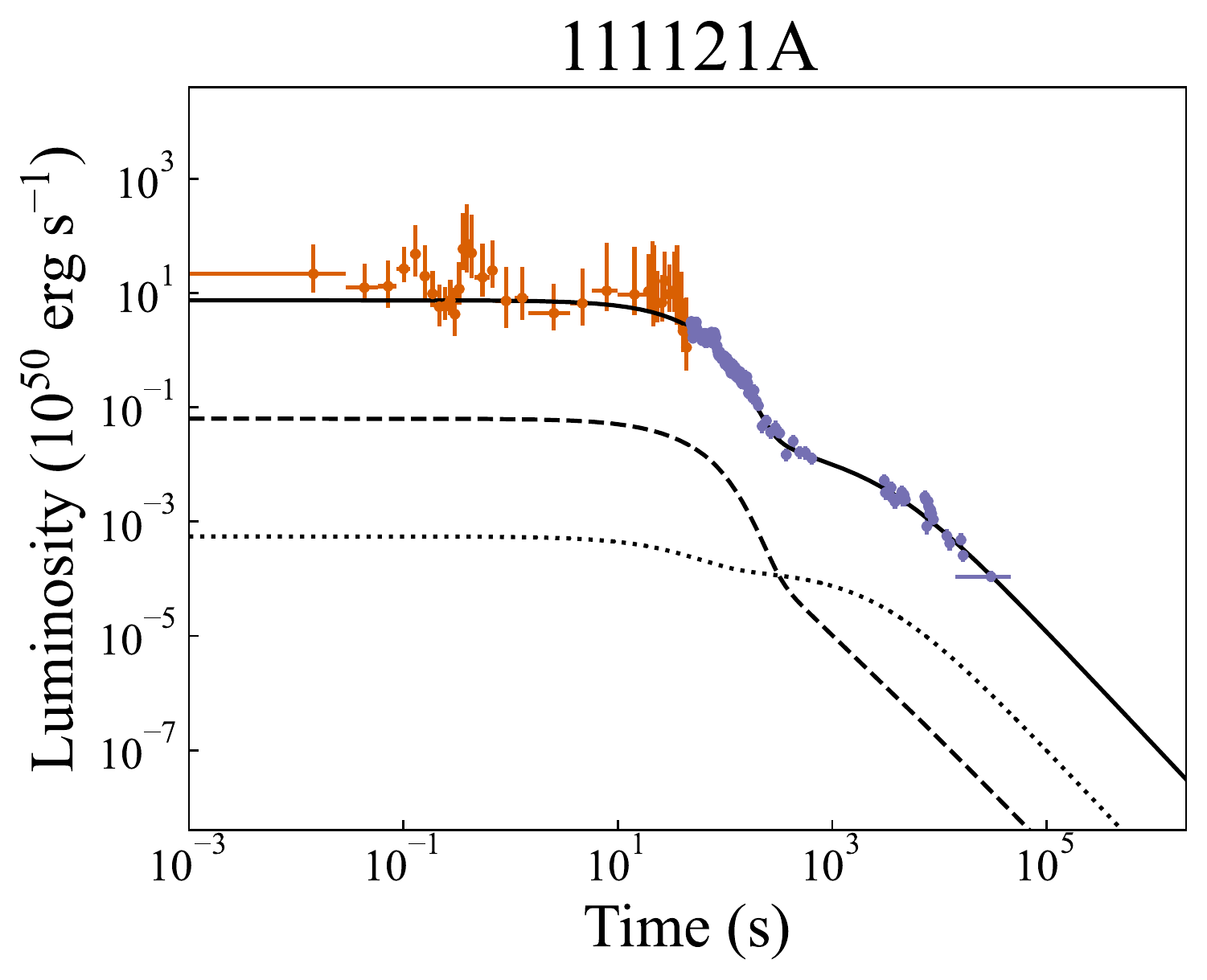} &
        \includegraphics[width=0.61\columnwidth]{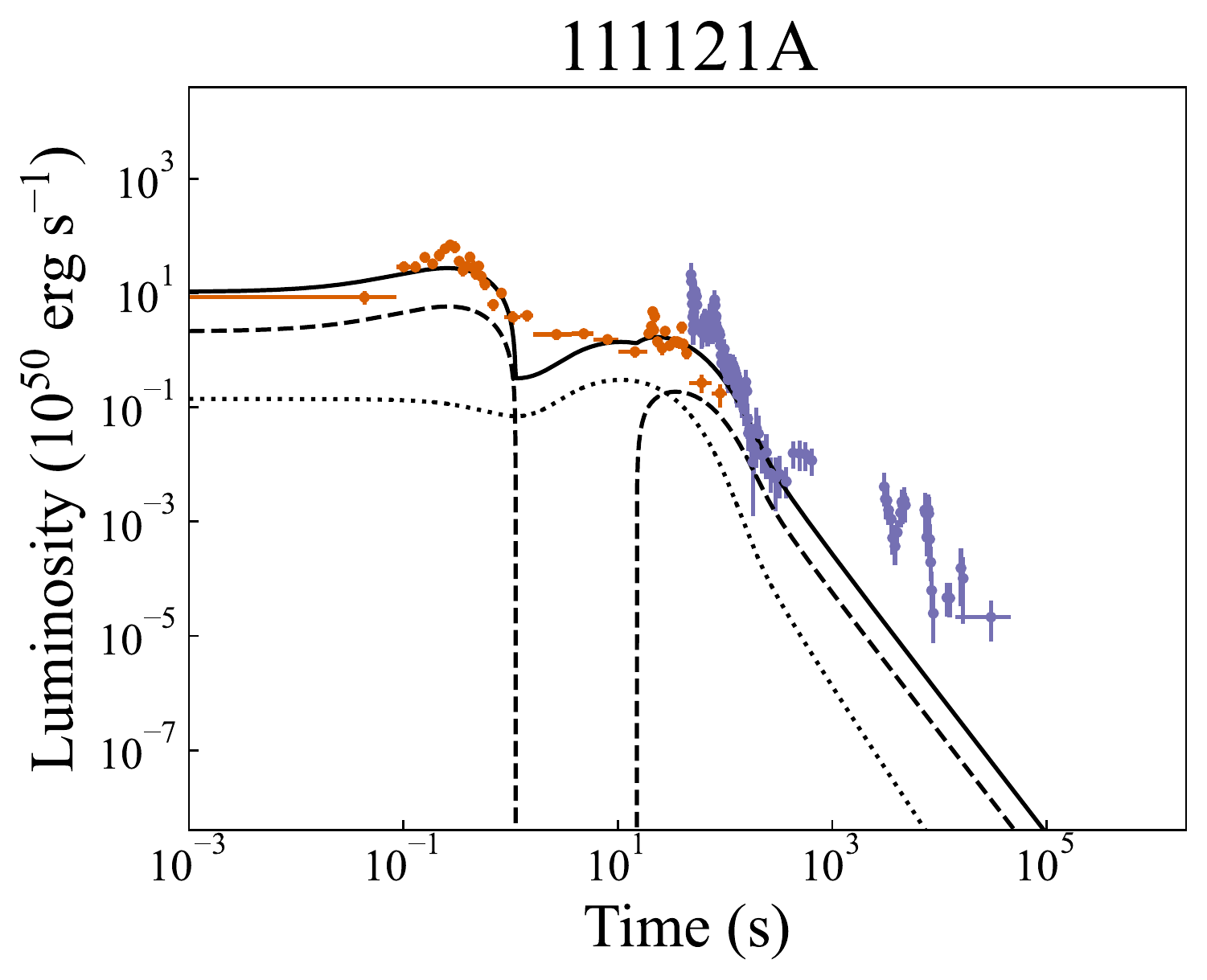} & 
        \includegraphics[width=0.61\columnwidth]{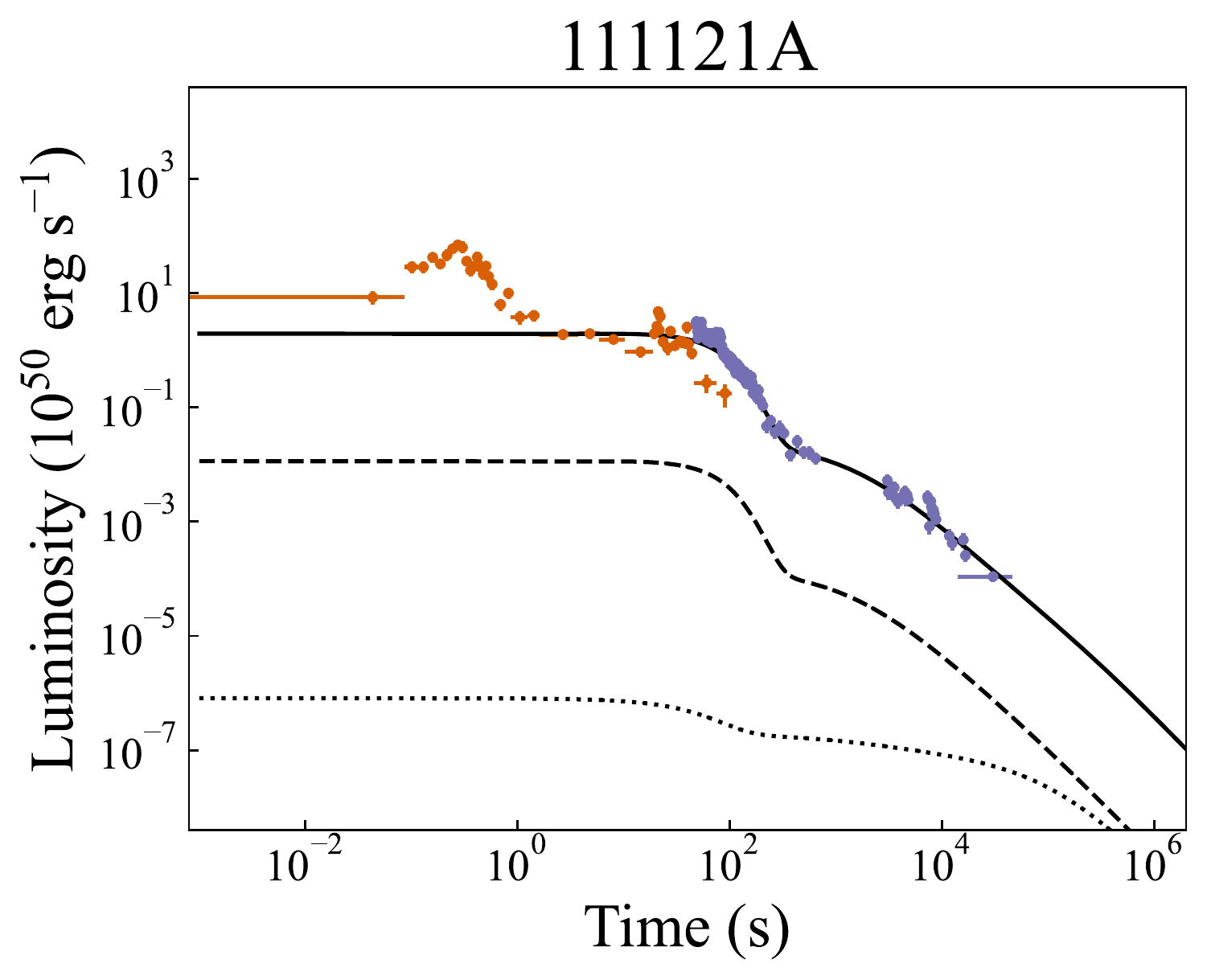}\\
        \includegraphics[width=0.61\columnwidth]{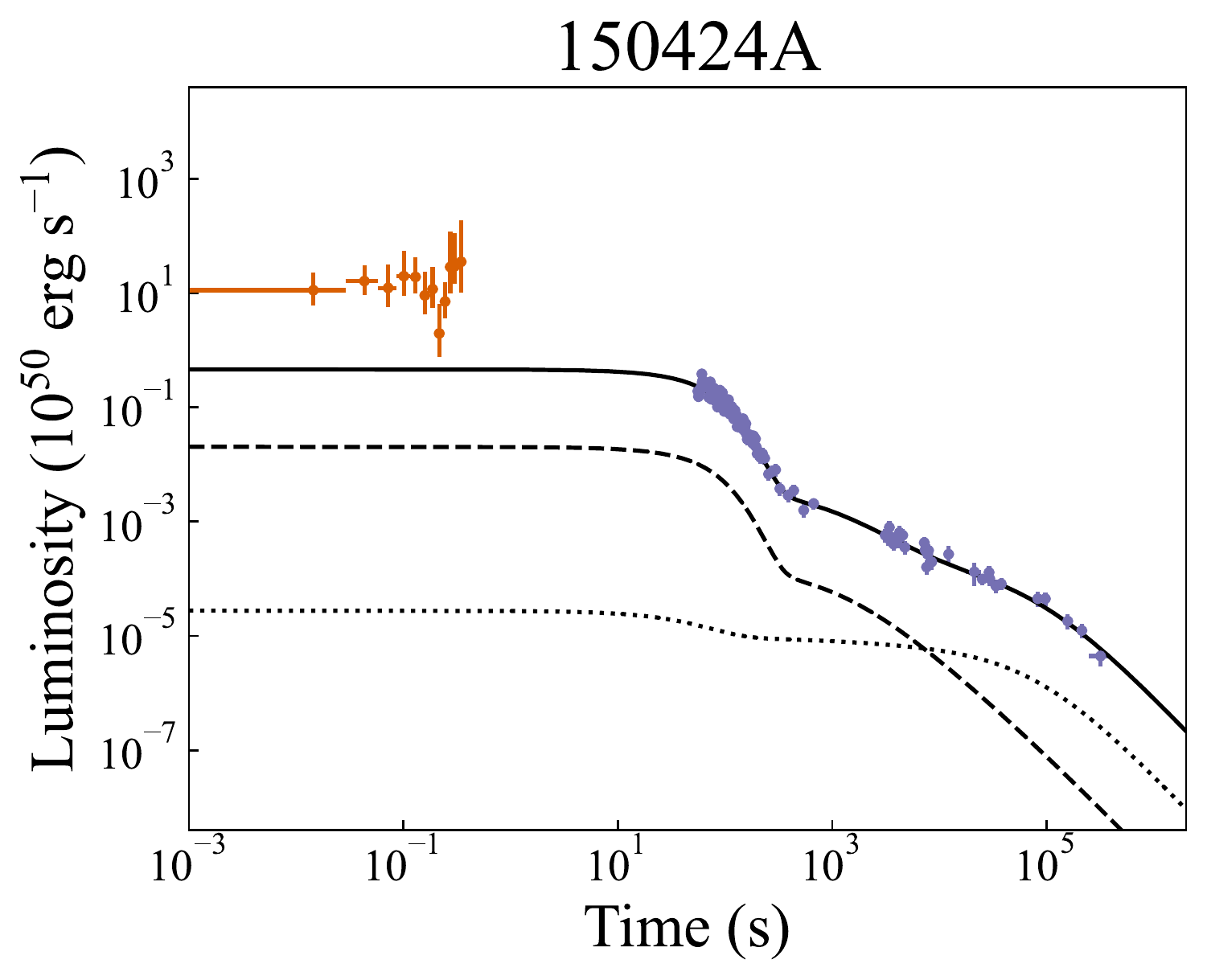} &
        \includegraphics[width=0.61\columnwidth]{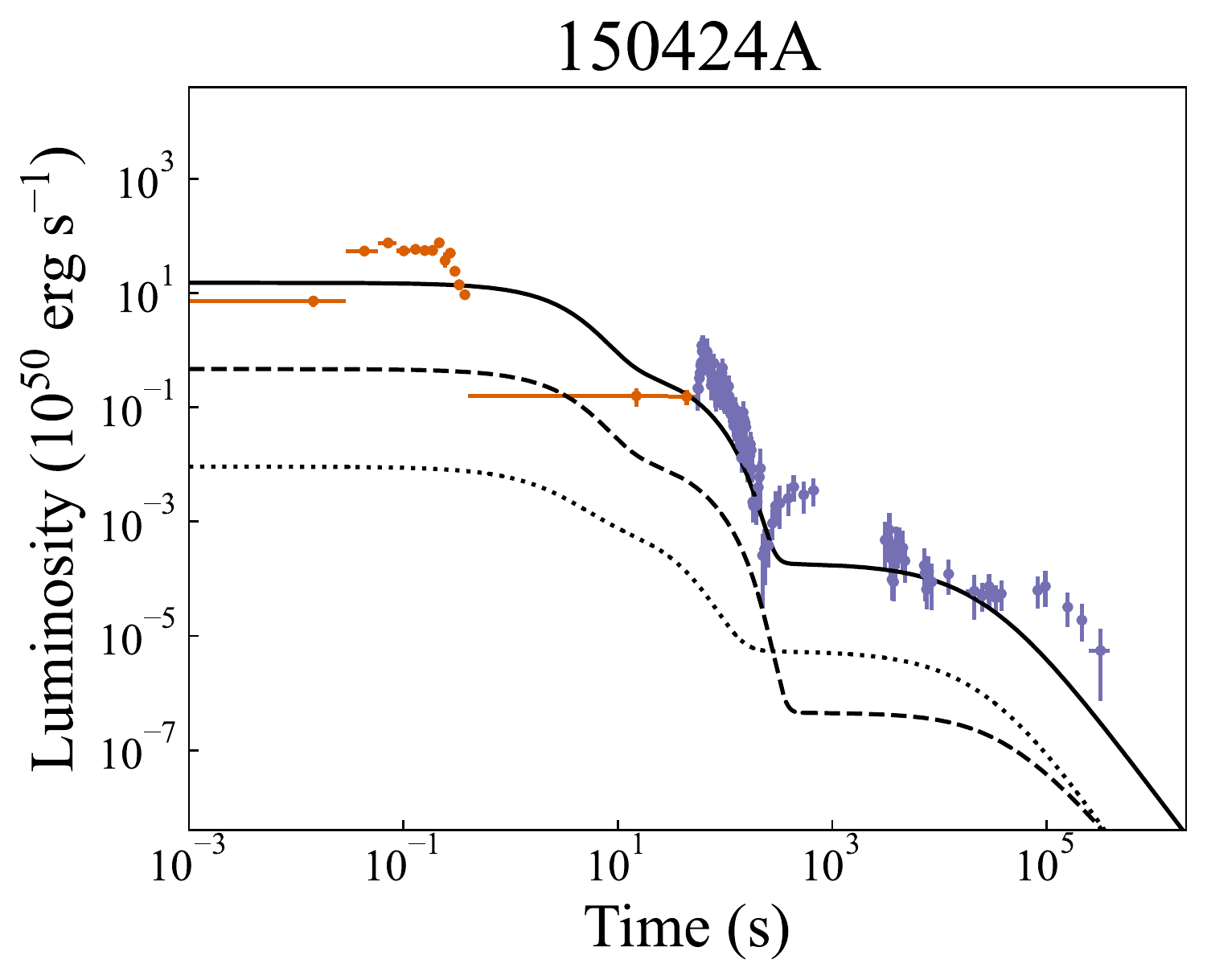} & 
        \includegraphics[width=0.61\columnwidth]{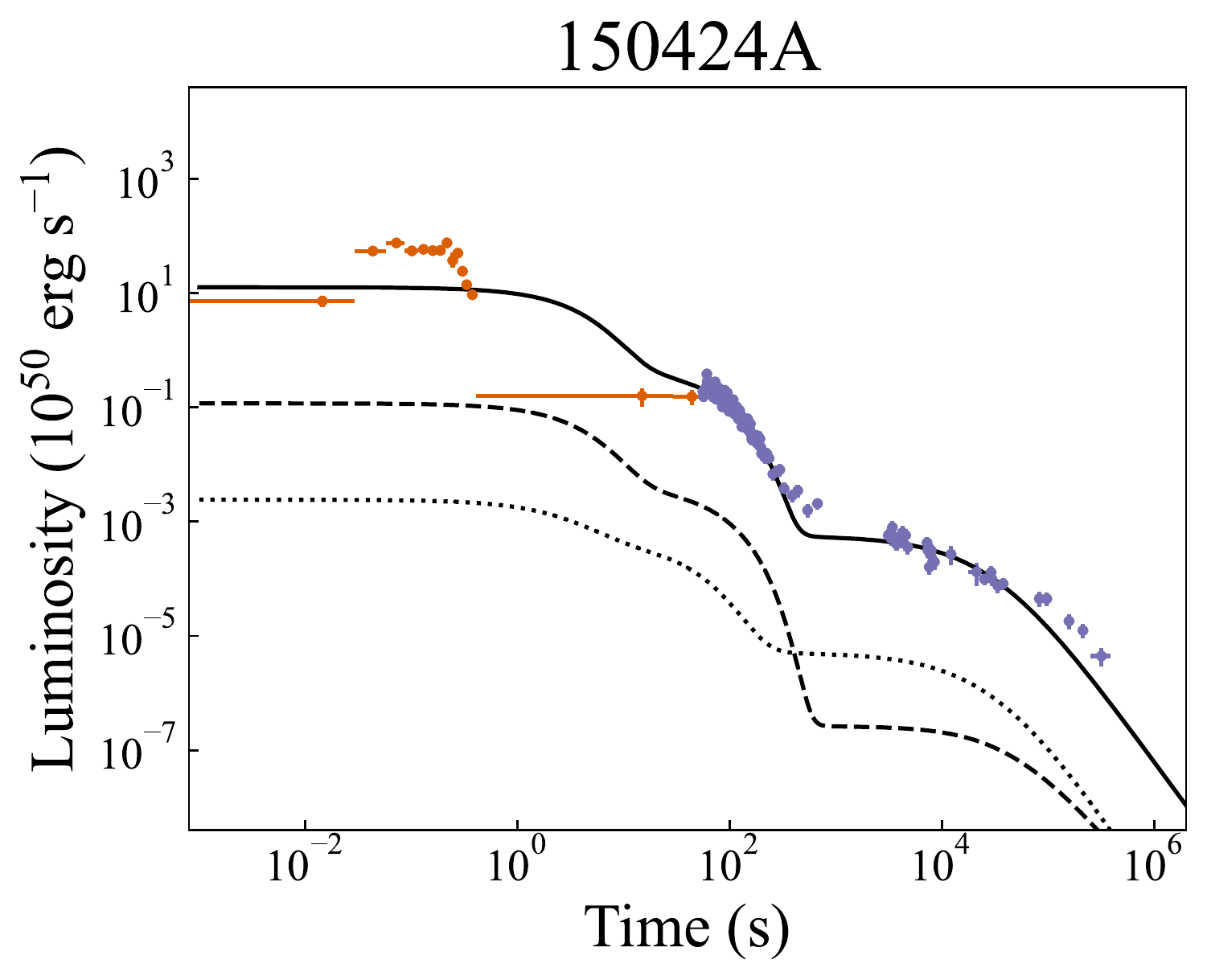}\\
        \includegraphics[width=0.61\columnwidth]{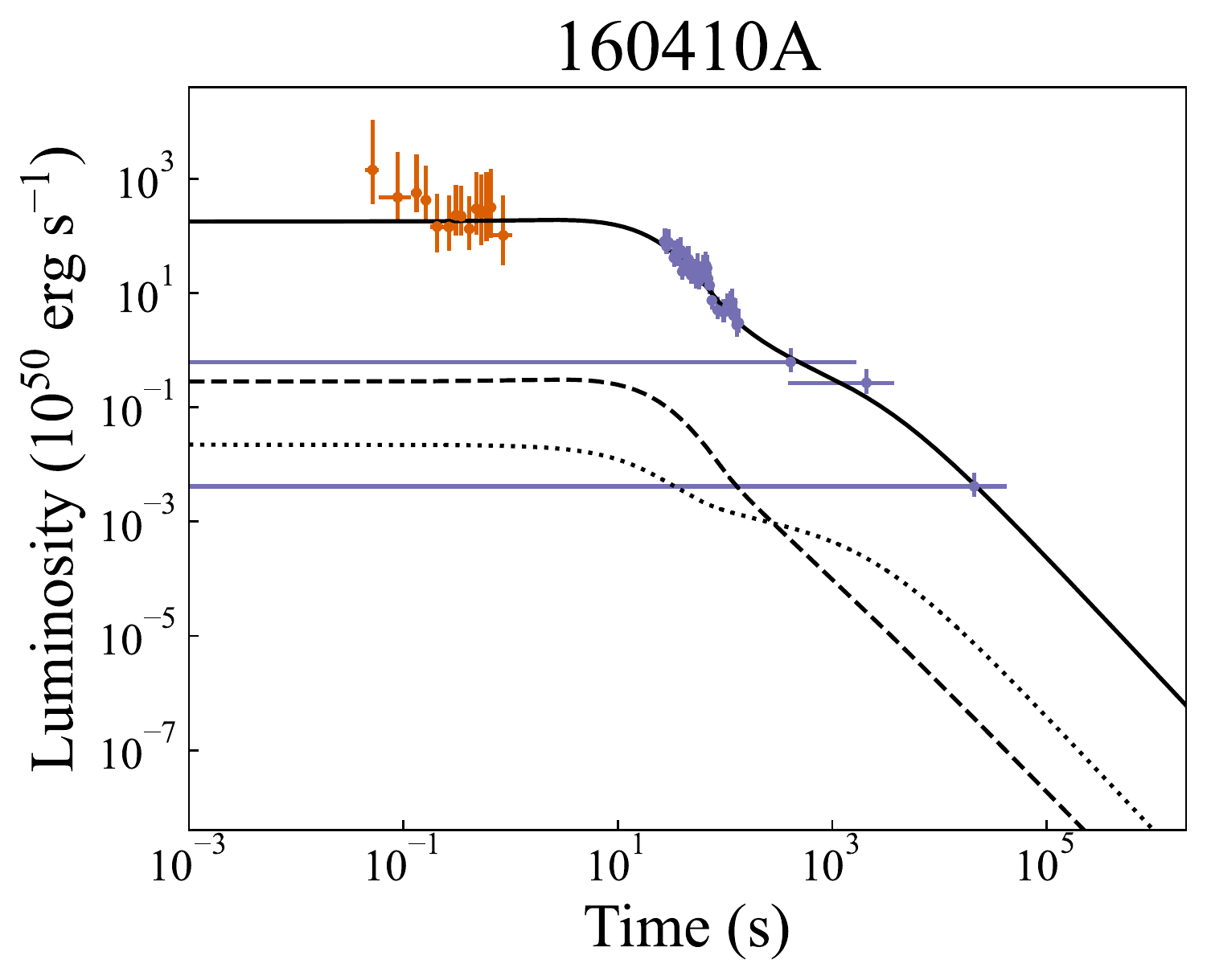} &
        \includegraphics[width=0.61\columnwidth]{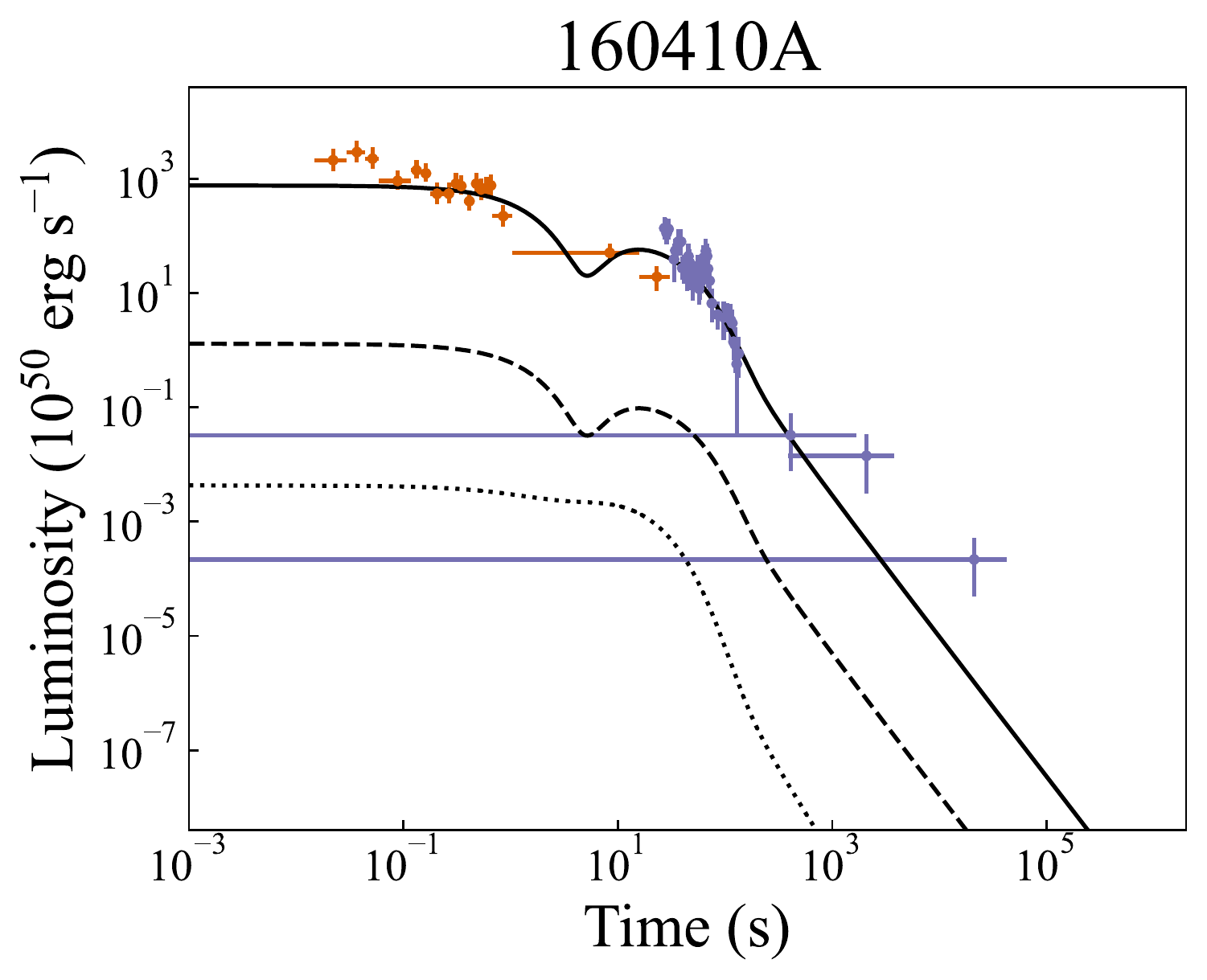} & 
        \includegraphics[width=0.61\columnwidth]{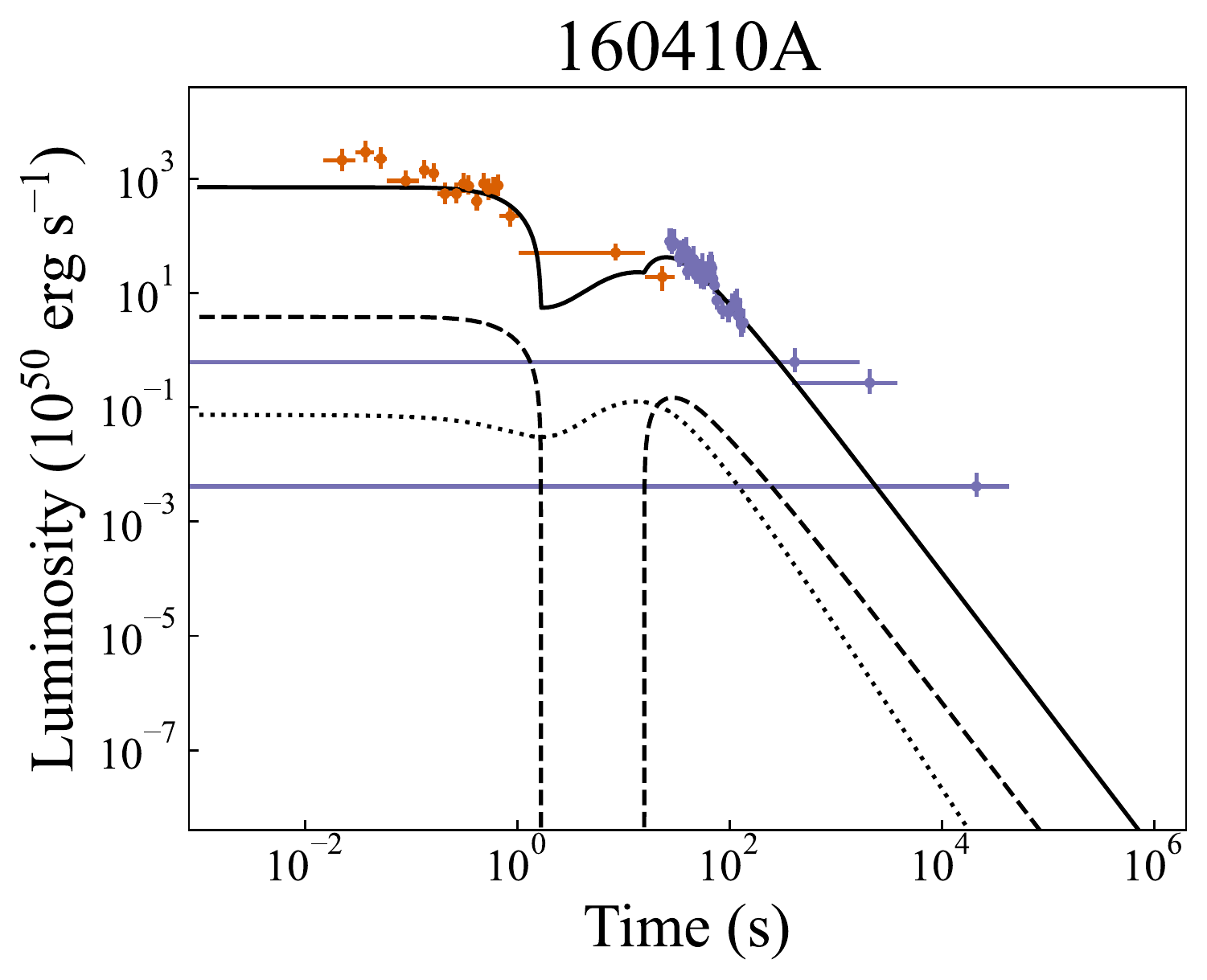}\\
    \end{tabular}
    \caption{Best fits to data with propagated conversion factor errors (cont.).}
    \label{fig:all_cf_lightcurves}
\end{figure*}

\section{Results and discussion}

\subsection{Light-curves}

Fig. \ref{fig:all_cf_lightcurves} shows the difference in the fits obtained to our GRB sample for 0.3-10 keV and 15-50 keV light-curves, as well as the best-fitting result when keeping both instruments' data in their native bandpasses. It is immediately apparent that the band chosen has a substantial impact on the observed data across our sample, as has been discussed previously for GRB100522A. Comparison with the previous work of \citet{Gibson1} also shows our data appear less volatile due to the increase in the sizes of the error bars. For several GRBs it is also clear that changing bands has affected the morphology of the best-fitting model light-curve, as can be seen with GRB111121A as an example - the 0.3-10 keV model sees the propeller remain switched on for the entire duration of the GRB with the dipole luminosity becoming dominant at late times; at 15-50 keV the propeller instead abruptly switches off and back on, but dominates the late-time emission. This rapid turn-off of the propeller allows for the fitting of a small dip in luminosity at $\sim 1$s which was unnecessary at 0.3-10 keV due to the larger errors on the BAT data.

Another notable difference between the fits obtained in the different bands across several GRBs is the fitting of the late-time XRT data, which is generally precise and accurate in the 0.3-10 keV results but often underestimated in the 15-50 keV fits. At 15-50 keV, the fractional errors on the BAT data are generally lower than for the XRT data, thus the MCMC fitting routine opts to prioritise improving the fit to the BAT data over marginal enhancements to the generally sparse XRT data points in the tail of the light-curve, as can be seen most prominently for GRB061006 and GRB100212A.

When additionally considering the native bandpass results, we can immediately see that the native bandpass data effectively mitigate the increased variability present in the BAT data at 0.3-10 keV and in the XRT data at 15-50 keV, and these data therefore appear to be the most robust. The native bandpass data also permit some exploration of the assumptions used when generating the bolometric luminosities - where BAT and XRT data temporally overlap, they generally agree fairly well as can be seen in GRB060614 for example, but other GRBs see inconsistency between them which may be indicative that a spectral break is indeed present, contradicting our assumption of a single power-law spectrum.

With the native bandpass best-fits it is clear that they highlight the sensitivity of the fits to the band used, as some native bandpass best-fitting light-curves closely mimic the morphology of the 0.3-10 keV best fit, and others closely match the 15-50 keV best fit, with no universal pattern regarding which band the native bandpass results will resemble. The light-curve for GRB051016B is indeed morphologically distinct from either band's best fits. These results are generally a compromise between the 0.3-10 keV and 15-50 keV fits. We still see the MCMC ignoring BAT data in certain light-curves, and late-time XRT data being underestimated similarly to in some 15-50 keV fits. The reduction in the errors on the data overall also did not allow the native bandpass fits to more accurately capture substructure in the light-curves. While the native bandpass data appear to be the most robust, they do reveal that it is difficult to establish, out of the 0.3-10 keV and 15-50 keV results, which ones are the most reliable, and fits to combined BAT+XRT light-curves presented in the same bandpass should therefore be treated with caution.

\begin{table*}
    \centering
    \begin{tabular}{c | c | ccc | ccc | ccc}
        \hline
        GRB & \citet{Gibson1} & & 0.3-10 keV & & & 15-50 keV & & & Native & \\
         & &  BAT & XRT & Total & BAT & XRT & Total & BAT & XRT & Total \\
        \hline
        050724 & 1507 & 5.9 & 427.7 & 452.4 & 446 & 625 & 1090 & 572 & 598 & 1189\\
        051016B & 435 & 2.3 & 25.6 & 48.6 & 52.6 & 76.2 & 149.4 & 31.6 & 33.5 & 85.7\\
        051227 & 233 & 1.79 & 5.40 & 32.69 & 95.0 & 20.8 & 139.8 & 93.3 & 10.7 & 128.1\\
        060614 & 44709 & 525 & 903 & 1447 & 7826 & 1350 & 9195 & 7566 & 2766 & 10349\\
        061006 & 242 & 9.68 & 0.32 & 33.29 & 58.0 & 9.3 & 90.5 & 52.3 & 10.8 & 86.2\\
        061210 & - & 0.17 & 0.04 & - & 5.81 & 0.14 & - & 5.01 & 0.74 & -\\
        070714B & 1180 & 18.1 & 49.4 & 87.1 & 156 & 134 & 309 & 177.4 & 76.2 & 273.2\\
        071227 & 158 & 0.25 & 1.05 & 26.23 & 3.9 & 6.1 & 33.6 & 2.88 & 1.35 & 27.86\\
        080123 & 283 & - & 39.9 & 62.4 & 14.0 & 77.0 & 113.1 & 8.16 & 7.97 & 38.13\\
        080503 & 2294 & 23.8 & 48.7 & 91.4 & 233 & 337 & 590 & 233.9 & 73.3 & 325.1\\
        100212A & 7310 & 20 & 2927 & 2966 & 733 & 1311 & 2062 & 757 & 2927 & 3703\\
        100522A & 22184 & 67.1 & 73.0 & 158.8 & 1395 & 82 & 1496 & 1536 & 154 & 1709\\
        111121A & 1742 & 8.6 & 120.1 & 147.8 & 192 & 358 & 569 & 583 & 257 & 859\\
        150424A & 1334 & 13.6 & 105.0 & 138 & 230 & 280 & 529 & 274 & 185 & 478\\
        160410A & 359 & 1.3 & 14.2 & 37.2 & 30.1 & 31.2 & 82.7 & 25.6 & 27.0 & 74.0
    \end{tabular}
    \caption{AICc values for the best-fitting models. Results are presented for fits to data extrapolated to 0.3-10 keV and to 15-50 keV, as well as data kept in their native bandpasses, including a breakdown of the BAT data an the XRT data's contributions to the total AICc value. Results from \citet{Gibson1} are included for comparison. GRB061210 is excluded as the small number of data points (six from the BAT; four from the XRT) cause problems the AICc calculation with nine fitting parameters.}
    \label{tab:AICcvalue_cf}
\end{table*}

\subsection{AICc values}

\begin{figure}
    \centering
    \begin{tabular}{c}
    \includegraphics[width = 0.94\columnwidth]{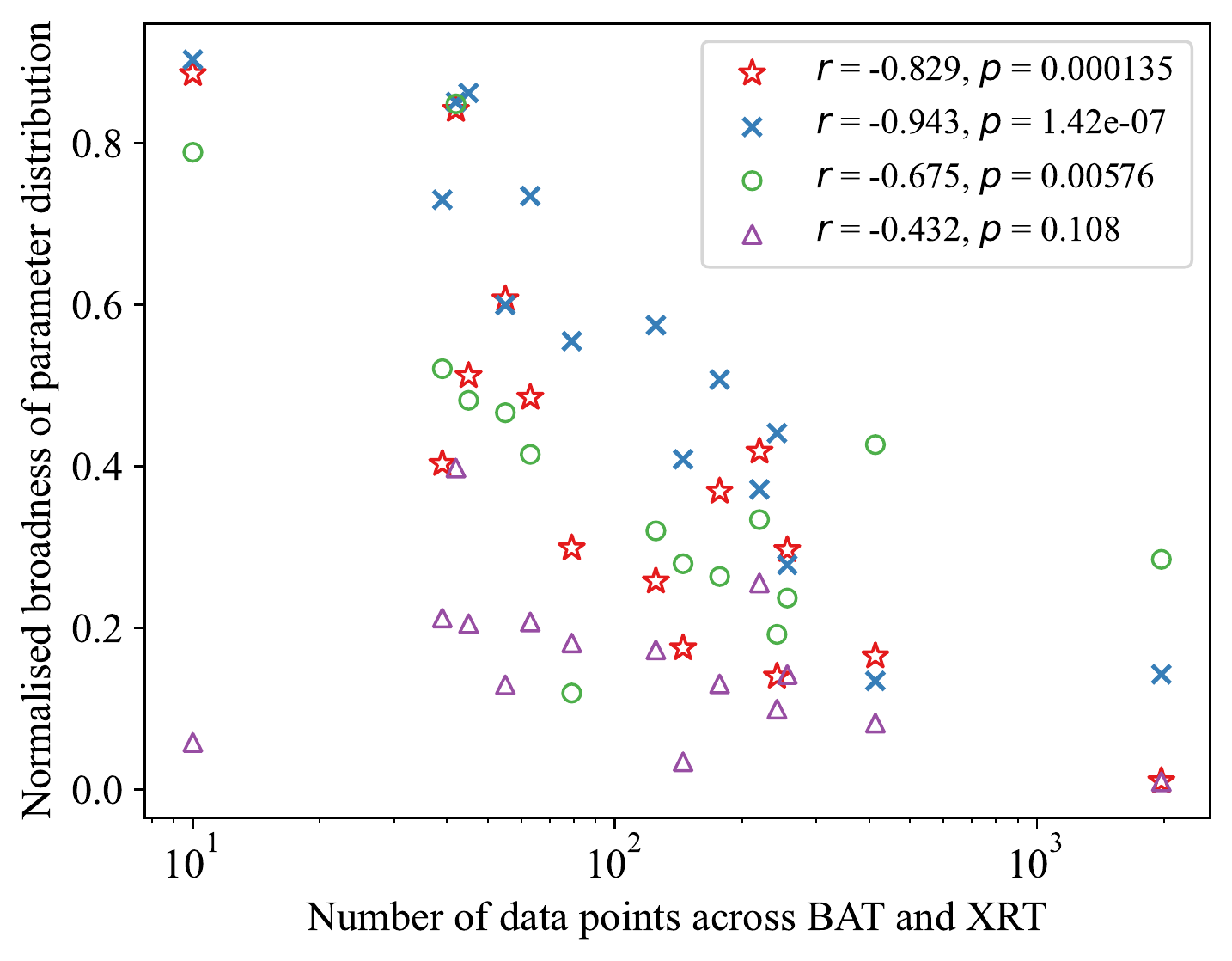}\\
    \includegraphics[width = 0.94\columnwidth]{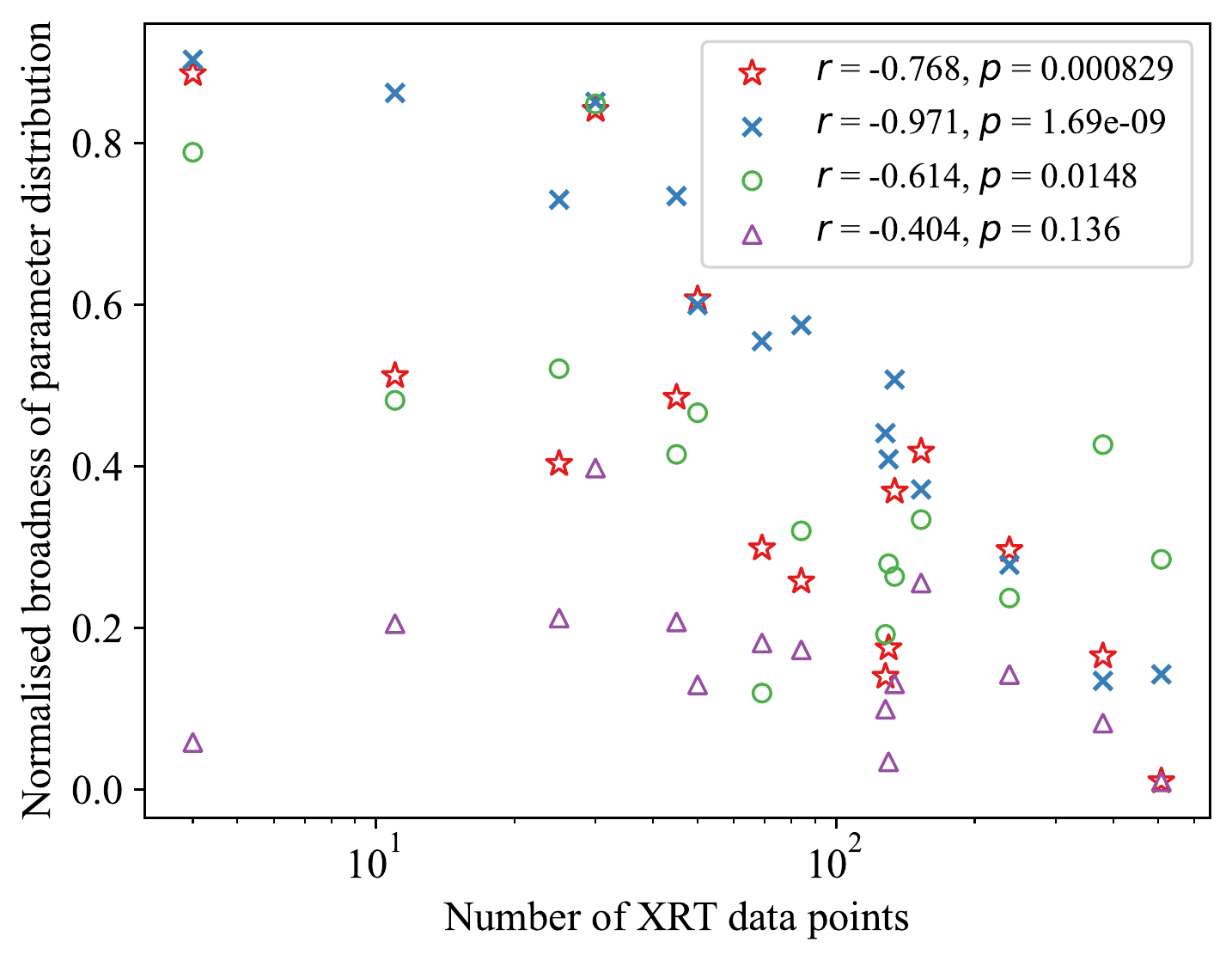}\\
    \includegraphics[width = 0.94\columnwidth]{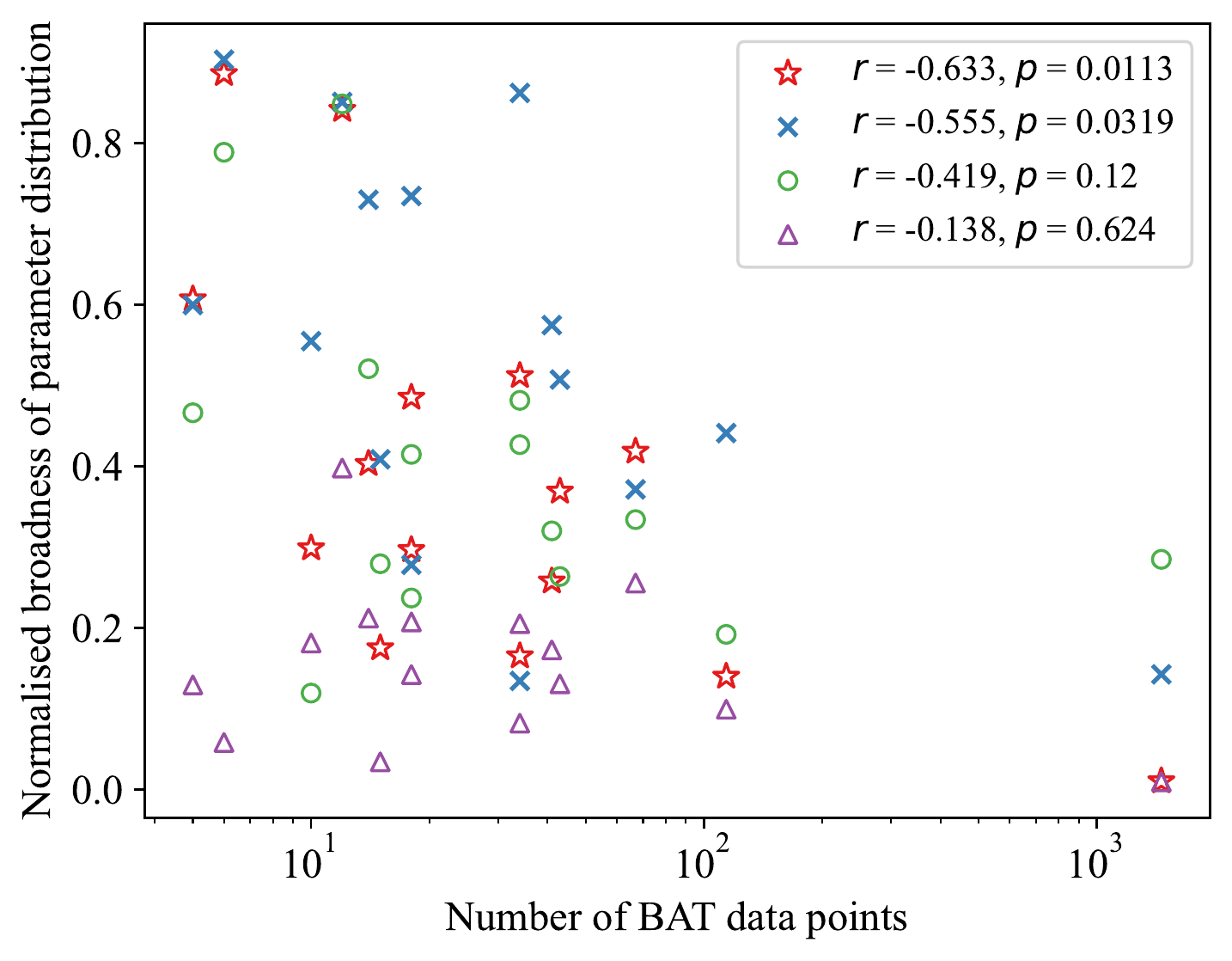}\\
    \end{tabular}
    \caption{Average broadness of each parameter distribution obtained from propagating the conversion factor errors, arranged on a per-GRB basis as a function of the number of XRT + BAT data points, the number of BAT data points, and the number of XRT data points. Red stars - native bandpass results; blue crosses - 0.3-10 keV extrapolation; green circles - 15-50 keV extrapolation; purple triangles - results from \citet{Gibson1}. Also included for each data set are the Spearman correlation coefficient $r$ and the $p$ value. Even though \citet{Gibson1} used a combined BAT+XRT dataset instead of treating each instrument's data separately, their data are included to serve as a control dataset.}
    \label{fig:broadness_vs_Ndata}
\end{figure}

The difference in where the errors are driving the morphology of the fits can be seen in Table \ref{tab:AICcvalue_cf}, which shows the best AICc values obtained by the MCMC for each GRB at each band, broken up by the contributions to the AICc made by the BAT and the XRT data. It is immediately obvious that our AICc values imply fits to the data that are more consistent than those obtained by \citet{Gibson1}, since our analysis gives the data larger error bars. The 0.3-10 keV fits are heavily dominated by the XRT data, as would be expected given that the XRT datapoints are more numerous and have smaller errors than the BAT datapoints. At 15-50 keV however, we generally see more equal contributions to the AICc.

The XRT data generally have more constrained ECFs than the BAT data, thus extrapolating XRT data to the BAT's native bandpass will naturally introduce less noise than vice versa. The greater number of XRT datapoints, combined with their larger (but not significantly so) errors at 15-50 keV, result in the BAT and XRT data generally contributing comparable amounts to the AICc at 15-50 keV, whereas the 0.3-10 keV fits are dominated by the XRT.

The native bandpass results are generally as would be expected here; the BAT data's and XRT data's contributions to the AICc are comparable to those of the BAT data at 15-50 keV and the XRT data at 0.3-10 keV respectively, since they are the same data. Perhaps surprising is that the BAT contributions are mostly higher than the XRT contributions for the native bandpass when most GRBs have more XRT data points, perhaps owing to the XRT data being less volatile and therefore easier to fit, as the model has proven well-suited to fitting continuum-like features rather than substructure. GRB100212A is a notable exception to this trend, owing to the XRT data here showing clear signs of substantial substructure.

Table \ref{tab:AICcvalue_cf} reflects the priorities of the MCMC walkers - most of the 0.3-10 keV fits have very low BAT contributions to the AICc, indicating that there is little incentive for the MCMC to attempt to improve fits to the early-time data. These generally have more to gain from attempting to better capture substructure in the XRT data. This is further illustrated by considering the only GRBs with a more significant BAT contribution to the AICc at 0.3-10 keV, GRB061006 and GRB061210, where not only are there large fluctuations in BAT luminosity which are difficult to model, but there are also few XRT data with relatively large errors.

Fig. \ref{fig:broadness_vs_Ndata} also provides insight into which part of the lightcurve dominates the fits. The broadness of the fitting parameter distributions obtained by the ensemble indicate how well-constrained the light-curve is. Tighter distributions imply that the MCMC found regions of parameter space that could precisely fit the observational data. Broader distributions, by contrast, indicate that the fits were not especially sensitive to the parameters.

More data points should naturally result in more constrained fits, thus the broadness of the parameter distributions should be negatively correlated with the number of data points to which we are fitting. As a corollary, if increasing the number of points within a subset of the dataset does not impact the broadness of the parameter distributions, then we may surmise that the subset in question is not driving the fits. Fig. \ref{fig:broadness_vs_Ndata} shows the correlation coefficients for the broadness of the parameter distributions against the number of BAT and XRT data points in each band. It is clear that strong negative correlations exist between the parameter broadness and the total number of data points, as well as with the number of XRT data points. However, these correlations are much weaker and less statistically significant with the number of BAT data points, indicating again that the fits are predominantly being driven by the XRT data.

Fig. \ref{fig:broadness_vs_Ndata} highlights that the native bandpass results are generally being driven by the entire dataset, not merely the BAT data or the XRT data as with the results in the other bands. The native bandpass results have the most significant relation between broadness and the number of BAT data points, and this is partly why the relation for the number of XRT data points is weaker than for the 0.3-10 keV results. We nonetheless see that the relation between the broadness of the parameter distributions and the total number of data points is somewhat weaker than for the 0.3-10 keV results. This may well be an artefact of the occasional inconsistencies between temporally overlapping BAT and XRT data points, forcing the MCMC to prioritise either the BAT or XRT data here.

\subsection{Parameter distributions}

\begin{figure*}
    \centering
    \addtolength{\tabcolsep}{-0.4em}
    \begin{tabular}{ccc}
        \includegraphics[width = 0.66\columnwidth]{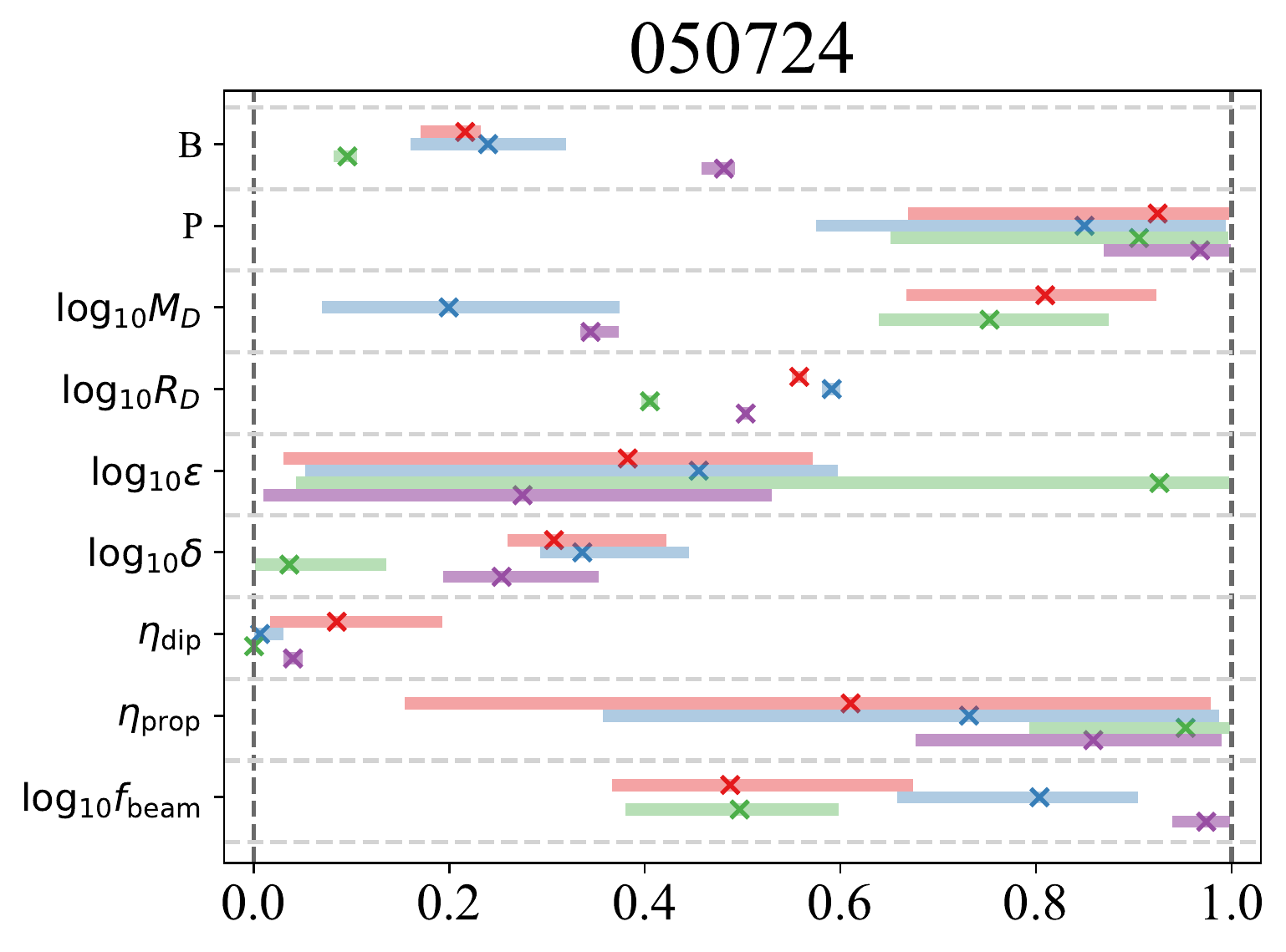} & \includegraphics[width = 0.66\columnwidth]{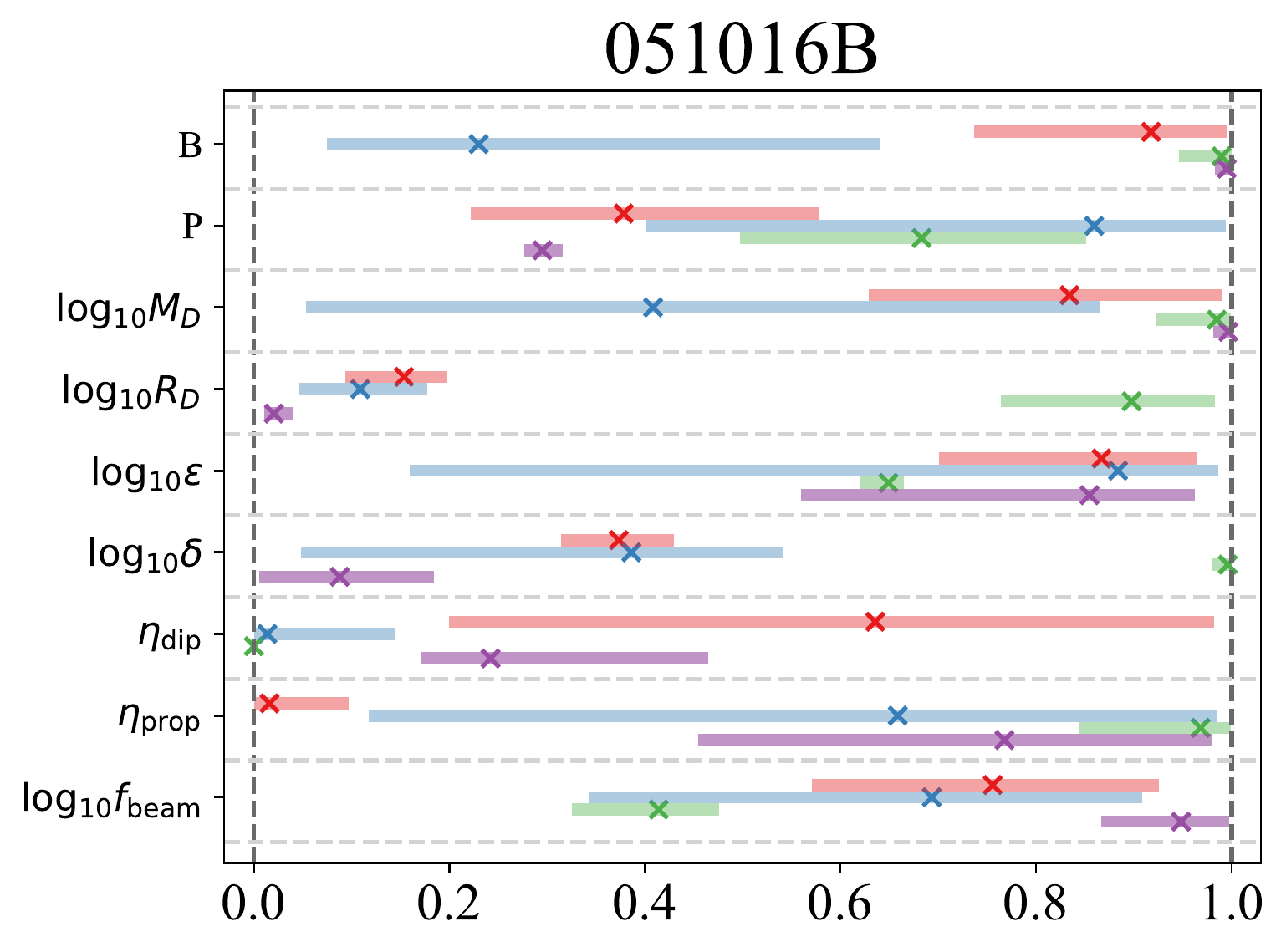} &
        \includegraphics[width = 0.66\columnwidth]{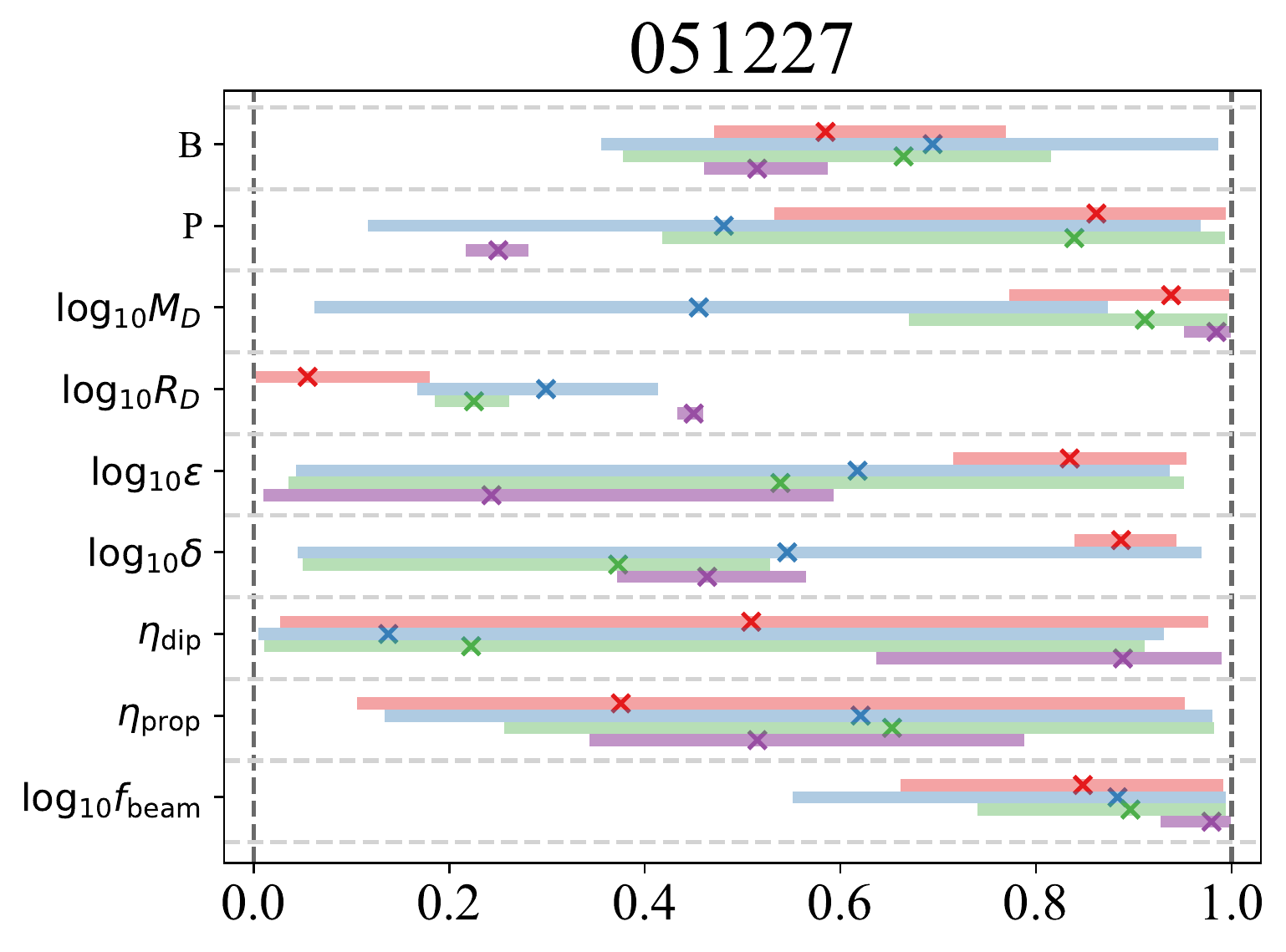}\\
        \includegraphics[width = 0.66\columnwidth]{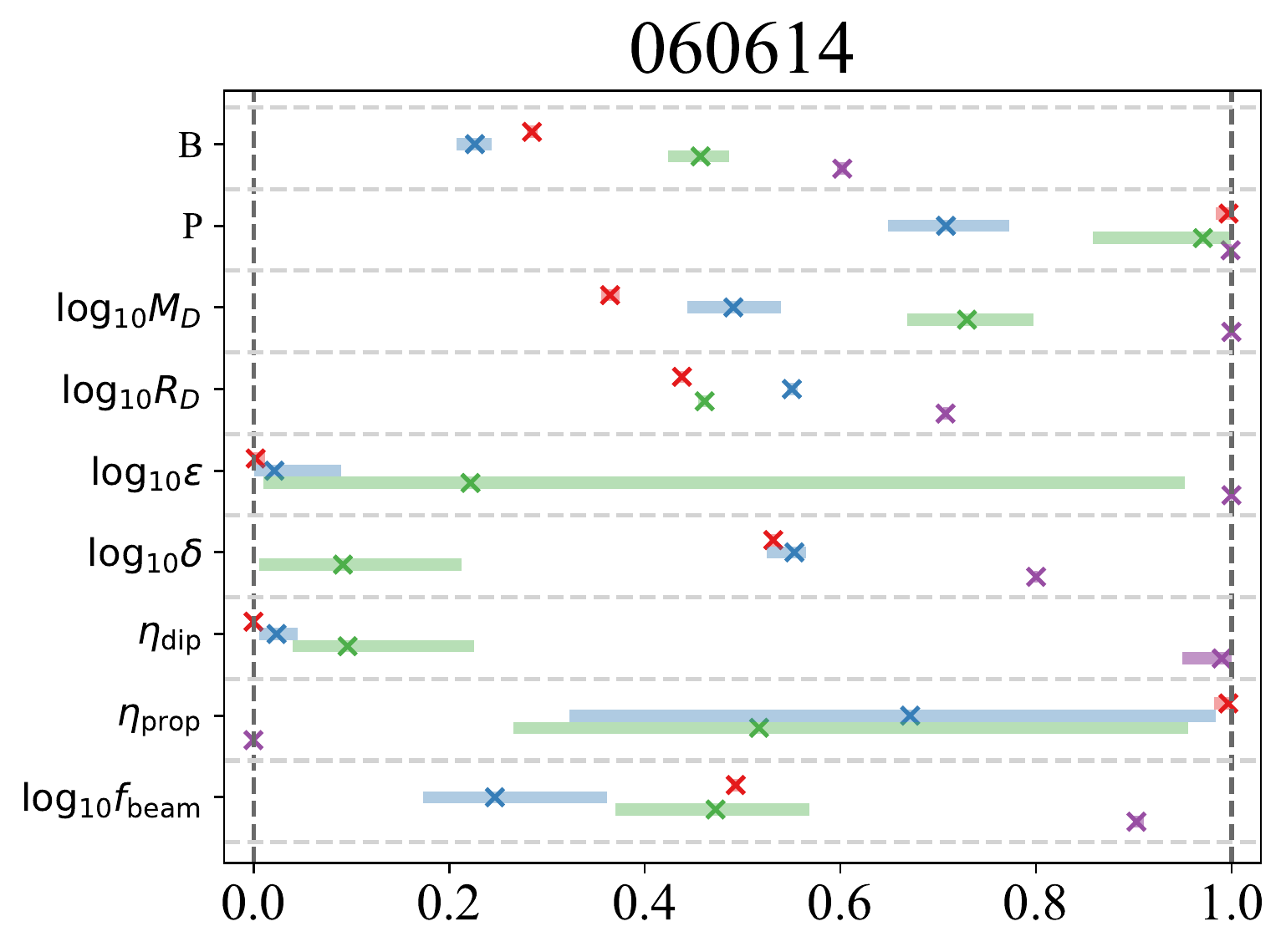} & \includegraphics[width=0.66\columnwidth]{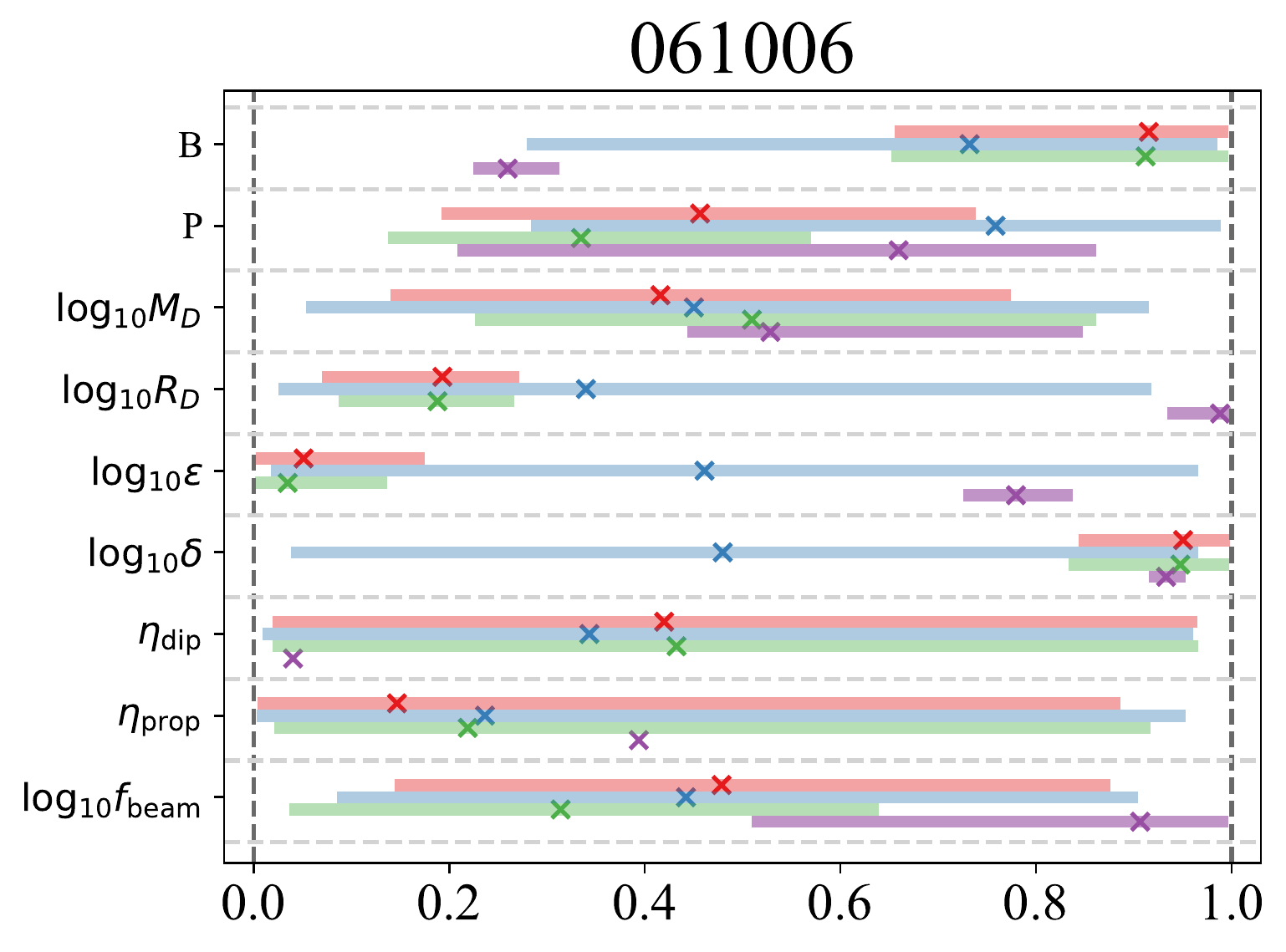} & \includegraphics[width = 0.66\columnwidth]{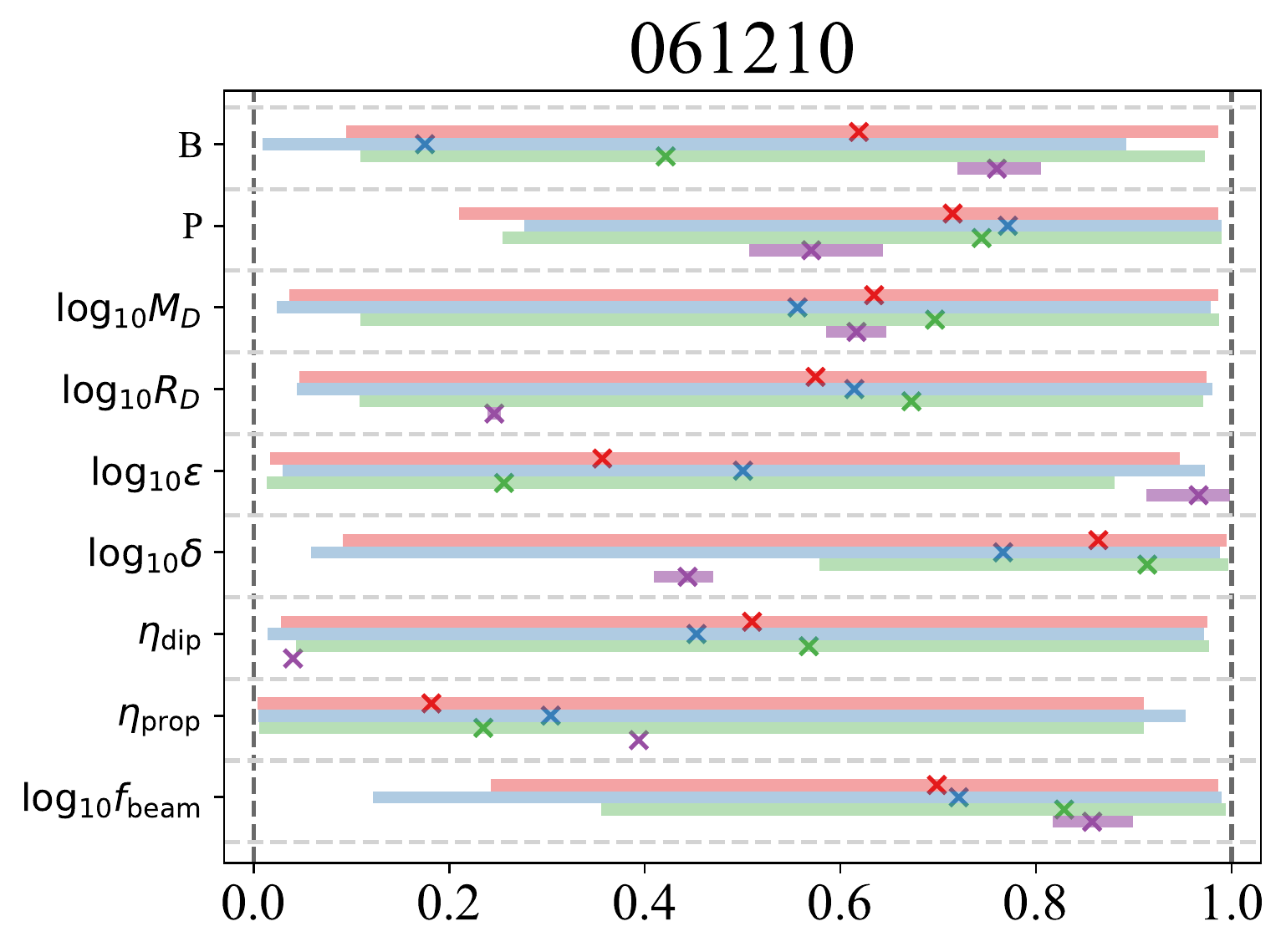}\\
        \includegraphics[width = 0.66\columnwidth]{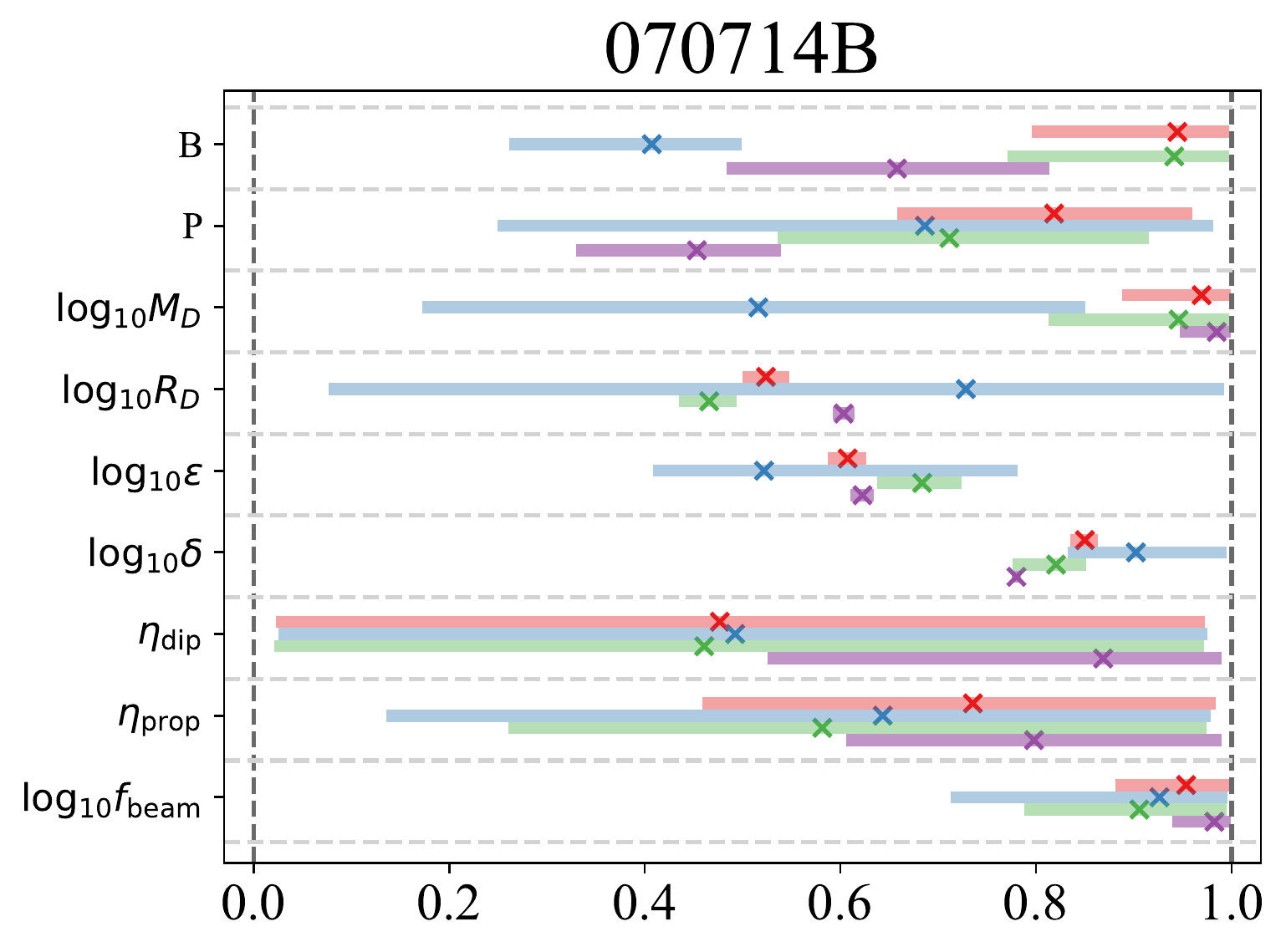} & \includegraphics[width = 0.66\columnwidth]{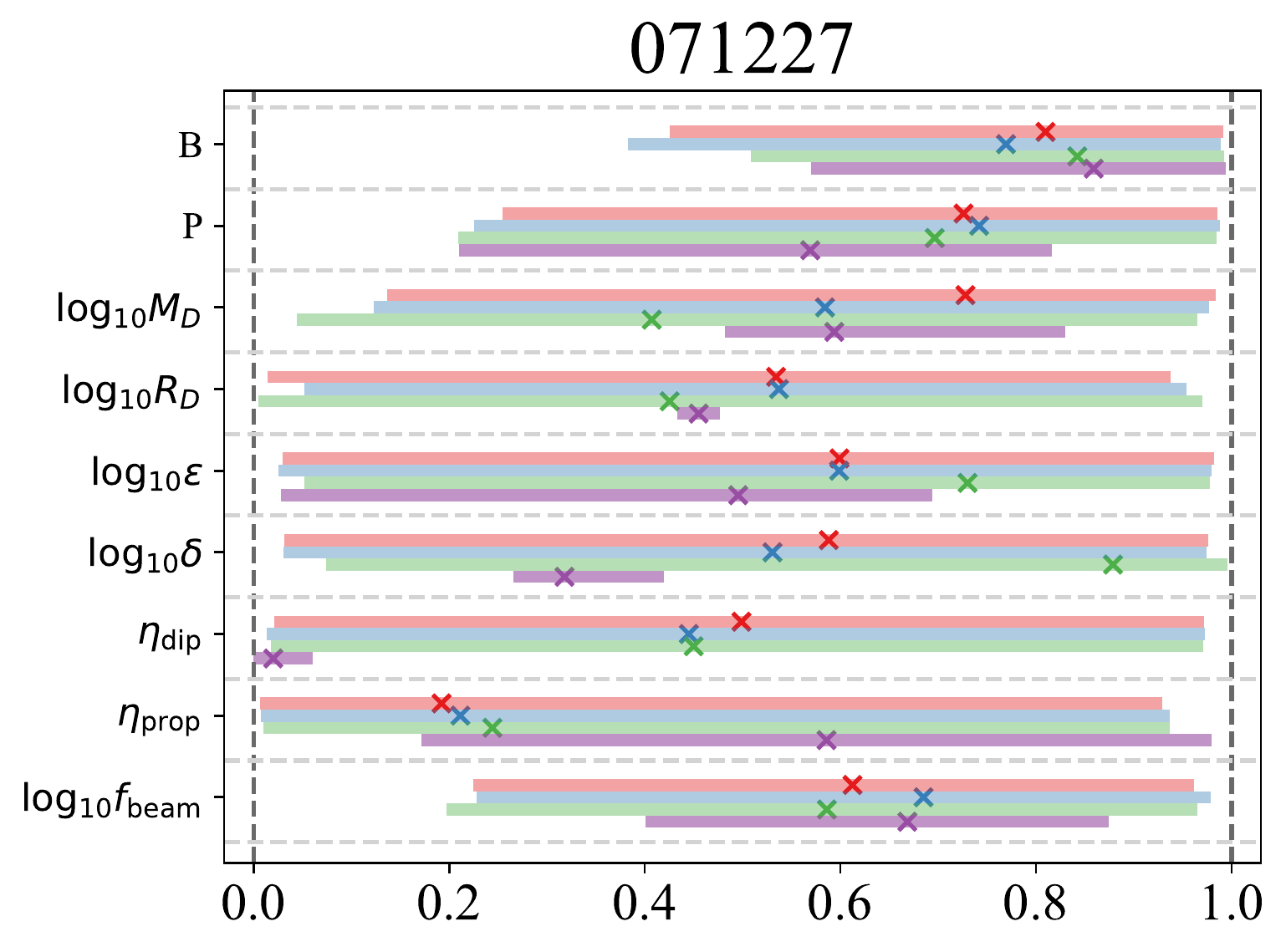} &
        \includegraphics[width=0.66\columnwidth]{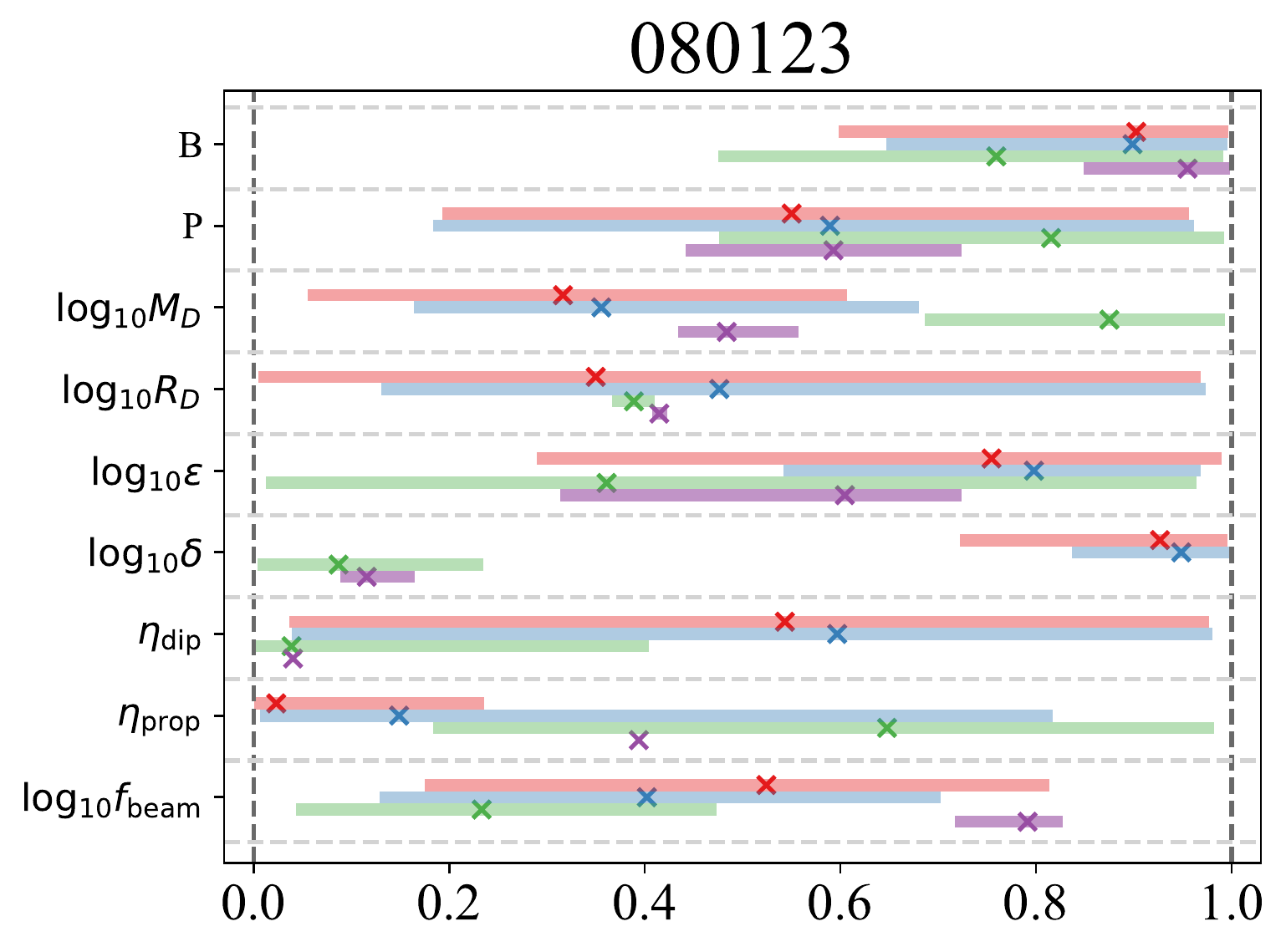}\\
        \includegraphics[width = 0.66\columnwidth]{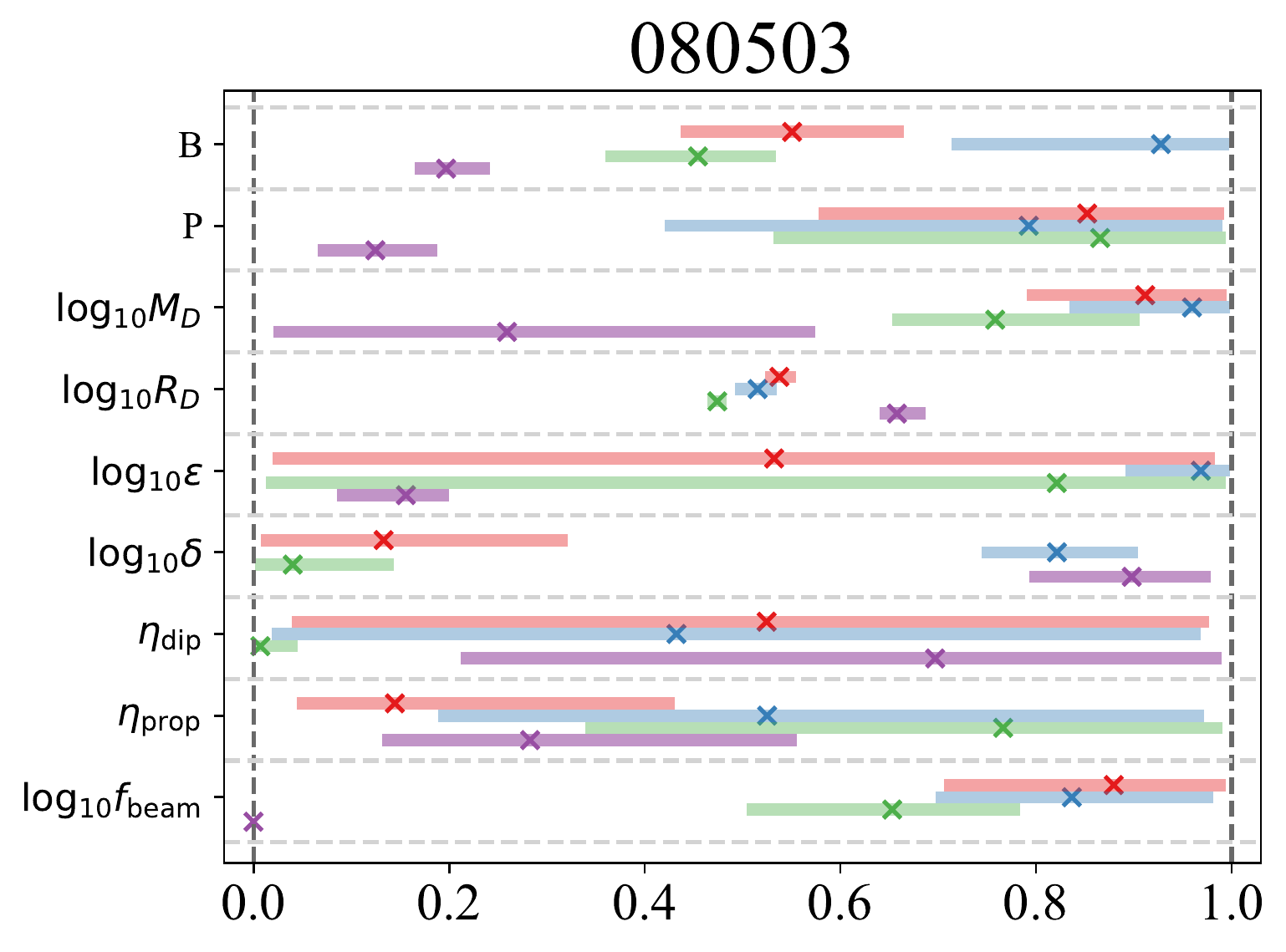} & \includegraphics[width = 0.66\columnwidth]{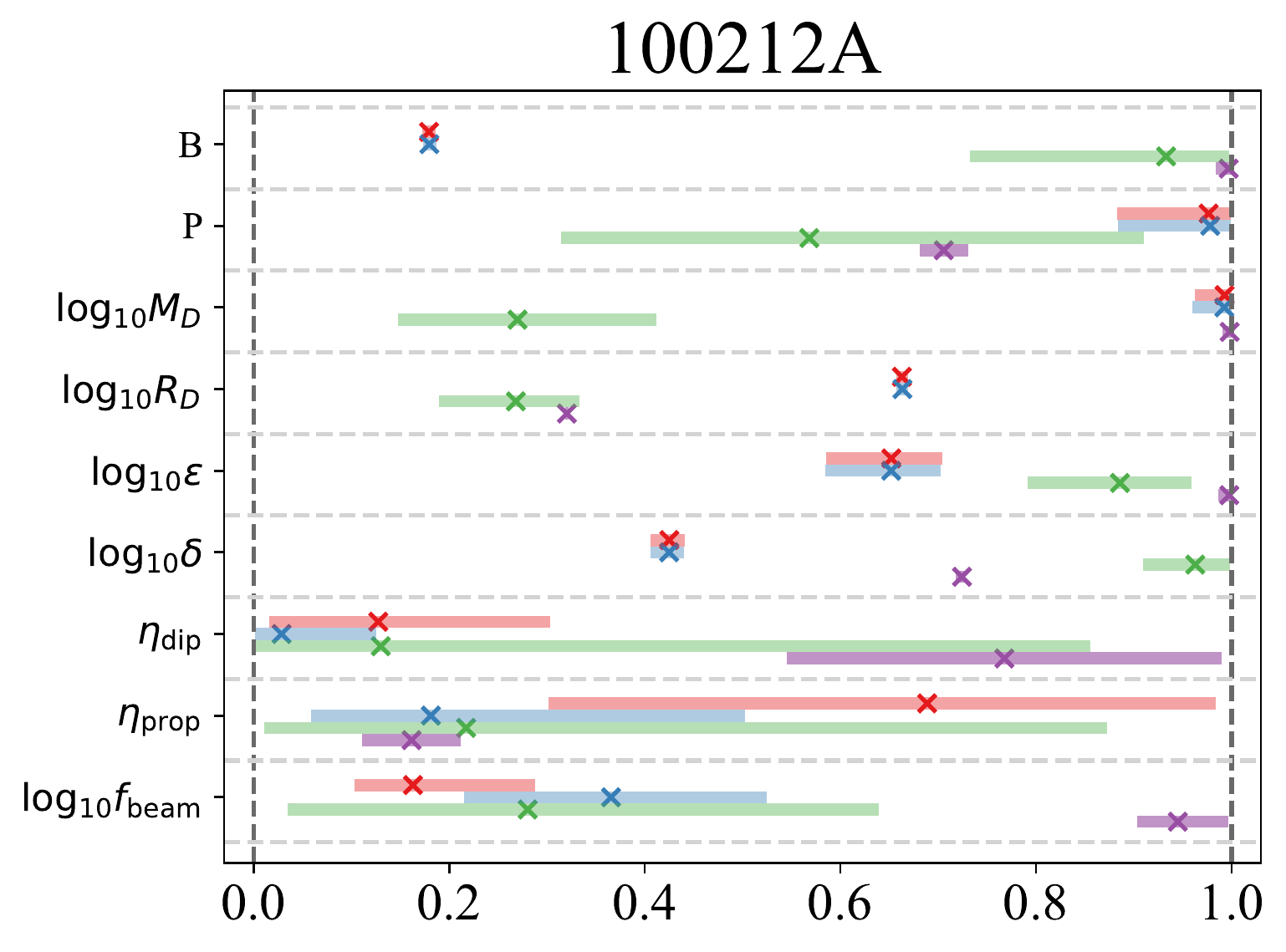} &
        \includegraphics[width=0.66\columnwidth]{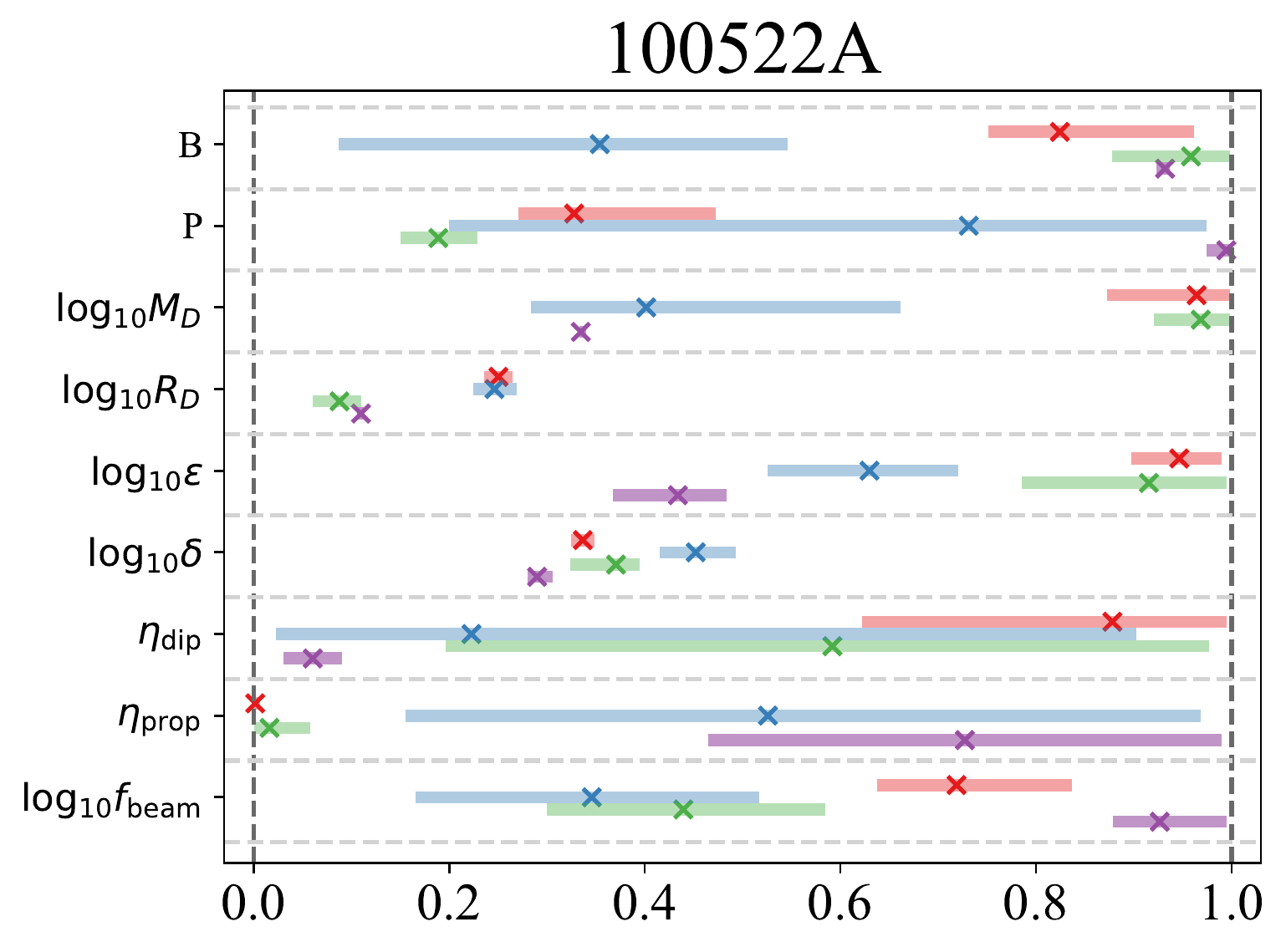}\\
        \includegraphics[width = 0.65\columnwidth]{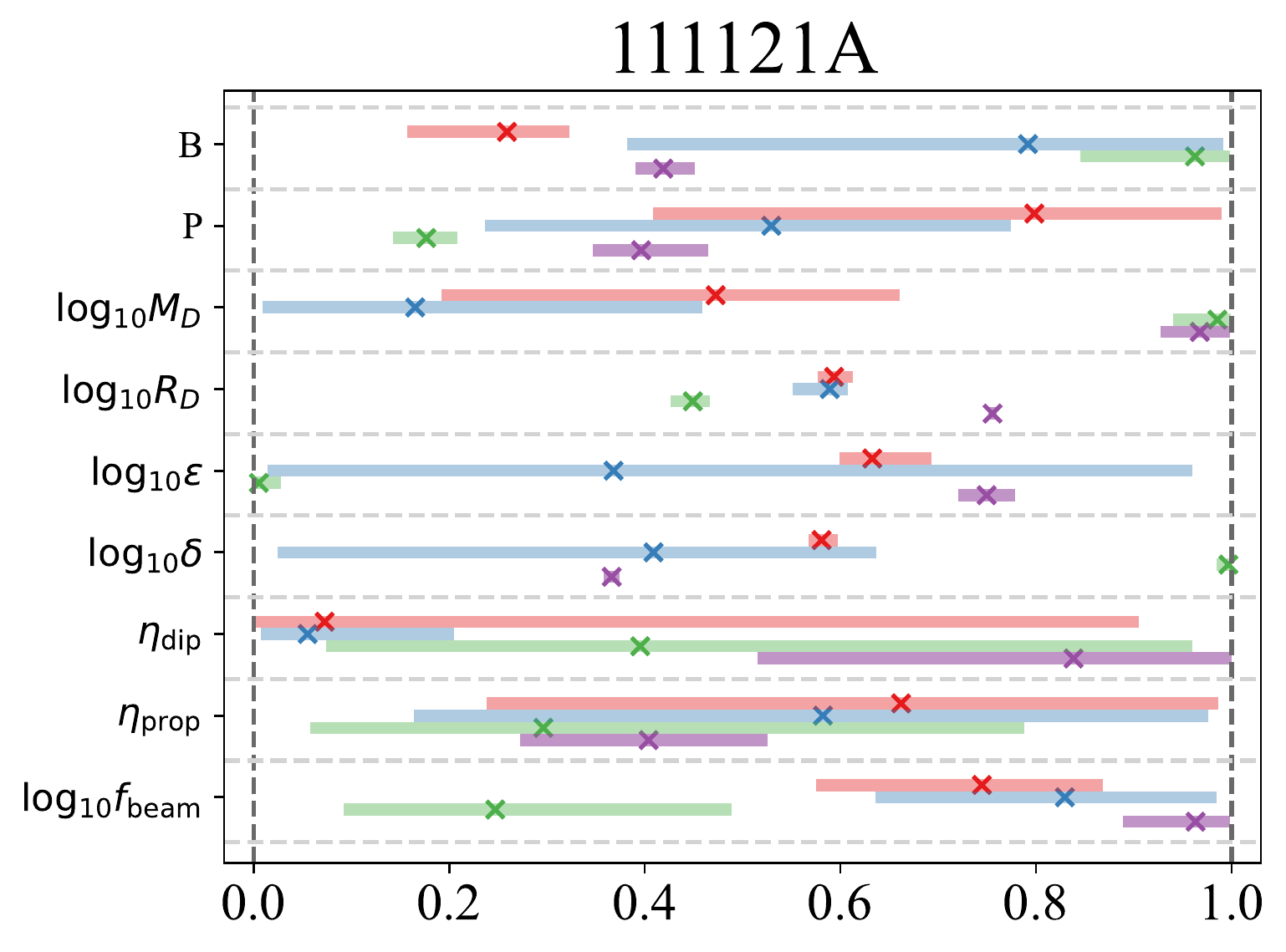} & \includegraphics[width = 0.66\columnwidth]{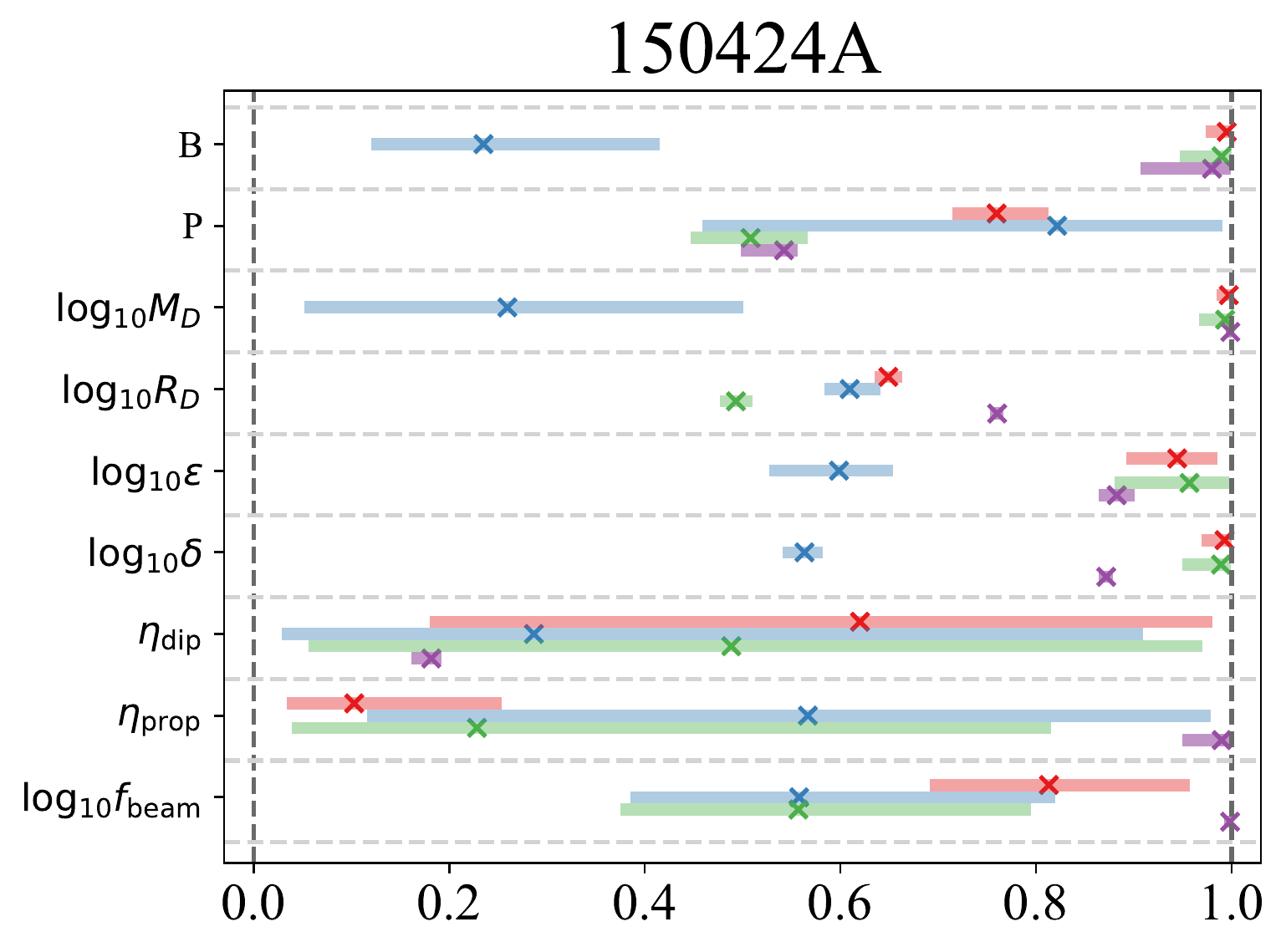} &
        \includegraphics[width=0.66\columnwidth]{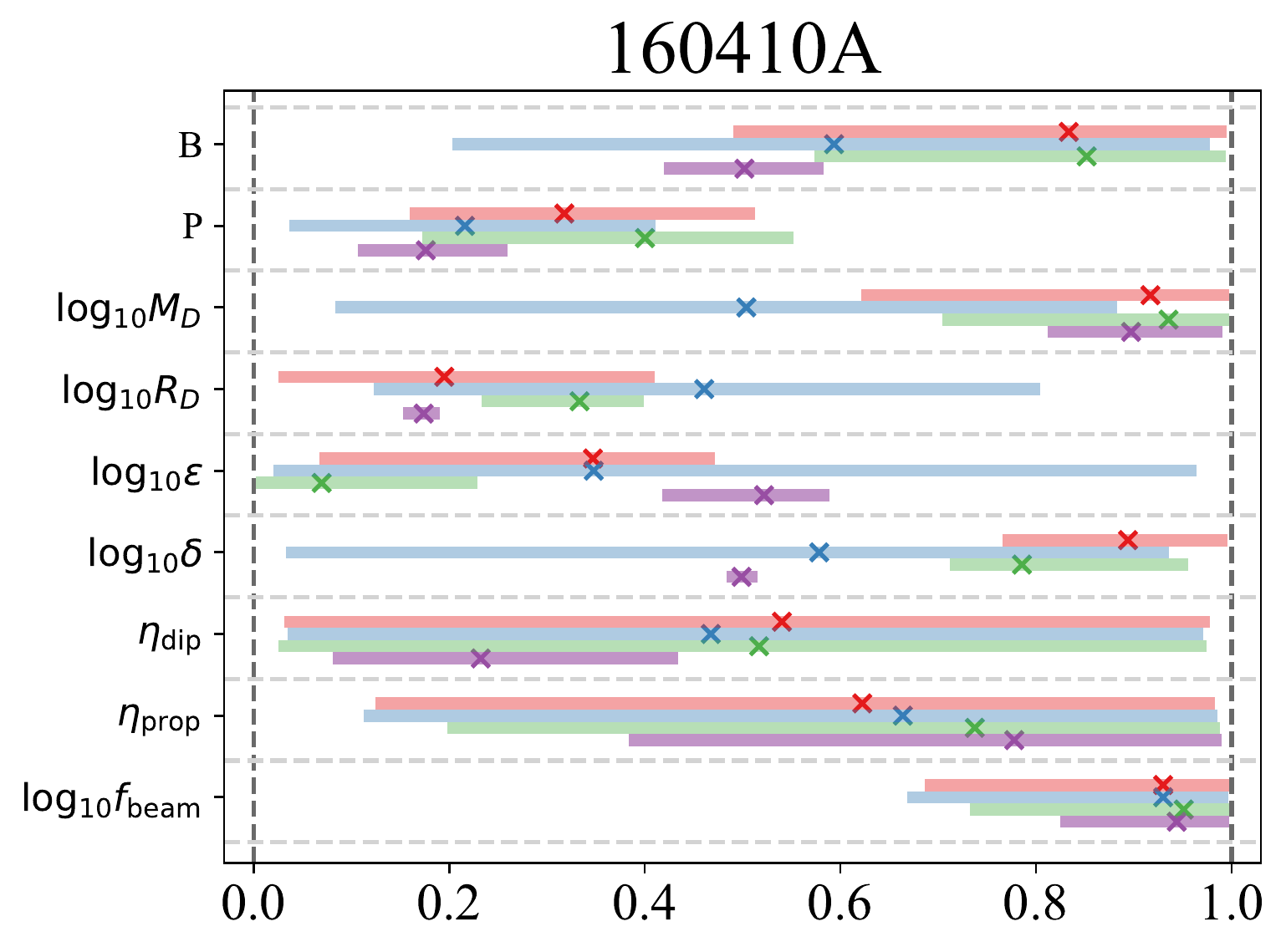}\\
    \end{tabular}
    \caption{Normalised parameter distributions, with 0 and 1 representing the lower and upper limits from Table \ref{tab:PropLimits} respectively, from the MCMC fitting routine for our sample of GRBs, arranged on a per-GRB basis. Red (topmost) data - native bandpass results; blue (second from top) data - 0.3-10 keV extrapolation; green (third from top) data - 15-50 keV extrapolation; purple (lowermost) data - results from \citet{Gibson1}.}
    \label{fig:cf_params_per_GRB}
\end{figure*}

Fig. \ref{fig:cf_params_per_GRB} shows the 95\% confidence regions for each (normalised) fitting parameter for each method of processing the data, arranged on a per-GRB basis. Clearly, our obtained 95\% regions are much broader than those of \citet{Gibson1}, reflecting the larger errors on our data. The hypothesis that this is driven by increased errors is also supported by the fact that the 0.3-10 keV fits generally have broader parameter distributions than the 15-50 keV fits, with the native bandpass results having the narrowest distributions.

What is also apparent is that some GRBs have clear inconsistencies in the best-fitting parameters, depending on the data processing used. GRB061006 and GRB071227 were the only GRBs in our sample where the 95\% regions for the 0.3-10 keV and 15-50 keV results were consistent with each other for all parameters, demonstrating that the nature of the data processing substantially altered the nature of the fits and the derived physical parameters. As we saw with the light-curves, the native bandpass results are generally consistent with either the 0.3-10 keV fits or the 15-50 keV fits, though rarely both, and we can sometimes see (e.g. GRB050724) the native bandpass results alternate between the 0.3-10 keV and 15-50 keV results for different parameters, suggesting that we may even be seeing some degeneracy in parameter space.

\begin{figure}
    \centering
    \begin{tabular}{c}
    \includegraphics[width=0.94\columnwidth]{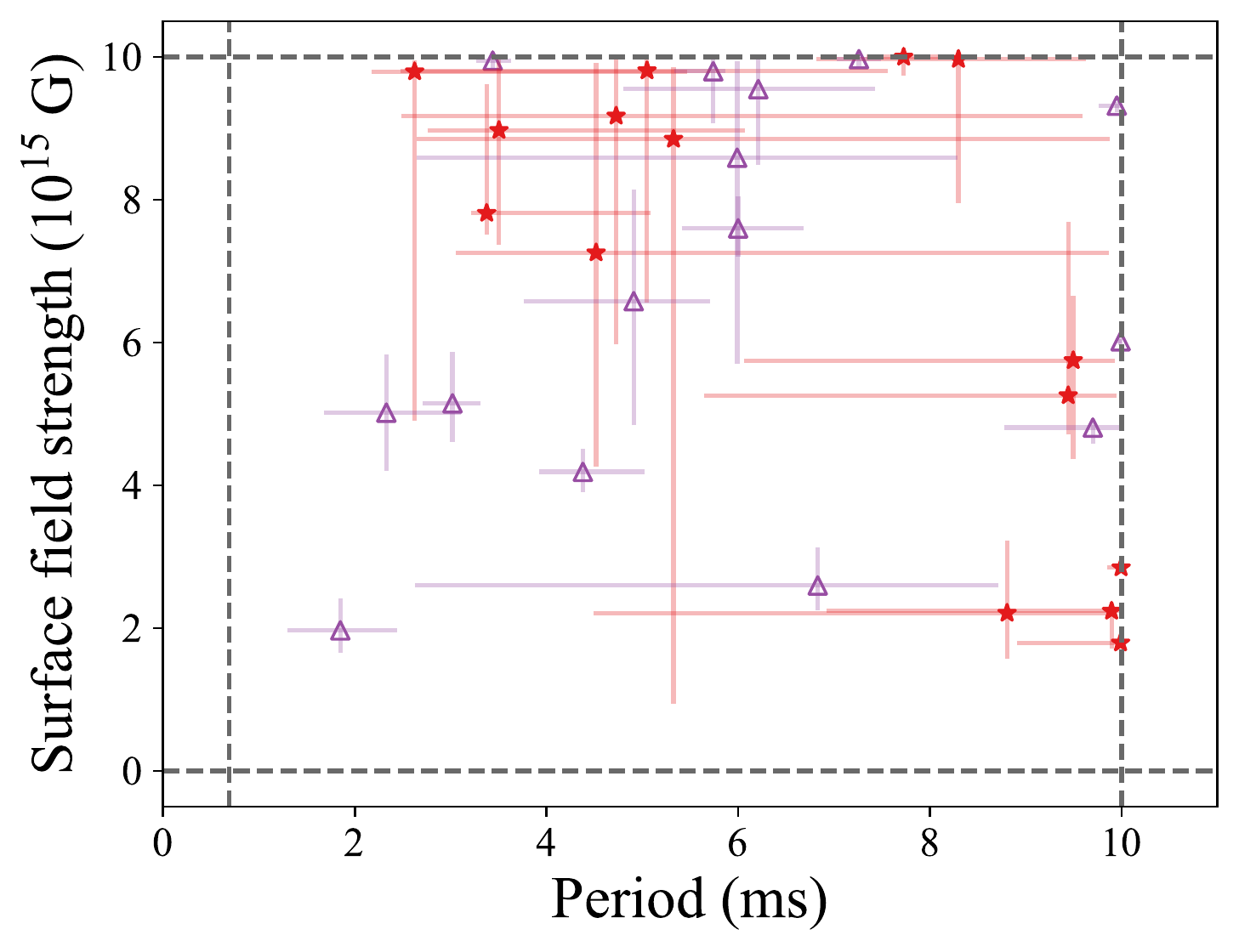}\\
    \includegraphics[width=0.94\columnwidth]{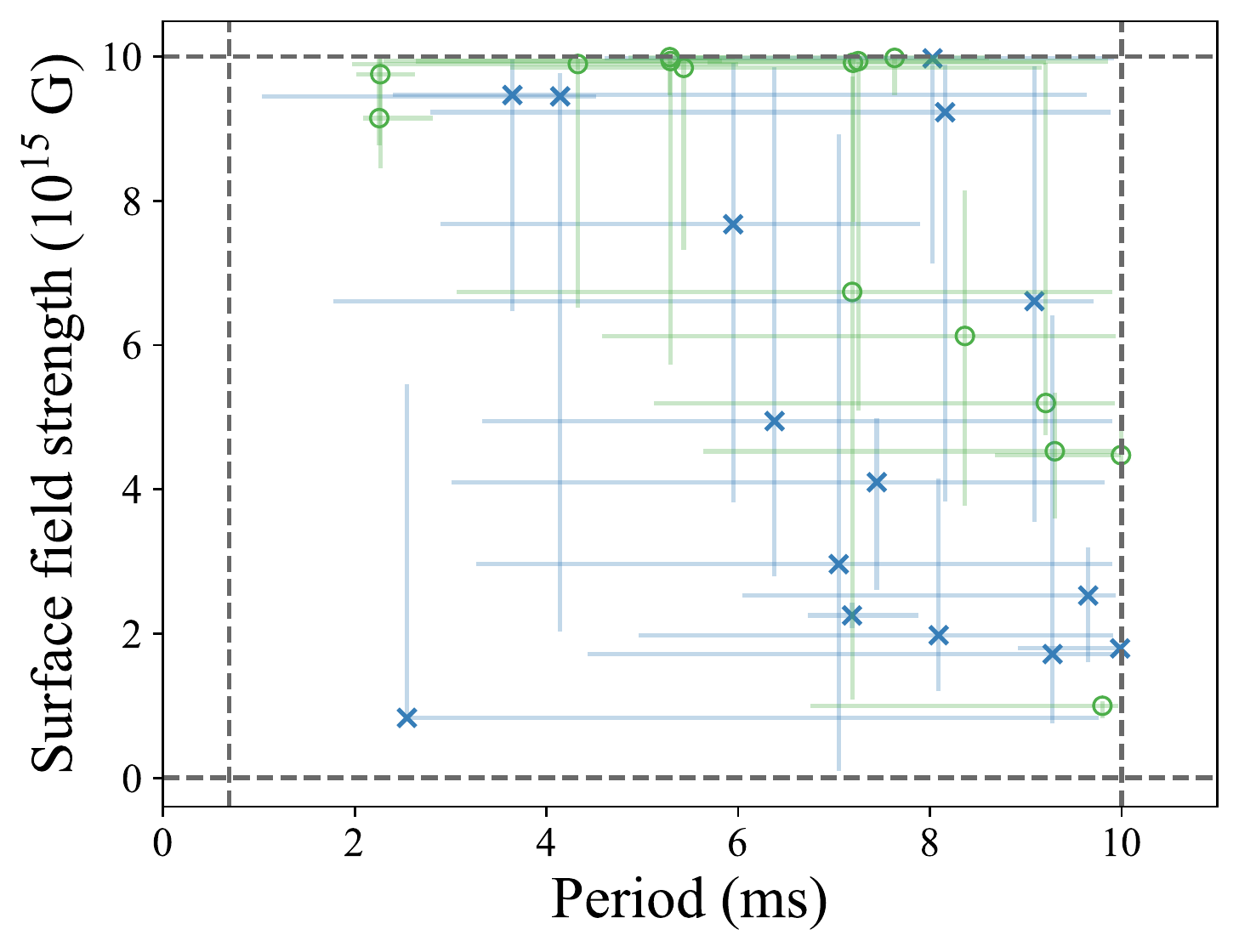}  
    \end{tabular}
    \caption{Magnetic field strength plotted against initial magnetar spin period with 95\% confidence intervals included. The dashed lines denote the limits of the permitted parameter space. Red stars - native bandpass results; blue crosses - 0.3-10 keV extrapolation; green circles - 15-50 keV extrapolation; purple triangles - results from \citet{Gibson1}.}
    \label{fig:BP_plot}
\end{figure}

Fig. \ref{fig:BP_plot} illustrates the difference in the physical parameters, with the best-fitting values for the magnetar's magnetic field strength and initial rotation period plotted together. A clear bunching up for high $B$ values can be seen for the 15-50 keV fits which is largely absent from the other fits. This, in conjunction with the underestimation of the late-time XRT data in some of these fits, may indicate that our best-fitting models at 15-50 keV are limited by their global energy budgets - a conclusion which would not be readily derived from the other fits.

\begin{figure}
    \centering
    \begin{tabular}{c}
    \includegraphics[width=0.94\columnwidth]{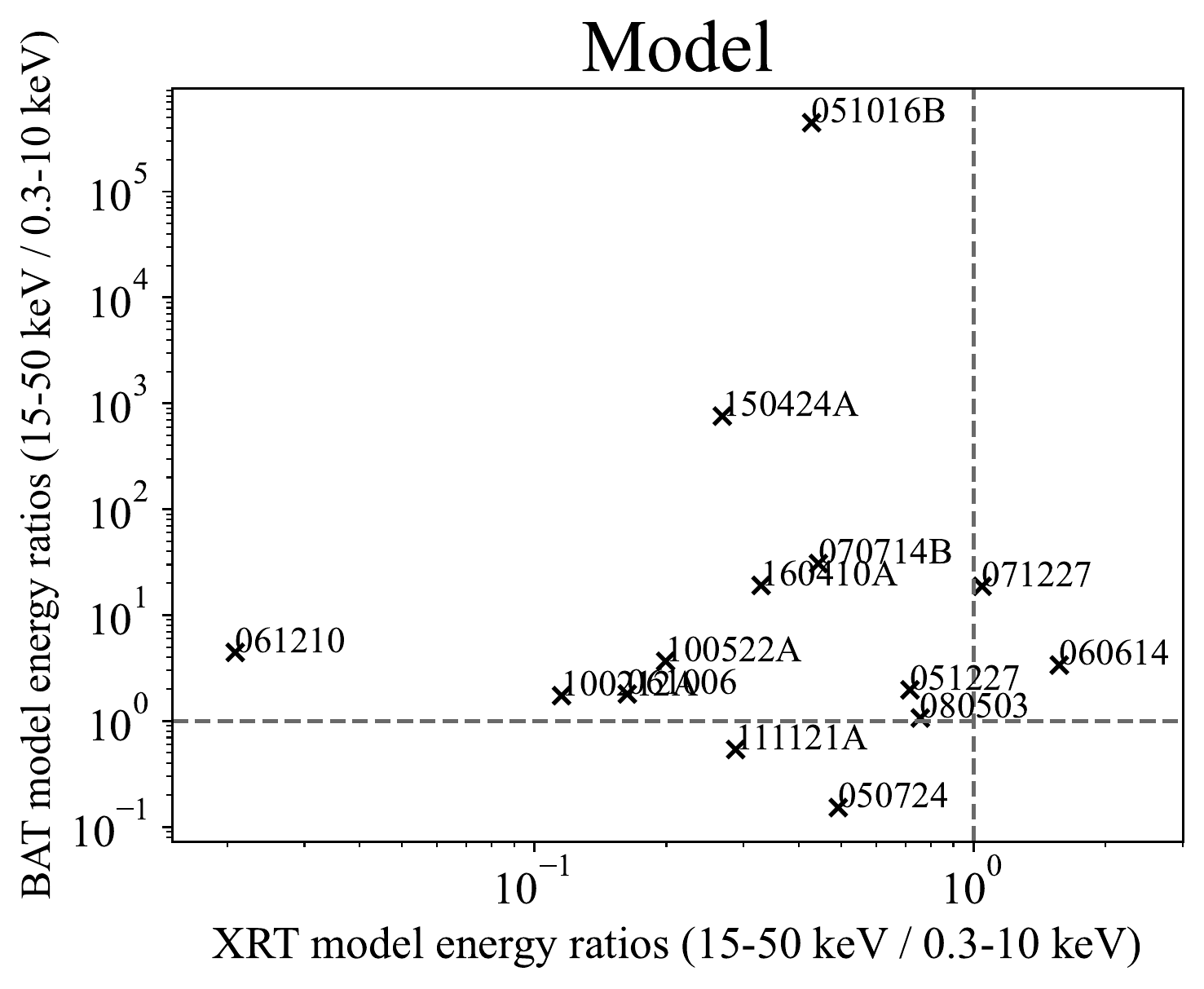}\\
    \includegraphics[width=0.94\columnwidth]{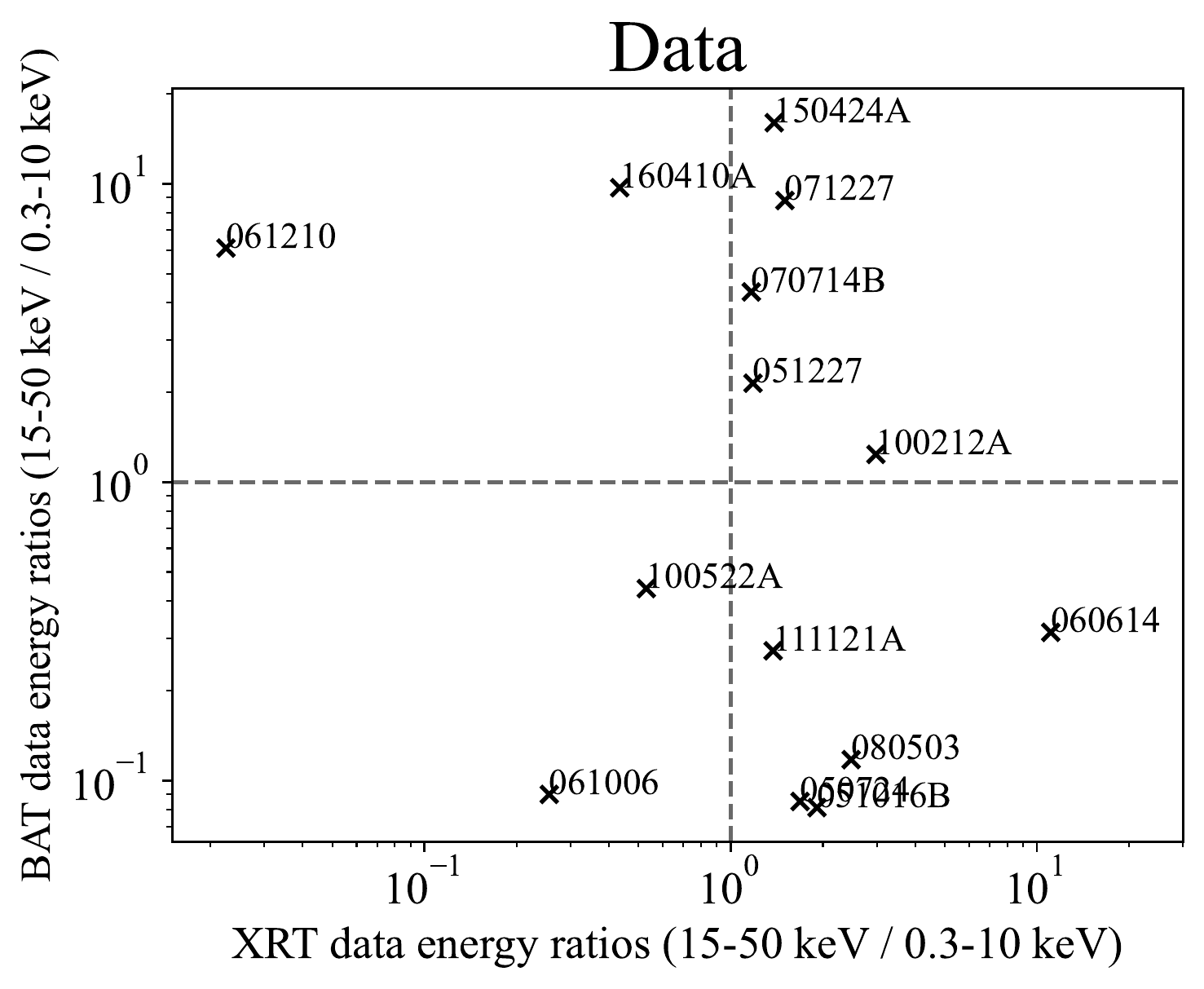}  
    \end{tabular}
    \caption{Ratios of each GRB's total integrated energies in the 15-50 keV extrapolation to in the 0.3-10 keV extrapolation within our model (top figure) and for total observed integrated energies (bottom figure). The modelled luminosity values were interpolated to the same $t$ values as the observed data.}
    \label{fig:energy_plots}
\end{figure}

We also investigated integrating the luminosities of the BAT and XRT data in each band to extract a total energy, as well as integrating the luminosities of the model light-curve interpolated to temporally match the observational data points. We plot the ratios of these results from each band in Fig. \ref{fig:energy_plots} which shows that while the observational data's total energies are highly comparable at 0.3-10 keV and 15-50 keV, the model fits generally skewed towards a greater energy budget at 0.3-10 keV for the XRT data compared to at 15-50 keV, while skewing towards a higher (sometimes substantially, as can be clearly seen in the light-curves) energy budget at 15-50 keV for the BAT data. Given that we've seen from Fig. \ref{fig:broadness_vs_Ndata} that the XRT data are the data driving the fits, we can see that the lower XRT total energies for the model at 15-50 keV may again be indicative of these fits being limited by their global energy budgets, with the issue being chiefly localised to the late-time data.

With generally broad and inconsistent parameter distributions across the two bands, we can state that the model of \citet{Gibson1} appears consistent with the observational data, but is not well constrained by them. The  best-fitting light-curves for each GRB generally fit to the observational data well, albeit sometimes failing to capture substructure, but the quality of the fits do not appear to be especially sensitive to the model's physical parameters due to the larger errors introduced by propagating ECF uncertainties. Further in-depth exploration of the physical parameters and the implications of their distributions may be unfeasible with ECF errors propagated.

\subsection{Data variability}

It should be noted that this propagation of ECF errors which are not necessarily independent of one another may wash out variability between the observational data, just as the lack of propagation can artificially inflate it. Our work in this way provides a maximum constraint on the uncertainties in the bolometric luminosities. Many large, statistically significant features of the observed light-curve do remain visible in both the 0.3-10 keV band and the 15-50 keV band even when we apply this propagation, such as the late-time flare of GRB050724 and the rebrightening in GRB100522A, so it is clear that the error propagation will not wash out all substructure outside of the continuum. Even if we only considered the native bandpass data with their reduced volatility, we here also see clear rebrightenings for other GRBs such as GRB100212A, though some smaller substructures in our GRB sample may still be artefacts of small deviations from the mean in the spectral hardness of the data. This inflation will however risk artificially increasing the uncertainty on our parameters.

A potential method to mitigate the issues inherent to our approach and those of previous works would be to alter the approach taken to fitting the data; instead of modifying the format of the observational data and propagating measurement uncertainties appropriately, the model lightcurves could be processed to take the same format as the data. The model yields a bolometric luminosity which can be converted into predicted photon count-rates if we assume a spectral shape and know the characterisation of \textit{Swift}'s instruments. These can then be passed through the \textit{Swift} pipeline as the observed data were, allowing for a like-for-like comparison which would circumvent the semi-systematic issues with the current approach when attempting to use non-independent data points. The computational demand of binning the model data for every configuration of parameters in the MCMC render this outside the scope of this work, however.

\section{Conclusions}

In this work, we have added error propagation on the count-rate-to-flux conversion factor to the flux measurements presented by the UK \textit{Swift} Science Data Centre. With this, we have reanalysed the results of \citet{Gibson1} with observational data extrapolated to a 0.3-10 keV band and a 15-50 keV band, and demonstrated that the derived results of fitting a magnetic propeller model to a sample of SGRBEEs are highly sensitive to the way in which the observational data are processed and propagated. There is therefore a need for works using UKSSDC data to be explicit about the way in which the data are processed to contextualise the behaviour of the observational data. The distributions of physical parameters, and also of global energy budgets, derived from models fitted to this data are also highly sensitive to the data processing methodology, which may have implications for the viability of the magnetic propeller model examined in this work and other, similar, model fits. The potential inconsistencies when fitting to the light-curves presented under these two different bandpasses may provide a benefit in helping to distinguish real features in the observed data from spurious ones. Our work highlights the potential pitfalls of extrapolating observational data to outside of their native bandpasses; even when comparing data from different \textit{Swift} instruments, it is most robust to keep data in their native bandpass as much as possible.

\section{Acknowledgements}

The authors thank the referees for their helpful comments. TRLM would like to acknowledge support from an STFC studentship. PAE acknowledges UKSA support. This work made use of data supplied by the UK Swift Science Data Centre at the University of Leicester. This research used the ALICE High Performance Computing Facility at the University of Leicester.

\section{Data Availability Statement}

No new data were generated in this research. Following the analysis performed in this work, the \textit{Swift} data processing package for Python \textit{swifttools}\footnote{\url{https://pypi.org/project/swifttools/}} now includes the option to automatically incorporate the ECF error propagation method that we have performed here.




\bibliographystyle{mnras}
\bibliography{mnras_template} 





\bsp	
\label{lastpage}
\end{document}